\documentclass[onecolumn]{IEEEtran} 
\usepackage{amsthm,amsmath,amssymb,amsfonts,bm}
\usepackage{mathtools}
\usepackage{tikz}
\usepackage{booktabs,multirow,nicematrix}
\usepackage[table,xcdraw]{xcolor}
\usepackage{adjustbox}
\usepackage{longtable,rotating}
\usepackage[ruled,linesnumbered]{algorithm2e}
\usepackage[shortlabels]{enumitem}
\usepackage{float,graphicx}
\usepackage[square, comma, sort&compress, numbers]{natbib}
\usepackage[colorlinks,linkcolor=blue]{hyperref} 

\newcommand{\rank}{{\mathrm{rank}}}
\newcommand{\wt}{{\mathrm{wt}}}

\newcommand{\F}{\mathbb{F}}

\renewcommand{\eqref}[1]{\textcolor{blue}{(\ref{#1})}}
\newcommand{\zero}{\boldsymbol{0}} 
 
\newcommand{\M}{\boldsymbol{M}}

\newcommand{\w}{\boldsymbol{w}}
\renewcommand{\v}{\boldsymbol{v}} 
\newcommand{\GRS}{\mathrm{GRS}} 

\newcommand{\A}{{\boldsymbol{A}}} 
\renewcommand{\L}{{\mathcal{L}}}
\renewcommand{\P}{{\mathcal{P}}}
\newcommand{\B}{{\boldsymbol{B}}} 
 \newcommand{\G}{{\boldsymbol{G}}}
\renewcommand{\H}{{\boldsymbol{H}}}
\newcommand{\C}{{\mathcal{C}}}

\renewcommand{\c}{\boldsymbol{c}}

\newtheorem{theorem}{Theorem}[section]
\newtheorem{lemma}[theorem]{Lemma}

\newtheorem{definition}[theorem]{Definition}
\newtheorem{example}[theorem]{Example}
\newtheorem{remark}[theorem]{Remark}
\newtheorem{openproblem}[theorem]{Open Problem}  

\allowdisplaybreaks
\begin{document}
	\title{Parity-check matrices and their MDS or NMDS properties for several classes of non-GRS type linear codes}
	\author{Zhonghao Liang, Chenlu Jia, Qunying Liao \thanks{Corresponding author: Qunying Liao. Emails: liangzhongh0807@163.com;chenluJiamath@163.com; qunyingliao@sicnu.edu.cn.}
		\thanks{
			Zhonghao Liang, Chenlu Jia and Qunying Liao are with College of Mathematical Sciences, Sichuan Normal University, Chengdu 610066, China.} 
		\thanks{This paper is supported by National Natural Science Foundation of China (12471494) and Natural Science Foundation of Sichuan Province (2024NS-
			FSC2051).}
	}
	
	\maketitle

\begin{abstract}
	Parity-check matrices and weight distributions are important for understanding the structure and performance of linear codes. However, obtaining explicit expressions for them is often challenging. In recent years, numerous families of non-GRS type linear codes have been constructed, motivated by applications in coding theory and cryptography. For many of these codes, explicit parity-check matrices remain unavailable, hindering further investigation of their duality properties, decoding methods, and weight distributions. In this paper, we first develop a method for constructing parity-check matrices using code extensions and Cramer's rule, and then apply this method or through direct constructions to obtain explicit parity-check matrices for column-twisted generalized Reed--Solomon codes (TGRS), modified GRS codes and so on. Second, we introduce the $(\L,\P)$-twisted generalized Roth--Lempel (TGRL) code $\C_{k,\ell}(\L,\P,\A_{m\times m},\B)$, obtain explicit parity-check matrices for three subclasses through direct constructions, and establish a necessary and sufficient condition for $\C_{k,\ell}(\L,\P,\A_{2\times 2},\B)$ with the most general $\L$, $\P$, and $\B$ to be maximum distance separable (MDS). For one-column TGRS codes, we prove that they are either MDS or near MDS (NMDS), furthermore, establish a necessary and sufficient condition for them to be NMDS, and determine the weight distribution of a family of these NMDS codes using subset sums in finite abelian groups. Finally, basing on known results for non-GRS properties of TGRS codes, we establish sufficient conditions for these codes considered in this paper to be non-GRS from the perspective of punctured codes.
\end{abstract}

  
	\section{Introduction} 
Let $\mathbb{F}_q$ be the $q$ elements finite field and $\mathbb{F}_{q}^{*}=\mathbb{F}_{q}\backslash\left\{0\right\}$, where $q$ is a prime power. A $k$-dimensional linear subspace of $\mathbb{F}_{q}^{n}$ with minimum (Hamming) distance $d$ is called an $[n, k,d]$ linear code over $\mathbb{F}_{q}$. For an $[n, k]_{q}$ linear code $\mathcal{C}$, if $n$ and $k$ are given, then the maximum value of the minimum (Hamming) distance $d$ is determined, i.e, $d\leq n-k+1$. Especially, if $d=n-k+1$, then $\mathcal{C}$ is maximum distance separable (in short, MDS); if $d=n-k$, then $\C$ is almost MDS (in short, AMDS); if the dual code of $\C$ is also AMDS, then $\C$ is near MDS (in short, NMDS). Let $A_i$ denote the number of codewords of weight $i$ in $\mathcal{C}$. The polynomial $\sum\limits_{i=0}^{n}A_{i}z^{i}$ is called the weight enumerator of $\mathcal{C}$, and the sequence $(A_0, A_1, ...,A_n)$
is referred to as the weight distribution of $\mathcal{C}$. Since the weight distribution contains some significant information including the error-correcting capability and error-detection probability, the weight distribution
of linear codes has always been one of hot topics in theory and practice in recent years \cite{Zhu2024extendedTGRS,ZhuSX2024NMDS,KaiextendNMDS,FanCLNMDS2024}.

Parity-check matrices play a central role in linear coding theory. They generate the dual code and provide the standard framework for syndrome decoding. More importantly, their structural complexity is closely linked to the difficulty of characterizing key properties of linear codes, including self-duality, self-orthogonality, linear complementary dual property, and minimum distance, as well as the decoding complexity. 

The generalized Reed-Solomon (in short, GRS) code, as a well-known class of MDS codes, is defined as
\begin{equation}\label{GRScodewords}
\mathrm{GRS}_{k}(\boldsymbol{\alpha})\triangleq\left\{\left(v_{1}f\left(\alpha_{1}\right), \ldots,v_{n}f\left(\alpha_{n}\right)\right) | f(x) \in \mathbb{F}_{q}^{k}[x]\right\},
\end{equation}
where $ \boldsymbol{v}=\left(v_{1},\ldots,v_{n}\right)\in\left(\mathbb{F}_{q}^{*}\right)^{n}$, $\boldsymbol{\alpha}=\left(\alpha_{1}, \ldots, \alpha_{n}\right) \in \mathbb{F}_{q}^{n}$ with $\alpha_{i} \neq \alpha_{j}(i \neq j)$, and 
$$\mathbb{F}_{q}^{k}[x]=\left\{f(x)=\sum_{i=0}^{k-1} f_{i} x^{i}\mid f_{i} \in \mathbb{F}_{q}, 0 \leq i \leq k-1\right\}.$$ A non-GRS linear code refers to a linear code inequivalent to any GRS code. 

In recent years, motivated by the fundamental problem of MDS codes classification and their cryptographic applications such as resistance to the Sidelnikov–Shestakov and Wieschebrink attacks, the construction of non-GRS type linear codes has attracted considerable attention\cite{ChennonRSAG,LiangGRL,Hu2025LPTGRS,Liu2026ColumnTGRS,Bhagat2026RowColumnTGRS,WanFDGRS,Ma2026extendedTGRS,Cheng2026TGRL,ZhangJHRS,WangMGRS2026,LiuHWnonRS,LiFnonRScyclic,JinLFnonRS2026,ZhuSXDRL,LiY2025extendedTGRS,LuoJQnonRScurves,LiuSnonRS2026tit,LiY2025TcomnonGRS,LiY2026nonGRSdeephole,Zhang2026extendedTGRS,WangYanTJTGRL2026}
. To date, the MDS property of these non-GRS type linear codes has been thoroughly characterized. Nevertheless, much less attention has been paid to their other structural properties such as self-duality, self-orthogonality, the LCD property and so on, owing to the absence of explicit parity-check matrix constructions. So far, non-GRS type linear codes  are mainly constructed by performing the following three types of operations on the generator matrix of GRS codes. 
\begin{itemize}
\item {\bf Add the column vector.} In 1989, Roth and Lempel \cite{RL1989} constructed the first class of non-GRS type MDS codes by adding two special column vectors to the generator matrix of RS codes, i.e., changed \begin{equation}\label{GRSf(x)}
\left(v_{1}f\left(\alpha_{1}\right), \ldots,v_{n}f\left(\alpha_{n}\right)\right), f(x) \in \mathbb{F}_{q}^{k}[x]
\end{equation} 
to be $$\left(v_{1}f\left(\alpha_{1}\right), \ldots,v_{n}f\left(\alpha_{n}\right),f_{k-1} ,f_{k-2}+\delta f_{k-1} \right), f(x) \in \mathbb{F}_{q}^{k}[x].$$ 
In 2025, Li et al. \cite{LiFWGRL2025} generalized the construction of Roth and Lempel. In 2026, Liu et al. \cite{Liu2026ColumnTGRS} constructed a class of non-GRS type linear codes by adding two column vectors, which differs from those used by Roth and Lempel, to the generator matrix of RS codes, i.e., changed Equation \eqref{GRSf(x)} to be
\begin{equation}\label{columTGRSf(x)}
\left(v_{1}f\left(\alpha_{1}\right), \ldots,v_{n}f\left(\alpha_{n}\right),f(b)-\lambda_1f(c),\ldots,f(b)-\lambda_sf(c)\right), s\in\left\{1,2\right\},f(x) \in \mathbb{F}_{q}^{k}[x],
\end{equation} established a necessary and
sufficient condition
for the corresponding code to be MDS and gave a complex parity-check matrix;
Recently, Gu et al., \cite{GuchangCTGRS2026} generalized the construction of Liu et al. \cite{Liu2026ColumnTGRS},  i.e., changed Equation \eqref{GRSf(x)} to be 
\begin{equation}\label{changecolumTGRSf(x)}
	\left(v_{1}f\left(\alpha_{1}\right), \ldots,v_{n}f\left(\alpha_{n}\right),f(b_1)-\lambda_1f(c),f(b_2)-\lambda_2f(c)\right), f(x) \in \mathbb{F}_{q}^{k}[x],
\end{equation}
gave a necessary and sufficient condition for the code to be MDS and the parity-check matrix.
Additionally, Wang et al. \cite{WangMGRS2026} introduced a class of non-GRS type linear codes and its extended codes by locally modifying column vectors, i.e., changed Equation \eqref{GRSf(x)} to be 
$$\left(v_{1}f\left(\alpha_{1}\right), \ldots,v_{n-1}f\left(\alpha_{n-1}\right),v_n\left(f(0)+\eta f_t\right)\right), f(x) \in \mathbb{F}_{q}^{k}[x].$$
\item {\bf Perform twisting operations.} In 2017, Beelen et al. \cite{BeelenTGRS2017} introduced twisted GRS (in short, TGRS) codes to construct non-GRS MDS codes by applying a twisting operation to the generator matrix of RS codes, i.e., changed Equation \eqref{GRSf(x)} to be $$\left(v_{1}f\left(\alpha_{1}\right), \ldots,v_{n}f\left(\alpha_{n}\right)\right),
f(x) \in \mathcal{V}_{k,t,h,\eta}=\left\{f(x)=\sum\limits_{i=0}^{k-1} f_{i} x^{i}+\eta f_{h} x^{k-1+t}\mid f_{i} \in \mathbb{F}_{q}\right\}.$$
After that, scholars generalized $\mathcal{V}_{k,t,h,\eta}$ and investigated the properties of the corresponding linear codes\cite{Hu2025LPTGRS,LaoTGRShull2026,ZhuCZFFATGRS,ZhuRTGRSdeconding,WangTGRSdecoding}. 

\item {\bf Add the column vector and perform twisting operations,  simultaneously.} In 2025, Li et al. \cite{LiY2025extendedTGRS} introduced two classes of non-GRS MDS codes by changing Equation \eqref{GRSf(x)} to be    
$$\left(v_{1}f\left(\alpha_{1}\right), \ldots,v_{n}f\left(\alpha_{n}\right), \eta f_{h}\right),f(x) \in \mathcal{V}_{k,1,h,\eta}$$
or
$$\left(v_{1}f\left(\alpha_{1}\right), \ldots,v_{n}f\left(\alpha_{n}\right), f_{k-1}\right),f(x) \in \mathcal{V}_{k,1,h,\eta},$$ and focused their non-GRS MDS properties, covering radii, and
deep holes. In 2026, Cheng et al. \cite{Cheng2026TGRL} and Ma et al. \cite{Ma2026extendedTGRS}, independently constructed a class of non-GRS MDS codes by changing Equation \eqref{GRSf(x)} to be    
$$\left(v_{1}f\left(\alpha_{1}\right), \ldots,v_{n}f\left(\alpha_{n}\right), f_{k-1}, f_{k-2}+\delta f_{k-1} \right),f(x) \in \mathcal{V}_{k,1,h,\eta};$$ Bhagat et al. \cite{Bhagat2026RowColumnTGRS} introduced a class of non-GRS MDS codes by generalizing the construction of Liu eta al. \cite{Liu2026ColumnTGRS},  i.e., changed Equation  \eqref{columTGRSf(x)} to   be   
$$\left(v_{1}f\left(\alpha_{1}\right), \ldots,v_{n}f\left(\alpha_{n}\right),f(b)-\lambda_1f(c)\right),f(x) \in \mathcal{V}_{k,1,h,\eta};$$ 
Recently, Gu et al., \cite{GuchangCTGRS2026} generalized the construction of Bhagat et al. \cite{Bhagat2026RowColumnTGRS},  i.e., changed $f(x)$ of Equation \eqref{changecolumTGRSf(x)} to be $\mathcal{V}_{k,1,h,\eta}$, gave a necessary and sufficient condition for the code to be MDS and the parity-check matrix. Zhang et al.\cite{Zhang2026extendedTGRS} introduced two classes of non-GRS MDS codes by changing Equation \eqref{GRSf(x)} to be    
$$\left(v_{1}f\left(\alpha_{1}\right), \ldots,v_{n}f\left(\alpha_{n}\right), f_{h}\right),f(x) \in \mathcal{V}_{k,2,h,\eta},0\leq h\leq k-2,$$
or
$$\left(v_{1}f\left(\alpha_{1}\right), \ldots,v_{n}f\left(\alpha_{n}\right), f_{h},f_{h}+\delta f_{k-1} \right),f(x) \in \mathcal{V}_{k,2,h,\eta},0\leq h\leq k-2;$$
Lei et al.\cite{LeiTGRL2026} introduced a class of non-GRS codes by changing Equation \eqref{GRSf(x)} to be    
$$\left(v_{1}f\left(\alpha_{1}\right), \ldots,v_{n}f\left(\alpha_{n}\right), f_{k-1},f_{k-2}+\delta f_{k-1}\right),f(x) \in \mathcal{V}_{k,1,h,\eta},0\leq h\leq k-1,$$
investigated the minimum distance and gave a necessary and
sufficient condition for such code to be MDS or NMDS. Wang et al. \cite{WangYanTJTGRL2026} introduced two classes of non-GRS MDS codes by changing Equation \eqref{GRSf(x)} to be    
$$\left(v_{1}f\left(\alpha_{1}\right), \ldots,v_{n}f\left(\alpha_{n}\right), f_{k-1},f_{k-2} \right),f(x) \in \mathcal{V}_{k,3,h,\eta}$$
or
$$\left(v_{1}f\left(\alpha_{1}\right), \ldots,v_{n}f\left(\alpha_{n}\right), f_{k-1},f_{k-2},f_{k-3}+\delta f_{k-1} \right),f(x) \in \mathcal{V}_{k,3,h,\eta}.$$   
\end{itemize}

It is worth noting that, among the non-GRS-type linear codes mentioned above, only a few codes have explicit parity-check matrices, and the related works are summarized in the following Table \ref{summary1ofH}.
\begin{table}[H]\label{summary1ofH}
\centering
\caption{Research progress on the parity-check matrix of non-GRS linear codes}
\begin{tabular}{|c|c|c|c|}
\hline
References for codes&  Parity-check matrix& Codes and References& Parity-check matrix\\
	\hline
\cite{LiFWGRL2025}&\checkmark  &\cite{Cheng2026TGRL}& \checkmark\\
	\hline
The first class of \cite{Liu2026ColumnTGRS}&$\times$  &The first class of \cite{LiY2025extendedTGRS}&$\times$  \\
\hline
The second class of \cite{Liu2026ColumnTGRS}&\checkmark  &The second class of \cite{LiY2025extendedTGRS}&$\times$  \\
	\hline
\cite{GuchangCTGRS2026}& \checkmark &\cite{Bhagat2026RowColumnTGRS}&$\times$\\
	\hline
\cite{WangMGRS2026}&$\times$ &\cite{LeiTGRL2026}& $\times$ \\
	\hline
The first class of \cite{Zhang2026extendedTGRS}& $\times$ &The first class of \cite{WangYanTJTGRL2026}&$\times$\\
	\hline
The second class of \cite{Zhang2026extendedTGRS}& $\times$ &The second class of \cite{WangYanTJTGRL2026}&$\times$\\
	\hline
\cite{Ma2026extendedTGRS}&$\times$&\cite{Hu2025LPTGRS}& \checkmark\\
\hline	
	\end{tabular}
\end{table} 

In this paper, we mainly provide the corresponding parity-check matrices for the known non-GRS type linear codes. Specifically, firstly, we introduced the $(\L,\P)$-twisted generalized Roth-Lempel (in short, $(\L,\P)$-TGRL) code, which generalize the definitions of linear codes given by Cheng et al. \cite{Cheng2026TGRL}, Ma et al. \cite{Ma2026extendedTGRS} and Wang et al. \cite{Wang2026TGRL}, give the parity-check matrices for several classes of $(\L,\P)$-TGRL codes, and a necessary and
sufficient condition   for a class of $(\L,\P)$-TGRL codes to be MDS; Secondly, we prove that a class of non-GRS type linear codes introduced by Liu et al. \cite{Liu2026ColumnTGRS} is either MDS or NMDS. Finally, we provide explicit parity-check matrices for numerous non-GRS-type linear codes from two perspectives, i.e., the direct construction and extended codes. 

This paper is organized as follows. In Section \ref{sec2}, we recall and introduce some necessary definitions and lemmas. In Section \ref{sec3}, we present the corresponding parity-check matrix for five classes of non-GRS type linear codes. In Section \ref{sec4}, we recall the definitions of two classes of extended codes and related results, introduce a method to construct a parity-check matrix of non-GRS type linear codes from the generator matrix of the original codes and apply this method to a series of non-GRS type linear codes.  In Section \ref{sec5}, we give a necessary and
sufficient condition for a class of $(\L,\P)$-TGRL codes to be MDS, and by proving $(1)$-column TGRS codes either MDS or NMDS, give a necessary and
sufficient condition for $(1)$-column TGRS codes to be NMDS. In Section \ref{sec6}, we discuss the non-GRS properties for several classes of linear codes  basing on the non-GRS property of $(\L,\P)$-TGRS codes.

\section{Preliminaries}\label{sec2}
Throughout this paper, for the convenience, we fix some notations as the following.
\begin{itemize}
	\item $q$ is a power of an odd prime $p$.
	\item $\F_{q}^{n}$ denote the $n$-dimension vector space over $\F_q$.
	\item $\F_q[x]$ is the polynomial ring over $\F_q$ and $\F_q^k[x]\triangleq\left\{f(x)\in\F_q[x]|\deg(f(x))\leq k-1\right\}$.
\item $\boldsymbol{\alpha}\triangleq\left(\alpha_{1}, \ldots, \alpha_{n}\right) \in \mathbb{F}_{q}^{n}$ with $\alpha_{i} \neq \alpha_{j}(i \neq j)$, $\boldsymbol{v}=\left(v_{1}, \ldots, v_{n}\right) \in\left(\mathbb{F}_{q}^{*}\right)^{n}$.
\item For any positive integer $k$, $[k]\triangleq \left\{0,1,\ldots,k-1\right\}$ and $\# L$ denotes the cardinality of a set $L$.
\item $\zero_{s\times t}$ denotes the $s\times t$ zero matrix, $\boldsymbol{I}_k$ denotes the $k\times k$ identity matrix, and $\boldsymbol{1}$ denotes the all-ones vector.
\item $\mathrm{GL}_{m}(\F_q)$ denotes the set of $m\times m$ invertible matrices over $\F_q$.
\item For a subset $\left\{\alpha_{1}, \ldots, \alpha_{n}\right\}\subseteq\mathbb{F}_{q}$, $u_{i}=\prod\limits_{j=1, j \neq i}^{n}\left(\alpha_{i}-\alpha_{j}\right)^{-1}$ for $1 \leq i \leq n$.
\end{itemize}

In this section, we recall the definitions for several classes of non‑GRS type linear codes. For some of these code families, we extend their definitions and present auxiliary lemmas that will be used throughout the paper.

\subsection{Several classes of non-GRS linear codes}

In this subsection, basing the known non-GRS type linear codes, we recall or introduce the definitions for five classes of linear codes.

The first class of non‑GRS type linear codes, i.e., $\left(\L,\P\right)$-twisted generalized Reed-Solomon codes introduced by Hu et al. in 2026 as follows. 
\begin{definition}\label{LPTGRSdefinition}{\rm(\cite{Hu2025LPTGRS}, Definition 2)}
	Let $n$ and $k$ be integers with $0<k\leq n$, $\B=(b_{i,j})_{k\times n-k}$, $\L\subseteq[n-k]$, $\# \L=\ell+1$ with $0\leq \ell\leq n-k-1$ and $\P\subseteq [k]$. The $(\L,\P)$-twisted generalized Reed-Solomon (in short, $(\L,\P)$-TGRS) code is defined as 
	$$\C\left(\L,\P,\B\right)\triangleq\left\{\left(v_{1}f\left(\alpha_{1}\right), \ldots,v_{n}f\left(\alpha_{n}\right)\right) | f(x) \in \mathcal{F}\left(\L,\P,\B\right)\right\},
	$$
	where  \begin{equation}\label{PolynomialF}
	\mathcal{F}\left(\L,\P,\B\right)=\left\{\sum\limits_{i=0}^{k-1} f_{i} x^{i}+\sum\limits_{i\in\P}f_{i} \sum\limits_{j\in\L}b_{i,j} x^{k+j}|f_{i} \in \F_{q}, 0 \leq i \leq k-1\right\}.
	\end{equation}
And the matrix $\B$ is called the coefficient matrix of $\C\left(\B\right)$.
\end{definition} 
\begin{remark}\label{LPTGRSTGRSG}
	It is easy to know that $\C\left(\L,\P,\B\right)$ with $\L=[\ell+1]$ and $\P=[k]$ has the generator matrix
	$$
	\G\left(\L,\P,\B\right)=\begin{pmatrix}
		v_{1}\left(1+\sum\limits_{j=0}^{\ell}b_{0,j}\alpha_{1}^{k+j}\right)&\cdots&v_{n}\left(1+\sum\limits_{j=0}^{\ell}b_{0,j}\alpha_{n}^{k+j}\right)\\
		v_{1}\left(\alpha_{1}+\sum\limits_{j=0}^{\ell}b_{1,j}\alpha_{1}^{k+j}\right)&\cdots&v_{n}\left(\alpha_{n}+\sum\limits_{j=0}^{\ell}b_{1,j}\alpha_{n}^{k+j}\right)\\ 
		\vdots& &\vdots\\
		v_{1}\left(\alpha_{1}^{k-1}+\sum\limits_{j=0}^{\ell}b_{k-1,j}\alpha_{1}^{k+j}\right)&\cdots&v_{n}\left(\alpha_{n}^{k-1}+\sum\limits_{j=0}^{\ell}b_{k-1,j}\alpha_{n}^{k+j}\right)\\
	\end{pmatrix}_{k\times n}.
	$$
\end{remark}

The second class of non‑GRS type linear codes is constructed from the generator matrix of RS codes by simultaneously performing the twisting operations introduced by Hu et al. in 2026 and the adding operations introduced by Liang et al. in 2026 as follows. 
\begin{definition}\label{LPTGRLdefinition}
Let $n$ and $k$ be integers with $0<m\leq k\leq n$, $\B=(b_{i,j})_{k\times n-k}$, $\L\subseteq[n-k]$, $\# \L=\ell+1$ with $0\leq \ell\leq n-k-1$ and $\P\subseteq [k]$. The $(\L,\P)$-twisted generalized Roth-Lempel (in short, $(\L,\P)$-TGRL) code is defined as
$$
\C_{k,\ell}(\L,\P,\A_{m\times m},\B)\triangleq\left\{\left(v_1 f\left(\alpha_1\right), \ldots, v_n f\left(\alpha_n\right),\boldsymbol{\beta}\right) \mid f(x) \in \mathcal{F}\left(\L,\P,\B\right)\right\}, 
$$
where $\mathcal{F}\left(\L,\P,\B\right)$ is given in \eqref{PolynomialF} and $\A_{m\times m}=(a_{ij})\in\mathrm{GL}_{m}(\F_q)$ and 
$$\begin{aligned}  \boldsymbol{\beta}=&\left(f_{k-m},\ldots,f_{k-1}\right)\A_{m\times m}\\
	=&\left(a_{11}f_{k-m}+a_{21}f_{k-(m-1)}+\cdots+a_{m1}f_{k-1},\ldots,a_{1m}f_{k-m}+a_{2m}f_{k-(m-1)}+\cdots+a_{mm}f_{k-1}\right).
\end{aligned}$$
\end{definition}
\begin{remark}\label{GRLTGRSG}
$(1)$ It is easy to know that $\C_{k,\ell}(\L,\P,\A_{m\times m},\B)$ with $\L=[\ell+1]$ and $\P=[k]$ has the generator matrix
	$$
\G(\L,\P,\A_{m\times m},\B)=\begin{pmatrix}
\G\left(\L,\P,\B\right)&\begin{matrix}
\zero_{(k-m)\times m}\\
\A_{m\times m}
\end{matrix}
	\end{pmatrix}_{k\times (n+m)}.
	$$
	
$(2)$ By taking $m=2$, $\ell=0$, $\A_{2\times 2}=\begin{pmatrix}
	0&1\\
	1&\delta
\end{pmatrix}$ and $\B=\begin{pmatrix}
0&0&\cdots&0\\
\vdots&\vdots& &\vdots\\
0&0&\cdots&0\\
b_{k-1,0}&0&\cdots&0
\end{pmatrix}$ in Definition \ref{LPTGRLdefinition}, the corresponding $\C_{k,\ell}(\L,\P,\A_{m\times m},\B)$ is just the code independently introduced by Cheng et al. \cite{Cheng2026TGRL} and Ma et al \cite{Ma2026extendedTGRS}.
 
$(3)$ By taking $m=3$, $\A_{3\times 3}=\begin{pmatrix}
	0&0&1\\
	0&1&0\\
	1&0&\delta
\end{pmatrix}$ and $\B=\begin{pmatrix}
0&0&0&0&\cdots&0\\
\vdots&\vdots&\vdots&\vdots& &\vdots\\
0&0&0&0&\cdots&0\\
0&0&b_{k-1,2}&0&\cdots&0
\end{pmatrix}$ in Definition \ref{LPTGRLdefinition}, the corresponding $\C_{k,\ell}(\L,\P,\A_{m\times m},\B)$ is just the code introduced by Wang et al \cite{Wang2026TGRL}.
\end{remark} 

The third class of non-GRS linear codes is constructed by adding a column vector with exactly one nonzero entry for $\C\left(\L,\P,\B\right)$ with $\L=[\ell+1]$ and $\P=\left\{h\right\}$ as follows. 
\begin{definition}\label{extendedLPTGRSdefinition} 
	Let $n$, $k$, $\ell$ and $h$ be integers with $0<k\leq n$, $0\leq \ell\leq n-k-1$ and $0\leq h\leq k-1$. Let $\boldsymbol{b}=\left(b_0,b_1,\ldots,b_{\ell}\right)\in\F_{q}^{\ell+1}\backslash\left\{\zero_{\ell+1}\right\}$ and $\eta\in\F_q^*$. A class of the extended code for $\C\left(\L,\P,\B\right)$ with $\L=[\ell+1]$ and $\P=\left\{h\right\}$ is defined as 
	$$\C_{h}\left(\eta,\boldsymbol{b}\right)\triangleq\left\{\left(v_{1}f\left(\alpha_{1}\right), \ldots,v_{n}f\left(\alpha_{n}\right),\eta f_{h}\right) | f(x) \in \mathcal{W}\right\},
	$$
	where
\begin{equation}\label{PolynomialW}	\mathcal{W}=\left\{\sum\limits_{i=0}^{k-1} f_{i} x^{i}+f_{h}\sum\limits_{j=0}^{\ell}b_{j}x^{k+j}|f_{i} \in \mathbb{F}_{q}, 0 \leq i \leq k-1,\right\}.
\end{equation}
\end{definition}
\begin{remark}\label{onetwistextendedTGRSG}
$(1)$ It is easy to know that $\C_{h}\left(\eta,\boldsymbol{b}\right)$ has the generator matrix
	$$
	\boldsymbol{G}_{h}(\eta,\boldsymbol{b})=\begin{pmatrix}
\G\left(\B\right)&\boldsymbol{\eta}
	\end{pmatrix}_{k\times (n+1)}
$$
with $\B=\begin{pmatrix}
\zero_{h\times (\ell+1)}&\zero_{h\times (n-k-\ell-1)}\\
\boldsymbol{b}&\zero_{1\times (n-k-\ell-1)}\\
\zero_{(k-h-1)\times (\ell+1)}&\zero_{(k-h-1)\times (n-k-\ell-1)}
\end{pmatrix}$ and $\boldsymbol{\eta}=\left(0,\cdots,0,\eta,0,\cdots,0\right)^T$.

$(2)$ By taking $\ell=0$, $h=k-1$ and $\eta=1$, the corresponding $\C_{h}\left(\boldsymbol{\alpha},\eta,\boldsymbol{b}\right)$ is just the code introduced by Zhu and Liao \cite{Zhu2024extendedTGRS}.

$(3)$ By taking $\ell=0$ and $\eta=b_0\in\F_q^*$, or $\ell=1$, $b_0=0$, $b_1\in\F_q^*$ and $\eta=1$, the corresponding $\C_{h}\left(\boldsymbol{\alpha},\eta,\boldsymbol{b}\right)$ is just the code introduced by Li et al. \cite{LiY2025extendedTGRS} or Zhang et al. \cite{Zhang2026extendedTGRS}, respectively.
\end{remark} 

Furthermore, by adding another column vector with exactly two nonzero entries for $\C_{h}\left(\boldsymbol{\alpha},\eta,\boldsymbol{b}\right)$, we introduce the extended $\C_{h}\left(\eta,\boldsymbol{b},\delta\right)$ as follows.
\begin{definition}\label{twoextendedLPTGRSdefinition}
	Let $n$, $k$, $\ell$ and $h$ be integers with $0<k\leq n$, $0\leq \ell\leq n-k-1$ and $0\leq h\leq k-2$. Let $\boldsymbol{b}=\left(b_0,b_1,\ldots,b_{\ell}\right)\in\F_{q}^{\ell+1}\backslash\left\{\zero_{\ell+1}\right\}$, $\eta\in\F_q^*$ and $\delta\in\F_q$. A class of the twice extended code for $\C\left(\L,\P,\B\right)$ with $\L=[\ell+1]$ and $\P=\left\{h\right\}$ is defined as 
	$$\C_{h}\left(\eta,\boldsymbol{b},\delta\right)\triangleq\left\{\left(v_{1}f\left(\alpha_{1}\right), \ldots,v_{n}f\left(\alpha_{n}\right),\eta f_{h},\eta f_{h}+\delta f_{k-1}\right) | f(x) \in \mathcal{W}\right\},
	$$
	where $\mathcal{W}$ is given in \eqref{PolynomialW}.
\end{definition}
\begin{remark}\label{twotwistextendedTGRSG}
$(1)$ It is easy to know that $\C_{h}\left(\eta,\boldsymbol{b},\delta\right)$ has the generator matrix
	$$
	\boldsymbol{G}_{h}(\eta,\boldsymbol{b},\delta)=\begin{pmatrix}
\boldsymbol{G}_{h}(\eta,\boldsymbol{b})&\boldsymbol{\delta}
	\end{pmatrix}_{k\times (n+2)}
	$$
with $\boldsymbol{\delta}=\left(0,\ldots,0,\eta,0,\ldots,0,\delta\right)^T$.
  
$(2)$ When $h=k-2$ and $\delta\in\F_q^*$, the codes $\C_{h}\left(\eta,\boldsymbol{b},\delta\right)$ and $\C_{k,\ell}(\L,\P,\A_{2\times 2},\B)$ with $\A_{2\times 2}=\begin{pmatrix}
	\eta&\eta\\
	\delta&0
\end{pmatrix}$ and $$\B=\begin{pmatrix}
\zero_{h\times (\ell+1)}&\zero_{h\times (n-k-\ell-1)}\\
\boldsymbol{b}&\zero_{1\times (n-k-\ell-1)}\\
\zero_{(k-h-1)\times (\ell+1)}&\zero_{(k-h-1)\times (n-k-\ell-1)}
\end{pmatrix}$$ are permutation equivalent to each other.
\end{remark}

The fourth class of non-GRS type linear codes, i.e., $(s)$-column TGRS codes, which is the generation of the code introduced by Liu et al. \cite{Liu2026ColumnTGRS} and Gu et al. \cite{GuchangCTGRS2026} as follows.
\begin{definition}\label{columnTGRSdefinition} 
	Let $n$ and $k$ be integers with $0<k\leq n$. Let $\boldsymbol{b}=\left(b_1,\ldots,b_{s}\right)\in\F_{q}^{s}$ with $b_i\neq c$ and $\boldsymbol{\lambda}=\left(\lambda_{1}, \ldots, \lambda_{s}\right) \in \mathbb{F}_{q}^{s}$ with $\lambda_{i} \neq \lambda_{j}(i \neq j)$. The $(s)$-column TGRS code is defined as 
	$$\C_{s}\left(\boldsymbol{\lambda},\boldsymbol{b},c\right)\triangleq\left\{\left(v_{1}f\left(\alpha_{1}\right), \ldots,v_{n}f\left(\alpha_{n}\right),f(b_1)-\lambda_1f(c),\ldots,f(b_s)-\lambda_sf(c)\right) | f(x) \in \F_{q}^k[x]\right\},
	$$ 
\end{definition}
\begin{remark}
$(1)$ It is easy to know that $\C_{s}\left(\boldsymbol{\lambda},b,c\right)$ has the generator matrix
$$
\G_{s}(\boldsymbol{\lambda},\boldsymbol{b},c)=\begin{pmatrix}
	\GRS_{k}(\boldsymbol{\alpha},\v)&\boldsymbol{\lambda}_1&\cdots&\boldsymbol{\lambda}_s 
\end{pmatrix}_{k\times (n+s)}
$$
with $\boldsymbol{\lambda}_i=\left(1-\lambda_{i},b_{i}-\lambda_{i}c,\ldots,b_{i}^{k-1}-\lambda_{i}c^{k-1}\right)^T$ and $\GRS_{k}(\boldsymbol{\alpha},\v)$ is the generator matrix of an $[n,k]$ GRS code with evaluation points $\boldsymbol{\alpha}$. 

$(2)$ By taking $s=2$, $b_1=b_2=b$ and   $\boldsymbol{v}=\boldsymbol{1}$ in Definition \ref{columnTGRSdefinition}, and $\alpha_{1},\ldots,\alpha_{n},b, c$ are pairwise distinct, the corresponding $\C_{s}\left(\boldsymbol{\lambda},\boldsymbol{b},c\right)$ is just the code introduced in \cite{Liu2026ColumnTGRS}. By taking $s=2$ and   $\boldsymbol{v}=\boldsymbol{1}$ in Definition \ref{columnTGRSdefinition}, the corresponding $\C_{s}\left(\boldsymbol{\lambda},\boldsymbol{b},c\right)$ is just Definition 2.3 in \cite{GuchangCTGRS2026}. 
\end{remark}

The fifth class of non-GRS type linear codes is constructed by performing a twisted operation on $(s)$-column TGRS codes which is the generation of the code introduced by Bhagat et al. \cite{Bhagat2026RowColumnTGRS} and Gu et al. \cite{GuchangCTGRS2026} as follows.   
\begin{definition}\label{rowcolumnTGRSdefinition} 
Let $n$, $k$ and $\ell$ be integers with $0<k\leq n$ and $0\leq \ell\leq n-k-1$, $\B=(b_{i,j})_{k\times n-k}$, $\L\subseteq\left[n-k\right]$ and $\P\subseteq[k]$. Let $c\in\F_q$, $\boldsymbol{b}=\left(b_1,\ldots,b_{s}\right)\in\F_{q}^{s}$ with $b_i\neq c$ and   $\boldsymbol{\lambda}=\left(\lambda_{1}, \ldots, \lambda_{s}\right) \in \F_{q}^{s}$ with $\lambda_{i} \neq \lambda_{j}(i \neq j)$. The $(\L,\P, s)$-row column TGRS codes code is defined as 
	$$\C_{s}\left(\boldsymbol{\lambda},\boldsymbol{b},c,\L,\P,\B\right)\triangleq\left\{\left(v_{1}f\left(\alpha_{1}\right), \ldots,v_{n}f\left(\alpha_{n}\right),f(b_1)-\lambda_1f(c),\ldots,,f(b_s)-\lambda_sf(c)\right) | f(x) \in\mathcal{F}\left(\L,\P,\boldsymbol{B}\right)\right\}, 
	$$
	where $\mathcal{F}\left(\L,\P,\boldsymbol{B}\right)$ is given in \eqref{PolynomialF}.
\end{definition}
\begin{remark}
	$(1)$ It is easy to know that $\C_{s}\left(\boldsymbol{\lambda},\boldsymbol{b},c,\L,\P,\B\right)$ with $\L=[\ell+1]$ and $\P=[k]$ has the generator matrix
	$$
	\boldsymbol{G}_{s}(\boldsymbol{\lambda},\boldsymbol{b},c,\L,\P,\B)=\begin{pmatrix}
		\boldsymbol{G}\left(\L,\P,\boldsymbol{B}\right)&\boldsymbol{\mu}_1&\cdots&\boldsymbol{\mu}_s 
	\end{pmatrix}_{k\times (n+s)}
	$$
	with $$\begin{aligned}
	&\boldsymbol{\mu}_i\\
	=&\left(1-\lambda_{i}+\sum\limits_{s=0}^{\ell}b_{0,s}\left(b_{i}^{k+s}-\lambda_{i}c^{k+s}\right),b_{i}-\lambda_{i}c+\sum\limits_{s=0}^{\ell}b_{1,s}\left(b_{i}^{k+s}-\lambda_{i}c^{k+s}\right),\ldots,b_{i}^{k-1}-\lambda_{i}c^{k-1}+\sum\limits_{s=0}^{\ell}b_{k-1,s}\left(b_{i}^{k+s}-\lambda_{i}c^{k+s}\right)\right)^T.
	\end{aligned}$$
	
$(2)$ By taking $s=1$, $\ell=t-1$, $\L=[t]$, $\P=\left\{h\right\}$ and $\boldsymbol{v}=\boldsymbol{1}$ with $1\leq t\leq n-k$, $b_{h,t-1}\in\F_q^*$ and $$\B=\begin{pmatrix}
		\zero_{h\times (t-1)}&\zero_{h\times 1}&\zero_{h\times (n-k-t)}\\
		\zero_{1\times (t-1)}&b_{h,t-1}&\zero_{1\times(n-k-t)}\\
		\zero_{(k-h-1)\times (t-1)}&\zero_{(k-h-1)\times 1}&\zero_{(k-h-1)\times (n-k-t)}
	\end{pmatrix}$$ in Definition \ref{rowcolumnTGRSdefinition}, the corresponding $(\L,\P,s)$-row column TGRS codes is just the row column TGRS code introduced in \cite{Bhagat2026RowColumnTGRS}.
	
$(3)$ By taking $s=2$, $\boldsymbol{v}=\boldsymbol{1}$, $\ell=0$,
$\L=\{0\}$ and $\P=\{h\}$, with $b_{h,0}\in\F_q^*$
as the only nonzero entry of $\B$, gives the corresponding
row-column TGRS code in
\cite[Definition~2.3]{GuchangCTGRS2026}.
\end{remark}

The sixth class of non-GRS type linear codes, i.e., modified generalized Reed-Solomon codes introduced by Wang et al. \cite{WangMGRS2026} as follows.
	\begin{definition}\label{MGRSdefinition}{\rm( \cite{WangMGRS2026}, Definition 3)}
	Let $2\leq k\leq n\leq q$, $1\leq t\leq k-1$, $\eta\in\F_q$, $\boldsymbol{\alpha}=\left(\alpha_{1}, \ldots, \alpha_{n}\right) \in \mathbb{F}_{q}^{n}$ with $\alpha_{i} \neq \alpha_{j}(i \neq j)$ and $\boldsymbol{v}=$ $\left(v_{1}, \ldots, v_{n}\right) \in\left(\mathbb{F}_{q}^{*}\right)^{n}$. The modified generalized Reed-Solomon (in short, MGRS) code $\mathrm{MGRS}_{n,k}\left(\boldsymbol{\alpha},\boldsymbol{v},\eta,t\right)$ is defined as
	$$
	\mathrm{MGRS}_{n,k}\left(\boldsymbol{\alpha},\boldsymbol{v},\eta,t\right)\triangleq\left\{\left(v_{1} f\left(\alpha_{1}\right), \ldots, v_{n-1} f\left(\alpha_{n-1}\right),v_{n}\left(f\left(0\right)+\eta f_{t}\right)\right) | f(x) \in \mathbb{F}_{q}^{k}[x]\right\},
	$$
	where $f_t$ denotes the coefficient of $x^{t}$ in $f(x)$.
\end{definition}  
\subsection{Some lemmas}
In this subsection, we review some basic notations and results about the elementary symmetric polynomials, the complete symmetric polynomials, a special determinant, the subset sum problem for finite abelian groups and MDS codes.

Firstly, we recall the definitions of the elementary symmetric polynomials and the complete symmetric polynomials, as well as the related results. 
\begin{definition}\label{elementarysymmetricpolynomial} {\rm(\cite{ADV2026GDGRS})}
	For any integer $t$, the $t$-th degree elementary symmetric polynomial in  $n$-variables is  defined as
	$$\sigma_{t}(x_{1},x_{2},\cdots,x_{n})=\begin{cases}
		0,&\text{if}\ t<0;\\
		1,&\text{if}\ t=0;\\
		\sum\limits_{1\leq j_{1}<j_{2}<\cdots<j_{t}\leq n}x_{j_1}x_{j_2}\cdots x_{j_t},&\text{if}\ 1\leq t\leq n;\\
		0,&\text{if}\ t>n,\\
		
	\end{cases}$$
	and denote $\sigma_{t}(x_{1},x_{2},\cdots,x_{n})$ by $\sigma_{t}$.
\end{definition}
\begin{definition}\label{completesymmetricpolynomial} {\rm(\cite{ADV2026GDGRS})}
	For any integer $t$, the $t^{th}$-degree complete symmetric polynomial in  $n$ variables is  defined as
	$$S_{t}(x_{1},x_{2},\cdots,x_{n})=\begin{cases}
		0,&\text{if}\ t<0;\\
		\sum\limits_{t_{1}+t_{2}+\cdots+t_{n}=t,t_{i}\geq 0}x_{1}^{t_{1}}x_{2}^{t_{2}}\cdots x_{n}^{t_{n}},&\text{if}\ t\geq 0,\\
	\end{cases}$$
	and denote $S_{t}(x_{1},x_{2},\cdots,x_{n})$ by $S_{t}$.
\end{definition}

\begin{lemma}\label{usapowersum} {\rm(\cite{ADV2026GDGRS}, Lemma 2.4)}
	Let $u_{i}=\prod\limits_{j=1, j \neq i}^{n}\left(\alpha_{i}-\alpha_{j}\right)^{-1}$ for $1 \leq i \leq n$.  Then for any subset $\left\{\alpha_{1}, \ldots, \alpha_{n}\right\}\subseteq\mathbb{F}_{q}$ with $n\geq 3$, we have
	$$
	\sum\limits_{i=1}^{n}u_{i}\alpha_{i}^{h}=\begin{cases}
		0,&\text{if}\ 0\leq h\leq n-2;\\
		S_{h-n+1}(\alpha_{1},\cdots,\alpha_{n}),&\text{if}\ h\geq n-1.
	\end{cases}
	$$ 
\end{lemma}
\begin{remark}\label{sigmaS}
	There is a fundamental relation between the
	elementary symmetric polynomial and the complete symmetric polynomial
	$$\sum\limits_{t=0}^{N}(-1)^{t}\sigma_{t}S_{N-t}=0,\ \text{for all}\ N\geq 1.$$
\end{remark} 

Next, we recall the formula for calculating a special determinant.
\begin{lemma}\label{DtVandermonde} 
For $1 \leq r \leq n-1$, let
$$ \M_n^{(r)} =
	\begin{pmatrix}
		1 & 1 & \dots & 1 \\
		\alpha_1 & \alpha_2 & \dots & \alpha_n \\
		\alpha_1^2 & \alpha_2^2 & \dots & \alpha_n^2 \\
		\vdots & \vdots & \dots & \vdots \\
		\alpha_1^{n-r-1} & \alpha_2^{n-r-1} & \dots & \alpha_n^{n-r-1} \\
		\alpha_1^{n-r+1} & \alpha_2^{n-r+1} & \dots & \alpha_n^{n-r+1} \\
		\vdots & \vdots & \dots & \vdots \\
		\alpha_1^{n-1} & \alpha_2^{n-1} & \dots & \alpha_n^{n-1} \\
		\alpha_1^{n} & \alpha_2^{n} & \dots & \alpha_n^{n}
	\end{pmatrix}_{n\times n}.
$$
	Then
$$
	\det(\M_n^{(r)}) =\prod\limits_{1\leq i<j\leq n}\left(\alpha_{j}-\alpha_{i}\right)\cdot \sigma_r(\alpha_1,\alpha_2,\dots,\alpha_n).
$$
\end{lemma}

The following lemma plays a key role in deriving the necessary and sufficient condition for the considered codes to be MDS  in Section \ref{sec5}.
\begin{lemma}\label{alphak+j}{\rm(\cite{Hu2025LPTGRS}, Lemma 4)}
	Let $\mathcal{T}=\{t_1,\dots,t_\tau\}\subseteq\{1,\dots,n\}$ with $\#\mathcal{T}=\tau$.
	Let $\prod\limits_{i=1}^{\tau}(x-\alpha_{t_i})=\sum\limits_{j=0}^{\tau}\mu_{j}^{(\mathcal{T})}x^{\tau-j}$
	and $\mu_{j}^{(\mathcal{T})}=0$ for $j>\tau$, where the coefficients $\mu_{j}^{(\mathcal{T})}$ are uniquely
	determined by the values of $\alpha_{t_1},\dots,\alpha_{t_\tau}$.
	For any $0\le\varepsilon\le n-\tau-1$, define $f_{\varepsilon,s}^{(\mathcal{T})}\in\mathbb{F}_q$
	for $0\le s\le\tau-1$ through the identity
	\begin{equation}
		\begin{aligned}
			\big(\alpha_{t_1}^{\tau+\varepsilon},\alpha_{t_2}^{\tau+\varepsilon},\dots,\alpha_{t_\tau}^{\tau+\varepsilon}\big)
			=\big(f_{\varepsilon,0}^{(\mathcal{T})},f_{\varepsilon,1}^{(\mathcal{T})},\dots,f_{\varepsilon,\tau-1}^{(\mathcal{T})}\big)
			\begin{pmatrix}
				1 & 1 & \cdots & 1\\
				\alpha_{t_1} & \alpha_{t_2} & \cdots & \alpha_{t_\tau}\\
				\vdots & \vdots & \ddots & \vdots\\
				\alpha_{t_1}^{\tau-1} & \alpha_{t_2}^{\tau-1} & \cdots & \alpha_{t_\tau}^{\tau-1}
			\end{pmatrix},
		\end{aligned}
	\end{equation}
	where
	\begin{equation}
		f_{\varepsilon,s}^{(\mathcal{T})}=-\sum\limits_{i=0}^{\min\{\varepsilon,s\}}
		\mu_{\,i+\tau-s}^{(\mathcal{T})}\,e_{\varepsilon-i}^{(\mathcal{T})},\quad 0\le s\le\tau-1,
	\end{equation}
	with $e_{0}^{(\mathcal{T})}=1$ and $e_{i}^{(\mathcal{T})}=-\sum\limits_{j=0}^{i-1}e_{j}^{(\mathcal{T})}\mu_{i-j}^{(\mathcal{T})}$ for $1\le i\le\varepsilon$. 
\end{lemma}
\begin{remark}
For convenience, when two sets $\mathcal{T}_1,\mathcal{T}_2$ with
$\tau_1=\#\mathcal{T}_1\neq\#\mathcal{T}_2=\tau_2$ appear simultaneously and each
cardinality determines the corresponding set uniquely in the current context,
we abbreviate the superscripts by the cardinalities, i.e., 
\[
\mu_j^{(\mathcal{T}_1)},\ f_{\varepsilon,s}^{(\mathcal{T}_1)},\ e_j^{(\mathcal{T}_1)} \to \mu_j^{(\tau_1)},\ f_{\varepsilon,s}^{(\tau_1)},\ e_j^{(\tau_1)},
\]
and similarly $\mu_j^{(\mathcal{T}_2)},f_{\varepsilon,s}^{(\mathcal{T}_2)},e_j^{(\mathcal{T}_2)}
\to \mu_j^{(\tau_2)},f_{\varepsilon,s}^{(\tau_2)},e_j^{(\tau_2)}$.
\end{remark}
	The following Lemma \ref{MDSorNMDSequivalent} provides a necessary and sufficient
condition for a linear code to be MDS or NMDS, respectively.
\begin{lemma}\label{MDSorNMDSequivalent}{\rm( \cite{WHFECC}, Theorem 2.4.3; \cite{NMDSsmall}, Definition 2.3)} 	
	Let $\mathcal{C}$ be an $[n, k]$ linear code over $\mathbb{F}_{q}$ with $k\geq 1$. Suppose that  $\boldsymbol{G}$ is the 
	generator matrix of $\mathcal{C}$. Then, the following two statements are true,
	
	$(1)$ $\mathcal{C}$ is $\mathrm{MDS}$ if and only if any $k$ columns of $\boldsymbol{G}$ are $\mathbb{F}_{q}$-linearly independent;	
	
	$(2)$ $\mathcal{C}$ is NMDS if and only if the following three conditions hold simultaneously,
	
	\ \ $(i)$ any $k-1$ columns of $\boldsymbol{G}$ are $\mathbb{F}_{q}$-linearly independent;
	
	\ \ $(ii)$ there exists $k$ columns of $\boldsymbol{G}$ which are $\mathbb{F}_{q}$-linearly dependent; 
	
	\ \ $(iii)$ Any $k+1$ columns of $\boldsymbol{G}$ are full rank.
\end{lemma}

Finally, we recall a classical counting formula for subset sums in finite abelian groups.
\begin{lemma}\label{subsumGneq0}
	Let $G$ be a finite abelian group of order $n$, $g\in G$, and $i\in\mathbb{Z}$ with $0\le i\le n-1$. For a subset $D\subseteq G$ and a non-negative integer $t$, we denote
	\[
	N(D,t,g)\triangleq\#\left\{ S\subseteq D:\#S=t,\;\sum_{s\in S} s = g \right\}.
	\]
	Then we have
	\[
	N(G\setminus\{0\},i,g)
	=\frac1n
	\sum_{s\mid \exp(G)}
	(-1)^{i+\left\lfloor \frac{i}{s}\right\rfloor}
	\binom{\frac{n}{s}-1}{\left\lfloor \frac{i}{s}\right\rfloor}
	\sum_{d\mid \gcd(s,e(g))}
	\mu\left(\frac{s}{d}\right)\#G[d],
	\]
	where $\exp(G)$ denotes the exponent of $G$,
	\[
	e(g)\triangleq\max\left\{d : d\mid \exp(G),\, g\in dG\right\},
	\]
	$G[d]\triangleq\{h\in G: dh=0\}$ is the $d$-torsion subgroup of $G$, and $\mu$ is the Möbius function.
\end{lemma}

\section{The direct constructions of the parity-check matrix for $\C_i(i=1,2,3,4)$}\label{sec3}
In this section, let
\begin{equation}\label{Bi}
	\B_i=\begin{pmatrix}
		0&\ldots&0&0&\ldots&0\\
		\vdots& &\vdots&\vdots& &\vdots\\
		0&\ldots&0&0&\ldots&0\\
		b_{k-i,0}&\ldots&b_{k-i,\ell}&0&\ldots&0\\ 
		\vdots& &\vdots&\vdots& &\vdots\\
		b_{k-1,0}&\ldots&b_{k-1,\ell}&0&\ldots&0\\
	\end{pmatrix}(i=1,2,3),
\end{equation}
for the following linear codes respectively,

$(1)$ $\C_{k,\ell}(\L,\P, \A_{m\times m},\B_1)\triangleq \C_1$ with $\L=[\ell+1]$, $\P=[k]$, $\A_{m\times m}\triangleq\A_{m}=\begin{pmatrix}
	0&0&\cdots&0&1\\
	0&0&\cdots&1&a_{2m}\\
	\vdots&\vdots& &\vdots&\vdots\\
	1&a_{m2}&\cdots&a_{m(m-1)}&a_{mm}
\end{pmatrix}$;

$(2)$  $\C_{k,\ell}(\L,\P, \A_{m\times m},\B_i)\triangleq \C_i(i=2,3)$ with $\L=[\ell+1]$, $\P=[k]$,  $\A_{m\times m}=\begin{pmatrix}
	0&1\\
	1&\delta
\end{pmatrix}$, 

$(3)$ $\C_{h}\left(\eta,\boldsymbol{b}\right)\triangleq \C_4,$\\
we directly construct the corresponding parity-check matrix. 
\subsection{The parity-check matrix of $\C_1$} 

\begin{theorem}\label{RL+k-1TGRSH}
	For \(0\le s\le\ell+m\), define
	\[
	d_{s,r}=
	\begin{cases}
		S_{s-\ell-m+r-1},&1\le r\le m-1,\\[2mm]
		S_{s-\ell-1}
		+\displaystyle\sum_{j=0}^{\ell}b_{k-1,j}S_{s-\ell+j},&r=m,
	\end{cases}
	\] 
	and for $1\leq r\leq m$, define
	\[
	k_{s,m-r+1}=-d_{s,r}
	-\sum\limits_{c=m-r+2}^{m}a_{r,c}k_{s,c}.
	\] 
	If \(n\ge k+\ell+2\), then \(\mathcal C_1\) has the following parity-check matrix
	\[
	\H(\A_m,\B)=
	\begin{pmatrix}
		\frac{u_1}{v_1}&\cdots&\frac{u_n}{v_n}&0&\cdots&0\\[1mm]
		\frac{u_1}{v_1}\alpha_1&\cdots&\frac{u_n}{v_n}\alpha_n&0&\cdots&0\\[1mm]
		\vdots&&\vdots&\vdots&&\vdots\\[1mm]
		\frac{u_1}{v_1}\alpha_1^{n-k-\ell-2}
		&\cdots&
		\frac{u_n}{v_n}\alpha_n^{n-k-\ell-2}
		&0&\cdots&0\\[2mm]
		\frac{u_1}{v_1}\alpha_1^{n-k-\ell-1}
		&\cdots&
		\frac{u_n}{v_n}\alpha_n^{n-k-\ell-1}
		&k_{0,1}&\cdots&k_{0,m}\\[2mm]
		\frac{u_1}{v_1}\alpha_1^{n-k-\ell}
		&\cdots&
		\frac{u_n}{v_n}\alpha_n^{n-k-\ell}
		&k_{1,1}&\cdots&k_{1,m}\\[2mm]
		\vdots&&\vdots&\vdots&&\vdots\\[2mm]
		\frac{u_1}{v_1}\alpha_1^{n-k+m-1}
		&\cdots&
		\frac{u_n}{v_n}\alpha_n^{n-k+m-1}
		&k_{\ell+m,1}&\cdots&k_{\ell+m,m}
	\end{pmatrix}_{(n+m-k)\times (n+m)}.
	\] 
\end{theorem}
\begin{proof}
	It is easy to know that $\C_1$ has the generator matrix 
	$$
\begin{aligned}
	&\G(\A_m,\B_1)\\
	=&\begin{pmatrix}
	v_1&\cdots&v_n&0&0&\cdots&0&0\\
	v_1\alpha_{1}&\cdots&v_n\alpha_{n}&0&0&\cdots&0&0\\
	\vdots& &\vdots&\vdots&\vdots&\iddots&\vdots&\vdots\\
	v_1\alpha_{1}^{k-m-1}&\cdots&v_n\alpha_{n}^{k-m-1}&0&0&\cdots&0&0\\
	v_1\alpha_{1}^{k-m}&\cdots&v_n\alpha_{n}^{k-m}&0&0&\cdots&0&1\\ 
	v_1\alpha_{1}^{k-m+1}&\cdots&v_n\alpha_{n}^{k-m+1}&0&0&\cdots&1&a_{2m}\\
	\vdots& &\vdots&\vdots&\vdots&\iddots&\vdots&\vdots\\
	v_1\alpha_{1}^{k-2}&\cdots&v_n\alpha_{n}^{k-2}&0&1&\cdots&a_{(m-1)(m-1)}&a_{(m-1)m}\\ 
	v_1\left(\alpha_{1}^{k-1}+\sum\limits_{j=0}^{\ell} b_{k-1,j}\alpha_{1}^{k+j} \right)&\cdots&v_n\left(\alpha_{n}^{k-1}+\sum\limits_{j=0}^{\ell} b_{k-1,j}\alpha_{n}^{k+j} \right)&1&a_{m2}&\cdots&a_{m(m-1)}&a_{mm}\\
\end{pmatrix}
\end{aligned}
	$$
	To prove that $\H(\A_m,\B_1)$ is the parity-check matrix of $\C_1$, we only need to check that the following two statements are true.
	
	$(1)$ $\rank(\H(\A_m,\B_1))=n+m-k$
	
	$(2)$ $\G(\A_m,\B_1)\H(\A_m,\B_1)^{T}=\zero$.
	
	{\bf For the statement (1),} it is easy to know that the $(n+m-k)\times n$ submatrix formed by the first $n$ columns of $\H(\A_m,\B_1)$ has full row rank, i.e., the row vectors of the $(n+m-k)\times n$ submatrix are $\F_q$-linearly independent. Since the row vectors of $\H(\A_m,\B_1)$ are extensions of these $\F_q$-linearly independent row vectors, they are also $\F_q$-linearly independent, thus, $\rank(\H(\A_m,\B_1))=n+m-k$.
	
	{\bf For the statement (2),} For convenience, we denote $\boldsymbol{g}_{i}^{(1)}$ and $\boldsymbol{h}_{j}^{(1)}$ be the $(i+1)$-th $(0\leq i\leq k-1)$ row and the $(j+1)$-th $(0\leq j\leq n-k+m-1)$ row of $\G(\A_m,\B_1)$ and $\H(\A_m,\B_1)$, respectively.
	
	{\bf Case 1.} For $0\leq i\leq k-m-1$ and $0\leq j\leq n-k+m-1$, or $k-m\leq i\leq k-1$ and $0\leq j\leq n-k-\ell-2$,  we have
	$$\boldsymbol{g}_{i}^{(1)}\left(\boldsymbol{h}_{j}^{(1)}\right)^{T}=\begin{cases}
		\sum\limits_{s=1}^{n}u_s\alpha_{s}^{i+j},&\ \text{if}\ 0\leq i\leq k-m-1\ \text{and}\ 0\leq j\leq n-k+m-1;\\
		&\ \ \ \ \text{or}\ k-m\leq i\leq k-2\ \text{and}\ 0\leq j\leq n-k-\ell-2;\\
		\sum\limits_{s=1}^{n}u_s\alpha_{s}^{i+j}+\sum\limits_{t=0}^{\ell} b_{k-1,t}\sum\limits_{s=1}^{n}u_s\alpha_{s}^{k+t+j} ,&\ \text{if}\ i=k-1\ \text{and}\ 0\leq j\leq n-k-\ell-2.\\
	\end{cases}$$ 
	It is easy to know that if $0\leq i\leq k-m-1$ and $0\leq j\leq n-k+m-1$, then 
	$$i+j\leq n-2;$$
	if $k-m\leq i\leq k-2$ and $0\leq j\leq n-k-\ell-2$, then 
	$$i+j\leq n-\ell-4<n-2;$$
	if $i=k-1$ and $0\leq j\leq n-k-\ell-2$, then 
	$$i+j\leq n-\ell-3<n-2,$$
	and
	$$k+t+j\leq n-2.$$
	Thus, by Lemma \ref{usapowersum}, for $0\leq i\leq k-1$ and $0\leq j\leq n-k-\ell-2$, or $0\leq i\leq k-m-1$ and $0\leq j\leq n-k+m-1$, we have 
	$\boldsymbol{g}_{i}^{(1)}\left(\boldsymbol{h}_{j}^{(1)}\right)^{T}=0.$
	
	{\bf Case 2.} For $k-m\leq i\leq k-1$ and $n-k-\ell-1\leq j\leq n-k+m-1$, we set $i=k-m-1+\vartheta$ with $1\leq \vartheta\leq m$ $j=n-k-\ell-1+\mu$ with $0\leq \mu\leq \ell+m$.
	
	On the one hand, for $1\leq \vartheta\leq m-1$ and $0\leq \mu\leq \ell+m$,  by Lemma \ref{usapowersum}, we have
	$$\begin{aligned}
		\boldsymbol{g}_{i}^{(1)}\left(\boldsymbol{h}_{j}^{(1)}\right)^{T}=&\sum\limits_{s=1}^{n}u_s\alpha_{s}^{i+j}+k_{\mu,m-\vartheta+1}+\sum\limits_{t=m-\vartheta+2}^{m}k_{\mu,t}a_{\vartheta,t}\\
		=&S_{\mu+\vartheta-\ell-m-1}+k_{\mu,m-\vartheta+1}+\sum\limits_{t=m-\vartheta+2}^{m}k_{\mu,t}a_{\vartheta,t} 
	\end{aligned}.$$
	Note that 
	$$S_{\mu+\vartheta-\ell-m-1}=d_{\mu,\vartheta}$$
	and
	$$k_{\mu,m-\vartheta+1}=-d_{\mu,\vartheta}-\sum\limits_{t=m-\vartheta+2}^{m}a_{\vartheta,t}k_{\mu,t},$$
	Thus, for $k-m\leq i\leq k-2$ and $n-k-\ell-1\leq j\leq n-k+m-1$, we have 
	$\boldsymbol{g}_{i}^{(1)}\left(\boldsymbol{h}_{j}^{(1)}\right)^{T}=0.$
	
	On the other hand, for $\vartheta=m$ and $0\leq \mu\leq \ell+m$,  by Lemma \ref{usapowersum},  we have
	$$\begin{aligned}
		\boldsymbol{g}_{i}^{(1)}\left(\boldsymbol{h}_{j}^{(1)}\right)^{T}=&\sum\limits_{s=1}^{n}u_s\alpha_{s}^{i+j}+\sum\limits_{t=0}^{\ell} b_{k-1,t}\sum\limits_{s=1}^{n}u_s\alpha_{s}^{k+t+j}+k_{\mu,m-\vartheta+1}+\sum\limits_{t=m-\vartheta+2}^{m}k_{\mu,t}a_{\vartheta,t}\\
		=&S_{\mu-\ell-1}+\sum\limits_{t=0}^{\ell} b_{k-1,t}S_{\mu+t-\ell}+k_{\mu,1}+\sum\limits_{t=2}^{m}k_{\mu,t}a_{\vartheta,t} 
	\end{aligned}.$$
	Note that
	$$S_{\mu-\ell-1}+\sum\limits_{t=0}^{\ell} b_{k-1,t}S_{\mu+t-\ell}=d_{\mu,m}$$
	and
	$$k_{\mu,1}=-d_{\mu,m}-\sum\limits_{t=2}^{m}a_{m,t}k_{\mu,t}.$$
	Thus, for $i=k-1$ and $n-k-\ell-1\leq j\leq n-k+m-1$, we have 
	$\boldsymbol{g}_{i}^{(1)}\left(\boldsymbol{h}_{j}^{(1)}\right)^{T}=0.$ 
\end{proof} 
\begin{remark}
	By taking $m=2$, $a_{2m}=\delta$, $\ell=0$ and $b_{k-1,0}=\lambda_2$ in Theorem \ref{RL+k-1TGRSH}, the corresponding result is just the case $\lambda_1=1$ of Theorem 2.8 in \cite{Cheng2026TGRL}. 
\end{remark}
 
\subsection{The parity-check matrix of $\C_2$}
\begin{theorem}\label{RL+k-2k-1TGRSH}
If $n\geq k+\ell+2$, then $\C_2$ has the following parity-check matrix 
$$\H_{1}(\delta,\B_2)=\begin{pmatrix}
	\frac{u_1}{v_1}&\cdots&\frac{u_n}{v_n}&0&0\\
	\frac{u_1}{v_1}\alpha_{1}&\cdots&\frac{u_n}{v_n}\alpha_{n}&0&0\\
	\vdots& &\vdots&\vdots&\vdots\\
	\frac{u_1}{v_1}\alpha_{1}^{n-k-\ell-2}&\cdots&\frac{u_n}{v_n}\alpha_{n}^{n-k-\ell-2}&0&0\\
	\frac{u_1}{v_1}\alpha_{1}^{n-k-\ell-1}&\cdots&\frac{u_n}{v_n}\alpha_{n}^{n-k-\ell-1}&\delta b_{k-2,\ell}- b_{k-1,\ell}&- b_{k-2,\ell}\\ 
	\frac{u_1}{v_1}\alpha_{1}^{n-k-\ell}&\cdots&\frac{u_n}{v_n}\alpha_{n}^{n-k-\ell}&\sum\limits_{j=0}^{1}\left(\delta b_{k-2,\ell-1+j}- b_{k-1,\ell-1+j}\right)S_{j}&-\sum\limits_{j=0}^{1} b_{k-2,\ell-1+j}S_{j}\\
	\frac{u_1}{v_1}\alpha_{1}^{n-k-\ell+1}&\cdots&\frac{u_n}{v_n}\alpha_{n}^{n-k-\ell+1}&\sum\limits_{j=0}^{2}\left(\delta b_{k-2,\ell-2+j}- b_{k-1,\ell-2+j}\right)S_{j}&-\sum\limits_{j=0}^{2} b_{k-2,\ell-2+j}S_{j}\\
	\vdots& &\vdots&\vdots&\vdots\\
	\frac{u_1}{v_1}\alpha_{1}^{n-k-1}&\cdots&\frac{u_n}{v_n}\alpha_{n}^{n-k-1}&\sum\limits_{j=0}^{\ell}\left(\delta b_{k-2,j}- b_{k-1,j}\right)S_{j}&-\sum\limits_{j=0}^{\ell} b_{k-2,j}S_{j}\\
	\frac{u_1}{v_1}\alpha_{1}^{n-k}&\cdots&\frac{u_n}{v_n}\alpha_{n}^{n-k}&\sum\limits_{j=0}^{\ell}\left(\delta b_{k-2,j}- b_{k-1,j}\right)S_{j+1}-1&-\sum\limits_{j=0}^{\ell} b_{k-2,j}S_{j+1}\\
	\frac{u_1}{v_1}\alpha_{1}^{n-k+1}&\cdots&\frac{u_n}{v_n}\alpha_{n}^{n-k+1}&\sum\limits_{j=0}^{\ell}\left(\delta b_{k-2,j}- b_{k-1,j}\right)S_{j+2}-S_{1}+\delta&-\sum\limits_{j=0}^{\ell} b_{k-2,j}S_{j+2}-1
\end{pmatrix}_{(n+2-k)\times (n+2)}. 
$$
\end{theorem}
\begin{proof}
It is easy to know that $\C_2$ has the generator matrix
$$
\G(\delta,\B_2)=\begin{pmatrix}
	v_1&\cdots&v_n&0&0\\
	v_1\alpha_{1}&\cdots&v_n\alpha_{n}&0&0\\
	\vdots& &\vdots&\vdots&\vdots\\
	v_1\alpha_{1}^{k-3}&\cdots&v_n\alpha_{n}^{k-3}&0&0\\
	v_1\left(\alpha_{1}^{k-2}+\sum\limits_{j=0}^{\ell} b_{k-2,j}\alpha_{1}^{k+j} \right)&\cdots&v_n\left(\alpha_{n}^{k-2}+\sum\limits_{j=0}^{\ell} b_{k-2,j}\alpha_{n}^{k+j} \right)&0&1\\ 
	v_1\left(\alpha_{1}^{k-1}+\sum\limits_{j=0}^{\ell} b_{k-1,j}\alpha_{1}^{k+j} \right)&\cdots&v_n\left(\alpha_{n}^{k-1}+\sum\limits_{j=0}^{\ell} b_{k-1,j}\alpha_{n}^{k+j} \right)&1&\delta\\
\end{pmatrix}
$$
To prove that $\H_{1}(\delta,\B_2)$ is the parity-check matrix of $\C_2$, we only need to check that the following two statements are true.

$(1)$ $\rank(\H_{1}(\delta,\B_2))=n+2-k$;

$(2)$ $\G(\delta,\B_2)\H_{1}(\delta,\B_2)^{T}=\zero$.

{\bf For the statement (1),} the proof can be completed by an argument similar to that for Theorem \ref{RL+k-1TGRSH} (1).

{\bf For the statement (2),} For convenience, we denote $\boldsymbol{g}_{i}^{(2)}$ and $\boldsymbol{h}_{j}^{(2)}$ be the $(i+1)$-th $(0\leq i\leq k-1)$ row and the $(j+1)$-th $(0\leq j\leq n-k+1)$ row of $\G(\delta,\B_2)$ and $\H_{1}(\delta,\B_2)$, respectively.

{\bf Case 1.} For $0\leq i\leq k-3$ and $0\leq j\leq n-k+1$,  we have
$$\boldsymbol{g}_{i}^{(2)}\left(\boldsymbol{h}_{j}^{(2)}\right)^{T}=\sum\limits_{s=1}^{n}v_s\alpha_{s}^{i}\cdot \frac{u_s}{v_s}\alpha_{s}^{j}=\sum\limits_{s=1}^{n}u_{s}\alpha_{s}^{i+j}.$$ 
By $0\leq i\leq k-3$ and $0\leq j\leq n-k+1$, we have $0\leq i+j\leq n-2$, thus, by Lemma \ref{usapowersum}, for $0\leq i\leq k-3$ and $0\leq j\leq n-k+1$,  we have 
 $\boldsymbol{g}_{i}^{(2)}\left(\boldsymbol{h}_{j}^{(2)}\right)^{T}=0.$

{\bf Case 2.} For $i\in\left\{k-2,k-1\right\}$ and $0\leq j\leq n-k-\ell-2$,  we have
$$\boldsymbol{g}_{i}^{(2)}\left(\boldsymbol{h}_{j}^{(2)}\right)^{T}=\sum\limits_{s=1}^{n}v_s\left(\alpha_{s}^{i}+\sum\limits_{t=0}^{\ell} b_{i,t}\alpha_{s}^{k+t} \right)\cdot \frac{u_s}{v_s}\alpha_{s}^{j}=\sum\limits_{s=1}^{n}u_{s}\alpha_{s}^{i+j}+\sum\limits_{s=1}^{n}\sum\limits_{t=0}^{\ell} b_{i,t}u_{s}\alpha_{s}^{k+t+j}.$$ 
By $i\in\left\{k-2,k-1\right\}$ and $0\leq j\leq n-k-\ell-2$, we have $$0\leq i+j\leq n-3-\ell<n-2$$
and 
$$k+t+j\leq k+\ell+n-k-\ell-2=n-2,$$ 
thus, by Lemma \ref{usapowersum}, for $i\in\left\{k-2,k-1\right\}$ and $0\leq j\leq n-k-\ell-2$,  we have  $\boldsymbol{g}_{i}^{(2)}\left(\boldsymbol{h}_{j}^{(2)}\right)^{T}=0.$

{\bf Case 3.} For $i=k-2$ and $n-k-\ell-1\leq j\leq n-k-1$, i.e., $j=n-k-\ell-1+\mu$ with $\mu\in\left\{0,1,\ldots,\ell\right\}$,  we have
$$\begin{aligned}
\boldsymbol{g}_{i}^{(2)}\left(\boldsymbol{h}_{j}^{(2)}\right)^{T}=&\sum\limits_{s=1}^{n}v_s\left(\alpha_{s}^{k-2}+\sum\limits_{t=0}^{\ell} b_{k-2,t}\alpha_{s}^{k+t} \right)\cdot \frac{u_s}{v_s}\alpha_{s}^{j}-\sum\limits_{t=0}^{\mu} b_{k-2,\ell-\mu+t}S_{t}\\
=&\sum\limits_{s=1}^{n}u_{s}\alpha_{s}^{k-2+j}+\sum\limits_{s=1}^{n}\sum\limits_{t=0}^{\ell} b_{k-2,t}u_{s}\alpha_{s}^{k+t+j}-\sum\limits_{t=0}^{\mu} b_{k-2,\ell-\mu+t}S_{t}.
\end{aligned}
$$ 

On the one hand, by $n-k-\ell-1\leq j\leq n-k-1$, we have $0\leq k-2+j\leq n-3<n-2$, thus, by Lemma \ref{usapowersum}, $$\sum\limits_{s=1}^{n}u_{s}\alpha_{s}^{k-2+j}=0.$$

On the other hand, by $j=n-k-\ell-1+\mu$, we can get $k+t+j=n-1+t+\mu-\ell,$  and when $0\leq t\leq \ell-\mu-1$,  $n-1+t+\mu-\ell\leq n-2$, thus, by Lemma \ref{usapowersum}, we can get 
\begin{equation}\label{H1B2case3}
\begin{aligned}
\sum\limits_{s=1}^{n}\sum\limits_{t=0}^{\ell} b_{k-2,t}u_{s}\alpha_{s}^{k+t+j}=&\sum\limits_{t=0}^{\ell} b_{k-2,t}\left(\sum\limits_{s=1}^{n}u_{s}\alpha_{s}^{n-1+t+\mu-\ell}\right)\\
=&\sum\limits_{t=\ell-\mu}^{\ell} b_{k-2,t}\left(\sum\limits_{s=1}^{n}u_{s}\alpha_{s}^{n-1+t+\mu-\ell}\right)\\
=&\sum\limits_{t=\ell-\mu}^{\ell} b_{k-2,t}S_{t+\mu-\ell}.
\end{aligned}
\end{equation}
Now, let $\hat{t}=t+\mu-\ell$ in \eqref{H1B2case3}, then $t=\hat{t}-\mu+\ell$ and $\hat{t}\in\left\{0,1,\ldots,\mu\right\}$, furthermore, 
$$\sum\limits_{s=1}^{n}\sum\limits_{t=0}^{\ell} b_{k-2,t}u_{s}\alpha_{s}^{k+t+j}=\sum\limits_{\hat{t}=0}^{\mu} b_{k-2,\hat{t}-\mu+\ell}S_{\hat{t}},$$
it means $\sum\limits_{s=1}^{n}\sum\limits_{t=0}^{\ell} b_{k-2,t}u_{s}\alpha_{s}^{k+t+j}-\sum\limits_{t=0}^{\mu} b_{k-2,\ell-\mu+t}S_{t}=0.$

And so, for $i=k-2$ and $n-k-\ell-1\leq j\leq n-k-1$, we have  $\boldsymbol{g}_{i}^{(2)}\left(\boldsymbol{h}_{j}^{(2)}\right)^{T}=0.$

{\bf Case 4.} For $i=k-2$ and $j\in\left\{n-k,n-k+1\right\}$, by Lemma \ref{usapowersum}, we have
$$\begin{aligned}
\boldsymbol{g}_{i}^{(2)}\left(\boldsymbol{h}_{j}^{(2)}\right)^{T}=&\begin{cases}
	\sum\limits_{s=1}^{n}v_s\left(\alpha_{s}^{k-2}+\sum\limits_{t=0}^{\ell} b_{k-2,t}\alpha_{s}^{k+t} \right)\cdot \frac{u_s}{v_s}\alpha_{s}^{j}-\sum\limits_{t=0}^{\ell} b_{k-2,t}S_{t+1}, & j=n-k;\\
	\sum\limits_{s=1}^{n}v_s\left(\alpha_{s}^{k-2}+\sum\limits_{t=0}^{\ell} b_{k-2,t}\alpha_{s}^{k+t} \right)\cdot \frac{u_s}{v_s}\alpha_{s}^{j}-\sum\limits_{t=0}^{\ell} b_{k-2,t}S_{t+2}-1, & j=n-k+1;
\end{cases}\\
=&\begin{cases}
	\sum\limits_{s=1}^{n}u_{s}\alpha_{s}^{k-2+j}+\sum\limits_{t=0}^{\ell} b_{k-2,t}\left(\sum\limits_{s=1}^{n}u_{s}\alpha_{s}^{k+t+j}\right)-\sum\limits_{t=0}^{\ell} b_{k-2,t}S_{t+1}, & j=n-k;\\
	\sum\limits_{s=1}^{n}u_{s}\alpha_{s}^{k-2+j}+\sum\limits_{t=0}^{\ell} b_{k-2,t}\left(\sum\limits_{s=1}^{n}u_{s}\alpha_{s}^{k+t+j}\right)-\sum\limits_{t=0}^{\ell} b_{k-2,t}S_{t+2}-1, & j=n-k+1;
\end{cases}\\
=&\begin{cases}
	\sum\limits_{s=1}^{n}u_{s}\alpha_{s}^{k-2+j}+\sum\limits_{t=0}^{\ell} b_{k-2,t}S_{t+1}-\sum\limits_{t=0}^{\ell} b_{k-2,t}S_{t+1}, & j=n-k;\\
	\sum\limits_{s=1}^{n}u_{s}\alpha_{s}^{k-2+j}+\sum\limits_{t=0}^{\ell} b_{k-2,t}S_{t+2}-\sum\limits_{t=0}^{\ell} b_{k-2,t}S_{t+2}-1, & j=n-k+1;
\end{cases}
\end{aligned}$$
and then $\boldsymbol{g}_{i}^{(2)}\left(\boldsymbol{h}_{j}^{(2)}\right)^{T}=0$ for $i=k-2$ and $j\in\left\{n-k,n-k+1\right\}$.

{\bf Case 5.} For $i=k-1$ and $n-k-\ell-1\leq j\leq n-k-1$, we have
$$\begin{aligned}
	\boldsymbol{g}_{i}^{(2)}\left(\boldsymbol{h}_{j}^{(2)}\right)^{T}=&\sum\limits_{s=1}^{n}v_s\left(\alpha_{s}^{k-1}+\sum\limits_{t=0}^{\ell} b_{k-1,t}\alpha_{s}^{k+t} \right)\cdot \frac{u_s}{v_s}\alpha_{s}^{j}+\sum\limits_{t=0}^{\mu} \left(\delta b_{k-2,\ell-\mu+t}-b_{k-1,\ell-\mu+t}\right)S_{t}-\delta\sum\limits_{t=0}^{\mu} b_{k-2,\ell-\mu+t}S_{t}\\
	=&\sum\limits_{s=1}^{n}u_{s}\alpha_{s}^{k-1+j}+\sum\limits_{s=1}^{n}\sum\limits_{t=0}^{\ell} b_{k-1,t}u_{s}\alpha_{s}^{k+t+j}-\sum\limits_{t=0}^{\mu} b_{k-1,\ell-\mu+t}S_{t}.
\end{aligned}
$$ 
Note that $k-1+j\leq n-2$, then in the similar proof as that of {\bf Case 3}, for $i=k-1$ and $n-k-\ell-1\leq j\leq n-k-1$, we have  $\boldsymbol{g}_{i}^{(2)}\left(\boldsymbol{h}_{j}^{(2)}\right)^{T}=0.$

{\bf Case 6.} For $i=k-1$ and $j\in\left\{n-k,n-k+1\right\}$, by Lemma \ref{usapowersum}, we have
$$
\begin{aligned}
&\boldsymbol{g}_{i}^{(2)}\left(\boldsymbol{h}_{j}^{(2)}\right)^{T}\\
=&\begin{cases}
	\sum\limits_{s=1}^{n}v_s\left(\alpha_{s}^{k-1}+\sum\limits_{t=0}^{\ell} b_{k-1,t}\alpha_{s}^{k+t} \right)\cdot \frac{u_s}{v_s}\alpha_{s}^{j}+\sum\limits_{t=0}^{\ell}\left( \delta b_{k-2,t}-b_{k-1,t}\right)S_{t+1}-1-\delta\sum\limits_{t=0}^{\ell} b_{k-2,t}S_{t+1}, & j=n-k;\\
	\sum\limits_{s=1}^{n}v_s\left(\alpha_{s}^{k-1}+\sum\limits_{t=0}^{\ell} b_{k-1,t}\alpha_{s}^{k+t} \right)\cdot \frac{u_s}{v_s}\alpha_{s}^{j}+\sum\limits_{t=0}^{\ell}\left( \delta b_{k-2,t}-b_{k-1,t}\right)S_{t+2}-S_{1}+\delta-\delta\left(\sum\limits_{t=0}^{\ell} b_{k-2,t}S_{t+2}+1\right), & j=n-k+1;
\end{cases}\\
=&\begin{cases}
\sum\limits_{s=1}^{n}u_{s}\alpha_{s}^{k-1+j}+\sum\limits_{t=0}^{\ell} b_{k-1,t}\left(\sum\limits_{s=1}^{n}u_{s}\alpha_{s}^{k+t+j}\right)-1-\sum\limits_{t=0}^{\ell} b_{k-1,t}S_{t+1},&j=n-k;\\
\sum\limits_{s=1}^{n}u_{s}\alpha_{s}^{k-1+j}+\sum\limits_{t=0}^{\ell} b_{k-1,t}\left(\sum\limits_{s=1}^{n}u_{s}\alpha_{s}^{k+t+j}\right)-\sum\limits_{t=0}^{\ell} b_{k-1,t}S_{t+2}-S_{1},&j=n-k+1;
\end{cases}\\
=&\begin{cases}
	S_{0}+\sum\limits_{t=0}^{\ell} b_{k-1,t}S_{t+1}-1-\sum\limits_{t=0}^{\ell} b_{k-1,t}S_{t+1},&j=n-k;\\
	S_{1}+\sum\limits_{t=0}^{\ell} b_{k-1,t}S_{t+2}-\sum\limits_{t=0}^{\ell} b_{k-1,t}S_{t+2}-S_{1},&j=n-k+1;
\end{cases}
\end{aligned}
$$
furthermore, by directly calculating, for $i=k-1$ and $j\in\left\{n-k,n-k+1\right\}$, we have  $\boldsymbol{g}_{i}^{(2)}\left(\boldsymbol{h}_{j}^{(2)}\right)^{T}=0.$ 

This completes the proof of Theorem \ref{RL+k-2k-1TGRSH}.
\end{proof}
\begin{remark}
By taking $b_{k-2,j}=0(j=0,1,\ldots,\ell)$ in Theorem \ref{RL+k-2k-1TGRSH}, the corresponding result is just the case $(m,a_{2m})=(2,\delta)$ of Theorem \ref{RL+k-1TGRSH}.
\end{remark}
\subsection{The parity-check matrix of $\C_3$}
In this subsection, for convenience, we fix the following notations.
\begin{itemize}
\item $\Phi_{i,j}=S_{i}+\sum\limits_{t=0}^{\ell}b_{k-j,t}S_{t+3}( i\in\left\{0,1,2\right\},j\in\left\{1,2,3\right\})$.
\item $\Psi_{i,j}^{(s)}=\sum\limits_{t=0}^{i}b_{k-j,\ell-t}S_{i-t+s}( i\in[\ell+1],j\in\left\{1,2,3\right\},s\in\left\{0,1,2\right\})$.
\item When $\Phi_{0,3}\in\F_q^*$, $\Theta_{i}=\frac{\sum\limits_{t=0}^{i}b_{k-3,\ell-t}S_{i-t}}{1+\sum\limits_{t=0}^{\ell}b_{k-3,t}S_{t+3}}=\Phi_{0,3}^{-1}\Psi_{i,3}^{(0)}(i\in[\ell+1]).$ 
\item When $\Phi_{0,3}\in\F_q^*$, $\Gamma_{j}=\frac{\sum\limits_{t=0}^{\ell}b_{k-3,\ell-t}S_{\ell-t+j}}{1+\sum\limits_{t=0}^{\ell}b_{k-3,t}S_{t+3}}=\Phi_{0,3}^{-1}\Psi_{\ell,3}^{(j)}(j\in\left\{1,2\right\}).$
\item $\Delta_{i}=\left(\Theta_{i}\Phi_{2,1}-\Psi_{i,1}^{(0)}\right)+\delta\left(\Psi_{i,2}^{(0)}-\Theta_{i}\Phi_{1,2}\right)(i\in[\ell+1])$.
\item $\widehat{\Delta_{i,j}}=\left(\Gamma_{j}\Phi_{2,1}-\Psi_{i,1}^{(j)}\right)-\delta\left(\Gamma_{j}\Phi_{1,2}-\Psi_{i,2}^{(j)}\right)(i\in[\ell+1],j\in\left\{1,2\right\})$.
\end{itemize}
\begin{theorem}\label{RL+k-3k-2k-1TGRSH}
If $\Phi_{0,3}\in\F_q^*$ and $k\geq 3$, then $\C_3$ has the following parity-check matrix 
	$$\H(\delta,\B_3)=\begin{pmatrix}
		\frac{u_1}{v_1}&\cdots&\frac{u_n}{v_n}&0&0\\
		\frac{u_1}{v_1}\alpha_{1}&\cdots&\frac{u_n}{v_n}\alpha_{n}&0&0\\
		\vdots& &\vdots&\vdots&\vdots\\
		\frac{u_1}{v_1}\alpha_{1}^{n-k-\ell-2}&\cdots&\frac{u_n}{v_n}\alpha_{n}^{n-k-\ell-2}&0&0\\
		\frac{u_1}{v_1}\left(\alpha_{1}^{n-k-\ell-1}-\Theta_{0}\alpha_{1}^{n-k+2}\right)&\cdots&\frac{u_n}{v_n}\left(\alpha_{n}^{n-k-\ell-1}-\Theta_{0}\alpha_{n}^{n-k+2}\right)&\Delta_{0}&\Theta_{0}\Phi_{1,2}-\Psi_{0,2}^{(0)}\\ 
		\frac{u_1}{v_1}\left(\alpha_{1}^{n-k-\ell}-\Theta_{1}\alpha_{1}^{n-k+2}\right)&\cdots&\frac{u_n}{v_n}\left(\alpha_{n}^{n-k-\ell}-\Theta_{1}\alpha_{n}^{n-k+2}\right)&\Delta_{1}&\Theta_{1}\Phi_{1,2}-\Psi_{1,2}^{(0)}\\
		\frac{u_1}{v_1}\left(\alpha_{1}^{n-k-\ell+1}-\Theta_{2}\alpha_{1}^{n-k+2}\right)&\cdots&\frac{u_n}{v_n}\left(\alpha_{n}^{n-k-\ell+1}-\Theta_{2}\alpha_{n}^{n-k+2}\right)&\Delta_{2}&\Theta_{2}\Phi_{1,2}-\Psi_{2,2}^{(0)}\\
		\vdots& &\vdots&\vdots&\vdots\\
		\frac{u_1}{v_1}\left(\alpha_{1}^{n-k-1}-\Theta_{\ell}\alpha_{1}^{n-k+2}\right)&\cdots&\frac{u_n}{v_n}\left(\alpha_{n}^{n-k-1}-\Theta_{\ell}\alpha_{n}^{n-k+2}\right)&\Delta_{\ell}&\Theta_{\ell}\Phi_{1,2}-\Psi_{\ell,2}^{(0)}\\
		\frac{u_1}{v_1}\left(\alpha_{1}^{n-k}-\Gamma_{1}\alpha_{1}^{n-k+2}\right)&\cdots&\frac{u_n}{v_n}\left(\alpha_{n}^{n-k}-\Gamma_{1}\alpha_{n}^{n-k+2}\right)&\widehat{\Delta_{\ell,1}}-1&\Gamma_{1}\Phi_{1,2}-\Psi_{\ell,2}^{(1)}\\
		\frac{u_1}{v_1}\left(\alpha_{1}^{n-k+1}-\Gamma_{2}\alpha_{1}^{n-k+2}\right)&\cdots&\frac{u_n}{v_n}\left(\alpha_{n}^{n-k+1}-\Gamma_{2}\alpha_{n}^{n-k+2}\right)&\widehat{\Delta_{\ell,2}}-S_1+\delta&\Gamma_{2}\Phi_{1,2}-\Psi_{\ell,2}^{(2)}-1
	\end{pmatrix}.
	$$ 
\end{theorem}
\begin{proof}
	It is easy to know that $\C_3$ has the generator matrix
	$$
	\G(\delta,\B_3)=\begin{pmatrix}
		v_1&\cdots&v_n&0&0\\
		v_1\alpha_{1}&\cdots&v_n\alpha_{n}&0&0\\
		\vdots& &\vdots&\vdots&\vdots\\
		v_1\alpha_{1}^{k-4}&\cdots&v_n\alpha_{n}^{k-4}&0&0\\
		v_1\left(\alpha_{1}^{k-3}+\sum\limits_{j=0}^{\ell} b_{k-3,j}\alpha_{1}^{k+j} \right)&\cdots&v_n\left(\alpha_{n}^{k-3}+\sum\limits_{j=0}^{\ell} b_{k-3,j}\alpha_{n}^{k+j} \right)&0&0\\
		v_1\left(\alpha_{1}^{k-2}+\sum\limits_{j=0}^{\ell} b_{k-2,j}\alpha_{1}^{k+j} \right)&\cdots&v_n\left(\alpha_{n}^{k-2}+\sum\limits_{j=0}^{\ell} b_{k-2,j}\alpha_{n}^{k+j} \right)&0&1\\ 
		v_1\left(\alpha_{1}^{k-1}+\sum\limits_{j=0}^{\ell} b_{k-1,j}\alpha_{1}^{k+j} \right)&\cdots&v_n\left(\alpha_{n}^{k-1}+\sum\limits_{j=0}^{\ell} b_{k-1,j}\alpha_{n}^{k+j} \right)&1&\delta\\
	\end{pmatrix}
	$$
To prove that $\H(\delta,\B_3)$ is the parity-check matrix of $\C_3$, we only need to check that the following two statements are true.

$(1)$ $\rank(\H(\delta,\B_3))=n+2-k$;

$(2)$ $\G(\delta,\B_3)\H(\delta,\B_3)^{T}=\zero$.

{\bf For the statement (1).} For the submatrix 
$$\H_{1}(\delta,\B_3)=\begin{pmatrix}
	\frac{u_1}{v_1}&\cdots&\frac{u_n}{v_n}\\
	\frac{u_1}{v_1}\alpha_{1}&\cdots&\frac{u_n}{v_n}\alpha_{n}\\
	\vdots& &\vdots\\
	\frac{u_1}{v_1}\alpha_{1}^{n-k-\ell-2}&\cdots&\frac{u_n}{v_n}\alpha_{n}^{n-k-\ell-2}\\
	\frac{u_1}{v_1}\left(\alpha_{1}^{n-k-\ell-1}-\Theta_{0}\alpha_{1}^{n-k+2}\right)&\cdots&\frac{u_n}{v_n}\left(\alpha_{n}^{n-k-\ell-1}-\Theta_{0}\alpha_{n}^{n-k+2}\right)\\ 
	\frac{u_1}{v_1}\left(\alpha_{1}^{n-k-\ell}-\Theta_{1}\alpha_{1}^{n-k+2}\right)&\cdots&\frac{u_n}{v_n}\left(\alpha_{n}^{n-k-\ell}-\Theta_{1}\alpha_{n}^{n-k+2}\right)\\
	\frac{u_1}{v_1}\left(\alpha_{1}^{n-k-\ell+1}-\Theta_{2}\alpha_{1}^{n-k+2}\right)&\cdots&\frac{u_n}{v_n}\left(\alpha_{n}^{n-k-\ell+1}-\Theta_{2}\alpha_{n}^{n-k+2}\right)\\
	\vdots& &\vdots\\
	\frac{u_1}{v_1}\left(\alpha_{1}^{n-k-1}-\Theta_{\ell}\alpha_{1}^{n-k+2}\right)&\cdots&\frac{u_n}{v_n}\left(\alpha_{n}^{n-k-1}-\Theta_{\ell}\alpha_{n}^{n-k+2}\right)\\
	\frac{u_1}{v_1}\left(\alpha_{1}^{n-k}-\Gamma_{1}\alpha_{1}^{n-k+2}\right)&\cdots&\frac{u_n}{v_n}\left(\alpha_{n}^{n-k}-\Gamma_{1}\alpha_{n}^{n-k+2}\right)\\
	\frac{u_1}{v_1}\left(\alpha_{1}^{n-k+1}-\Gamma_{2}\alpha_{1}^{n-k+2}\right)&\cdots&\frac{u_n}{v_n}\left(\alpha_{n}^{n-k+1}-\Gamma_{2}\alpha_{n}^{n-k+2}\right)
\end{pmatrix}
$$
formed by the first $n$ columns of matrix $\H(\delta,\B_3)$, note that $$\H_1(\delta,\B_3)=\boldsymbol{A}_{(n+2-k)\times n}\cdot\boldsymbol{V}_{n}\cdot\mathrm{diag}\left\{\frac{u_{1}}{v_{1}},\ldots,\frac{u_{n}}{v_{n}} \right\},$$
where
$$\small\boldsymbol{A}_{(n+2-k)\times n}=\begin{pmatrix}
	\boldsymbol{E}_{n+2-k}&\begin{matrix}
		0&0&\cdots&0\\
		\vdots&\vdots&\ddots&\vdots\\
		0&0&\cdots&0\\
		-\Theta_{0}&0&\cdots&0\\
		\vdots&\vdots&\ddots&\vdots\\
		-\Theta_{\ell}&0&\cdots&0\\
		-\Gamma_{1}&0&\cdots&0\\
		-\Gamma_{2}&0&\cdots&0
	\end{matrix}
\end{pmatrix}, \boldsymbol{V}_{n}=\begin{pmatrix} 
	1&1&\cdots&1\\
	\alpha_{1} & \alpha_{2} & \ldots & \alpha_{n} \\
	\vdots & \vdots & \ddots & \vdots \\   
	\alpha_{1}^{n-1} & \alpha_{2}^{n-1} & \ldots & \alpha_{n}^{n-1}
\end{pmatrix}.$$
It's easy to see that $\boldsymbol{E}_{n+2-k}$ is a $(n+2-k)\times (n+2-k)$ submatrix of $\boldsymbol{A}_{(n+2-k)\times n}$. And then
$$\rank\left(\H_{1}(\delta,\B_3)\right)=\rank\left(\boldsymbol{A}_{(n+2-k)\times n}\right)=n+2-k.$$ 
Hence, the row vectors of $\H_{1}(\delta,\B_3)$ are $\F_q$-linearly independent. It is easy to know that the row vectors of $\H(\delta,\B_3)$ are obtained by appending two coordinates to the corresponding row vector of $\H_{1}(\delta,\B_3)$. Since appending coordinates to linearly independent vectors preserves linearly independence, thus the row vectors of $\H(\delta,\B_3)$ are $\F_q$-linearly independent, i.e., $\rank\left(\H(\delta,\B_3)\right)=n+2-k.$

{\bf For the statement (2),} For convenience, we denote $\boldsymbol{g}_{i}^{(3)}$ and $\boldsymbol{h}_{j}^{(3)}$ be the $(i+1)$-th $(0\leq i\leq k-1)$ row and the $(j+1)$-th $(0\leq j\leq n-k+1)$ row of $\G(\delta,\B_3)$ and $\H(\delta,\B_3)$, respectively.

{\bf Case 1.} For $0\leq i\leq k-4$ and $0\leq j\leq n-k+1$,   we have
$$\boldsymbol{g}_{i}^{(3)}\left(\boldsymbol{h}_{j}^{(3)}\right)^{T}=\begin{cases}
\sum\limits_{s=1}^{n}u_{s}\alpha_{s}^{i+j}, & 0\leq j\leq n-k-\ell-2;\\
\sum\limits_{s=1}^{n}u_{s}\alpha_{s}^{i+j}-\Theta_{j-(n-k-\ell-1)}\sum\limits_{s=1}^{n}u_{s}\alpha_{s}^{i+n-k+2}, & n-k-\ell-1\leq j\leq n-k-1;\\
\sum\limits_{s=1}^{n}u_{s}\alpha_{s}^{i+j}-\Gamma_{j-(n-k-1)}\sum\limits_{s=1}^{n}u_{s}\alpha_{s}^{i+n-k+2}, & n-k\leq j\leq n-k+1.
\end{cases}$$ 
By $0\leq i\leq k-4$ and $0\leq j\leq n-k+1$, we can get $$0\leq i+j\leq n-3<n-2$$
and
$$i+n-k+2\leq n-2,$$ thus, by Lemma \ref{usapowersum}, for $0\leq i\leq k-4$ and $0\leq j\leq n-k+1$,  we have 
$\boldsymbol{g}_{i}^{(3)}\left(\boldsymbol{h}_{j}^{(3)}\right)^{T}=0.$

{\bf Case 2.} For $k-3\leq i\leq k-1$ and $0\leq j\leq n-k-\ell-2$, we have
$$\boldsymbol{g}_{i}^{(3)}\left(\boldsymbol{h}_{j}^{(3)}\right)^{T}=\sum\limits_{s=1}^{n}u_{s}\alpha_{s}^{i+j}+\sum\limits_{t=0}^{\ell}b_{i,t}\sum\limits_{s=1}^{n}u_{s}\alpha_{s}^{k+t+j}.$$
By $k-3\leq i\leq k-1$, $0\leq j\leq n-k-\ell-2$ and $0\leq t\leq\ell$, we can get $$i+j\leq n-\ell-3<n-2$$ and $$k\leq k+t+j\leq n-2,$$ thus, by Lemma \ref{usapowersum}, for $k-3\leq i\leq k-1$ and $0\leq j\leq n-k-\ell-2$,  we have 
$\boldsymbol{g}_{i}^{(3)}\left(\boldsymbol{h}_{j}^{(3)}\right)^{T}=0.$

{\bf Case 3.} For $i=k-3$ and $n-k-\ell-1\leq j\leq n-k-1$, we set $j=n-k-\ell-1+\mu$ with $0\leq \mu\leq \ell$, we have
$$
\begin{aligned}
\boldsymbol{g}_{i}^{(3)}\left(\boldsymbol{h}_{j}^{(3)}\right)^{T}=&\sum\limits_{s=1}^{n}u_{s}\left(\alpha_{s}^{k-3}+\sum\limits_{t=0}^{\ell}b_{k-3,t}\alpha_{s}^{k+t}\right)\left(\alpha_{s}^{j}-\Theta_{\mu}\alpha_{s}^{n-k+2}\right)\\
	=&\sum\limits_{s=1}^{n}u_{s}\alpha_{s}^{k-3+j}+\sum\limits_{t=0}^{\ell}b_{k-3,t}\sum\limits_{s=1}^{n}u_{s}\alpha_{s}^{k+t+j}-\Theta_{\mu}\left(\sum\limits_{s=1}^{n}u_{s}\alpha_{s}^{n-1}+\sum\limits_{t=0}^{\ell}b_{k-3,t}\sum\limits_{s=1}^{n}u_{s}\alpha_{s}^{n+t+2}\right).
\end{aligned}$$
Note that
$$k-3+j\leq n-4<n-2,$$
then by Lemma \ref{usapowersum} and $\Theta_{\mu}=\frac{\sum\limits_{t=0}^{\mu}b_{k-3,\ell-t}S_{\mu-t}}{1+\sum\limits_{t=0}^{\ell}b_{k-3,t}S_{t+3}}$, we have
$$\boldsymbol{g}_{i}^{(3)}\left(\boldsymbol{h}_{j}^{(3)}\right)^{T}=\sum\limits_{t=0}^{\ell}b_{k-3,t}S_{k+t+j-n+1}-\sum\limits_{t=0}^{\mu}b_{k-3,\ell-t}S_{\mu-t}=\sum\limits_{t=\ell-\mu}^{\ell}b_{k-3,t}S_{t+\mu-\ell}-\sum\limits_{t=0}^{\mu}b_{k-3,\ell-t}S_{\mu-t}.$$
Let $\hat{t}=\ell-t$, then $0\leq\hat{t}\leq\mu$ and $t+\mu-\ell=\mu-\hat{t}$, hence,
$$\boldsymbol{g}_{i}^{(3)}\left(\boldsymbol{h}_{j}^{(3)}\right)^{T}=\sum\limits_{\hat{t}=0}^{\mu}b_{k-3,\ell-\hat{t}}S_{\mu-\hat{t}}-\sum\limits_{t=0}^{\mu}b_{k-3,\ell-t}S_{\mu-t}=0.$$

{\bf Case 4.} For $i=k-3$ and $j\in\left\{n-k,n-k+1\right\}$, we set $j=n-k-1+\mu$ with $\mu\in\left\{1,2\right\}$, then
$$
\begin{aligned}
	\boldsymbol{g}_{i}^{(3)}\left(\boldsymbol{h}_{j}^{(3)}\right)^{T}=&\sum\limits_{s=1}^{n}u_{s}\left(\alpha_{s}^{k-3}+\sum\limits_{t=0}^{\ell}b_{k-3,t}\alpha_{s}^{k+t}\right)\left(\alpha_{s}^{j}-\Gamma_{\mu}\alpha_{s}^{n-k+2}\right)\\
=&\sum\limits_{s=1}^{n}u_{s}\alpha_{s}^{k-3+j}+\sum\limits_{t=0}^{\ell}b_{k-3,t}\sum\limits_{s=1}^{n}u_{s}\alpha_{s}^{k+t+j}-\Gamma_{\mu}\left(\sum\limits_{s=1}^{n}u_{s}\alpha_{s}^{n-1}+\sum\limits_{t=0}^{\ell}b_{k-3,t}\sum\limits_{s=1}^{n}u_{s}\alpha_{s}^{n+t+2}\right)\\
=&\sum\limits_{t=0}^{\ell}b_{k-3,t}S_{k+t+j-n+1}-\Gamma_{\mu}\left(S_0+\sum\limits_{t=0}^{\ell}b_{k-3,t}S_{t+3}\right)\\
=&\sum\limits_{t=0}^{\ell}b_{k-3,t}S_{t+\mu}-\sum\limits_{t=0}^{\ell}b_{k-3,\ell-t}S_{\ell-t+\mu}.
\end{aligned}$$
Let $\hat{t}=\ell-t$, then $t+\mu=\ell-\hat{t}+\mu$ and $0\leq\hat{t}\leq\ell$, and so
$$\boldsymbol{g}_{i}^{(3)}\left(\boldsymbol{h}_{j}^{(3)}\right)^{T}=\sum\limits_{\hat{t}=0}^{\ell}b_{k-3,\ell-\hat{t}}S_{\ell-\hat{t}+\mu}-\sum\limits_{t=0}^{\ell}b_{k-3,\ell-t}S_{\ell-t+\mu}=0.$$

{\bf Case 5.} For $i=k-2$ and $n-k-\ell-1\leq j\leq n-k-1$, we set $j=n-k-\ell-1+\mu$ with $0\leq \mu\leq \ell$, we have
$$
\begin{aligned}
&\boldsymbol{g}_{i}^{(3)}\left(\boldsymbol{h}_{j}^{(3)}\right)^{T}\\
=&\sum\limits_{s=1}^{n}u_{s}\left(\alpha_{s}^{k-2}+\sum\limits_{t=0}^{\ell}b_{k-2,t}\alpha_{s}^{k+t}\right)\left(\alpha_{s}^{j}-\Theta_{\mu}\alpha_{s}^{n-k+2}\right)+\Theta_{\mu}\Phi_{1,2}-\Psi_{\mu,2}^{(0)}\\
=&\sum\limits_{s=1}^{n}u_{s}\alpha_{s}^{k-2+j}+\sum\limits_{t=0}^{\ell}b_{k-2,t}\sum\limits_{s=1}^{n}u_{s}\alpha_{s}^{k+t+j}-\Theta_{\mu}\left(\sum\limits_{s=1}^{n}u_{s}\alpha_{s}^{n}+\sum\limits_{t=0}^{\ell}b_{k-2,t}\sum\limits_{s=1}^{n}u_{s}\alpha_{s}^{n+t+2}\right)+\Theta_{\mu}\Phi_{1,2}-\Psi_{\mu,2}^{(0)}.
\end{aligned}$$
Note that
$$k-2+j\leq n-3<n-2,$$
then by Lemma \ref{usapowersum} and $\Theta_{\mu}=\frac{\sum\limits_{t=0}^{\mu}b_{k-3,\ell-t}S_{\mu-t}}{1+\sum\limits_{t=0}^{\ell}b_{k-3,t}S_{t+3}}$, we have
$$\boldsymbol{g}_{i}^{(3)}\left(\boldsymbol{h}_{j}^{(3)}\right)^{T}=\sum\limits_{t=0}^{\ell}b_{k-2,t}S_{k+t+j-n+1}-\Theta_{\mu}\left(S_{1}+\sum\limits_{t=0}^{\ell}b_{k-2,t}S_{t+3}\right)+\Theta_{\mu}\Phi_{1,2}-\Psi_{\mu,2}^{(0)}.$$
By an argument similar to that for Theorem \ref{RL+k-3k-2k-1TGRSH} (2) {\bf Case 4} and the definitions of $\Psi_{\mu,2}^{(0)}$ and $\Phi_{1,2}$, the result is immediate.

{\bf Case 6.} For $i=k-2$ and $j\in\left\{n-k,n-k+1\right\}$, we set $j=n-k-1+\mu$ with $\mu\in\left\{1,2\right\}$, then
$$
\begin{aligned}
	&\boldsymbol{g}_{i}^{(3)}\left(\boldsymbol{h}_{j}^{(3)}\right)^{T}\\
	=&\sum\limits_{s=1}^{n}u_{s}\left(\alpha_{s}^{k-2}+\sum\limits_{t=0}^{\ell}b_{k-2,t}\alpha_{s}^{k+t}\right)\left(\alpha_{s}^{j}-\Gamma_{\mu}\alpha_{s}^{n-k+2}\right)+\Gamma_{\mu}\Phi_{1,2}-\Psi_{\ell,2}^{(\mu)}-S_{\mu-2}\\
=&\sum\limits_{s=1}^{n}u_{s}\alpha_{s}^{k-2+j}+\sum\limits_{t=0}^{\ell}b_{k-2,t}\sum\limits_{s=1}^{n}u_{s}\alpha_{s}^{k+t+j}-\Gamma_{\mu}\left(\sum\limits_{s=1}^{n}u_{s}\alpha_{s}^{n}+\sum\limits_{t=0}^{\ell}b_{k-2,t}\sum\limits_{s=1}^{n}u_{s}\alpha_{s}^{n+t+2}\right)+\Gamma_{\mu}\Phi_{1,2}-\Psi_{\ell,2}^{(\mu)}-S_{\mu-2}\\
=&S_{\mu-2}+\sum\limits_{t=0}^{\ell}b_{k-2,t}S_{\mu+t}-\Gamma_{\mu}\left(S_1+\sum\limits_{t=0}^{\ell}b_{k-2,t}S_{t+3}\right)+\Gamma_{\mu}\Phi_{1,2}-\Psi_{\ell,2}^{(\mu)}-S_{\mu-2}.
\end{aligned}$$
By an argument similar to that for Theorem \ref{RL+k-3k-2k-1TGRSH} (2) {\bf Case 3} and the definitions of $\Psi_{\ell,2}^{(\mu)}$ and $\Phi_{1,2}$, the result is immediate.

{\bf Case 7.} For $i=k-1$ and $n-k-\ell-1\leq j\leq n-k-1$, we set $j=n-k-\ell-1+\mu$ with $0\leq \mu\leq \ell$, we have
$$
\begin{aligned}
&\boldsymbol{g}_{i}^{(3)}\left(\boldsymbol{h}_{j}^{(3)}\right)^{T}\\
=&\sum\limits_{s=1}^{n}u_{s}\left(\alpha_{s}^{k-1}+\sum\limits_{t=0}^{\ell}b_{k-1,t}\alpha_{s}^{k+t}\right)\left(\alpha_{s}^{j}-\Theta_{\mu}\alpha_{s}^{n-k+2}\right)+\Delta_{\mu}+\delta\left(\Theta_{\mu}\Phi_{1,2}-\Psi_{\mu,2}^{(0)}\right)\\
=&\sum\limits_{s=1}^{n}u_{s}\alpha_{s}^{k-1+j}+\sum\limits_{t=0}^{\ell}b_{k-1,t}\sum\limits_{s=1}^{n}u_{s}\alpha_{s}^{k+t+j}-\Theta_{\mu}\left(\sum\limits_{s=1}^{n}u_{s}\alpha_{s}^{n+1}+\sum\limits_{t=0}^{\ell}b_{k-1,t}\sum\limits_{s=1}^{n}u_{s}\alpha_{s}^{n+t+2}\right)+\Delta_{\mu}+\delta\left(\Theta_{\mu}\Phi_{1,2}-\Psi_{\mu,2}^{(0)}\right)
\end{aligned}
$$
Note that
$$k-1+j\leq n-2,$$
then by Lemma \ref{usapowersum}, we have
$$
\boldsymbol{g}_{i}^{(3)}\left(\boldsymbol{h}_{j}^{(3)}\right)^{T}=\sum\limits_{t=0}^{\ell}b_{k-1,t}S_{t+\mu-\ell}-\Theta_{\mu}\left(S_{2}+\sum\limits_{t=0}^{\ell}b_{k-1,t}S_{t+3}\right)+\Delta_{\mu}+\delta\left(\Theta_{\mu}\Phi_{1,2}-\Psi_{\mu,2}^{(0)}\right)
$$
By an argument similar to that for Theorem \ref{RL+k-3k-2k-1TGRSH} (2) {\bf Case 3}, the definitions of $\Psi_{\mu,1}^{(0)}$, $\Phi_{1,2}$, $\Phi_{2,1}$ and
$$\Delta_{\mu}=\left(\Theta_{\mu}\Phi_{2,1}-\Psi_{\mu,1}^{(0)}\right)+\delta\left(\Psi_{\mu,2}^{(0)}-\Theta_{\mu}\Phi_{1,2}\right),$$ the result is immediate.

{\bf Case 8.} For $i=k-1$ and $j\in\left\{n-k,n-k+1\right\}$, we set $j=n-k-1+\mu$ with $\mu\in\left\{1,2\right\}$, then
$$
\begin{aligned}
	&\boldsymbol{g}_{i}^{(3)}\left(\boldsymbol{h}_{j}^{(3)}\right)^{T}\\
	=&\sum\limits_{s=1}^{n}u_{s}\left(\alpha_{s}^{k-1}+\sum\limits_{t=0}^{\ell}b_{k-1,t}\alpha_{s}^{k+t}\right)\left(\alpha_{s}^{j}-\Gamma_{\mu}\alpha_{s}^{n-k+2}\right)+\widehat{\Delta_{\ell,\mu}}-S_{\mu-1}+\delta S_{\mu-2}+\delta\left(\Gamma_{\mu}\Phi_{1,2}-\Psi_{\ell,2}^{(\mu)}-S_{\mu-2}\right)\\
=&\sum\limits_{s=1}^{n}u_{s}\alpha_{s}^{k-1+j}+\sum\limits_{t=0}^{\ell}b_{k-1,t}\sum\limits_{s=1}^{n}u_{s}\alpha_{s}^{k+t+j}-\Gamma_{\mu}\left(\sum\limits_{s=1}^{n}u_{s}\alpha_{s}^{n+1}+\sum\limits_{t=0}^{\ell}b_{k-1,t}\sum\limits_{s=1}^{n}u_{s}\alpha_{s}^{n+t+2}\right)\\
&+\widehat{\Delta_{\ell,\mu}}-S_{\mu-1}+\delta S_{\mu-2}+\delta\left(\Gamma_{\mu}\Phi_{1,2}-\Psi_{\ell,2}^{(\mu)}-S_{\mu-2}\right)\\
=&S_{\mu-1}+\sum\limits_{t=0}^{\ell}b_{k-1,t}S_{\mu+t}-\Gamma_{\mu}\left(S_2+\sum\limits_{t=0}^{\ell}b_{k-1,t}S_{t+3}\right)+\widehat{\Delta_{\ell,\mu}}-S_{\mu-1}+\delta S_{\mu-2}+\delta\left(\Gamma_{\mu}\Phi_{1,2}-\Psi_{\ell,2}^{(\mu)}-S_{\mu-2}\right).
\end{aligned}
$$	
By an argument similar to that for Theorem \ref{RL+k-3k-2k-1TGRSH} (2) {\bf Case 4}, the definitions of $\Psi_{\ell,2}^{(\mu)}$, $\Phi_{1,2}$, $\Phi_{2,1}$ and
$$\widehat{\Delta_{\ell,\mu}}=\left(\Gamma_{\mu}\Phi_{2,1}-\Psi_{\ell,1}^{(\mu)}\right)-\delta\left(\Gamma_{\mu}\Phi_{1,2}-\Psi_{\ell,2}^{(\mu)}\right),$$ the result is immediate.

This completes the proof of Theorem \ref{RL+k-3k-2k-1TGRSH}.	
\end{proof} 
\begin{remark}
By taking $b_{k-s,j}=0$ with $s\in\left\{2,3\right\}$ and $j\in[\ell+1]$, or $b_{k-3,j}=0$ with $j\in[\ell+1]$ in Theorem \ref{RL+k-3k-2k-1TGRSH}, the corresponding result is just Theorem \ref{RL+k-1TGRSH} or Theorem \ref{RL+k-2k-1TGRSH}, respectively.
\end{remark}
\subsection{The parity-check matrix of $\C_4$}  
\begin{theorem}\label{extendedhTGRSH}  
Let
$$
	\Lambda_{i,h}=\sum\limits_{j=0}^{n-1-h}(-1)^{j}\sigma_{j}\,\alpha_{i}^{n-1-h-j}.$$
Then
	\[
	\H_{h}(\boldsymbol{\alpha},\eta,\boldsymbol{b})=
	\begin{pmatrix}
		\frac{u_{1}}{v_{1}}&\cdots&\frac{u_{n}}{v_{n}}&-\eta^{-1}\sum\limits_{t=0}^{\ell}b_{t}\,S_{k+t-n+1}\\
		\frac{u_{1}}{v_{1}}\alpha_{1}&\cdots&\frac{u_{n}}{v_{n}}\alpha_{n}&-\eta^{-1}\sum\limits_{t=0}^{\ell}b_{t}\,S_{k+t-n+2}\\
		\vdots& &\vdots&\vdots\\
		\frac{u_{1}}{v_{1}}\alpha_{1}^{n-k-1}&\cdots&\frac{u_{n}}{v_{n}}\alpha_{n}^{n-k-1}&-\eta^{-1}\sum\limits_{t=0}^{\ell}b_{t}\,S_{t}\\
		\frac{u_{1}}{v_{1}}\Lambda_{1,h}&\cdots&\frac{u_{n}}{v_{n}}\Lambda_{n,h}&-\eta^{-1}
	\end{pmatrix}_{(n+1-k)\times(n+1)}
	\]
	is a parity-check matrix of $\C_4$.
\end{theorem}

\begin{proof}
	It is easy to know that $\C_4$ has the generator matrix
$$
\G_{h}(\boldsymbol{\alpha},\eta,\boldsymbol{b})=\begin{pmatrix}
	v_{1}&\cdots&v_{n}&0\\
	v_{1}\alpha_{1}&\cdots&v_{n}\alpha_{n}&0\\
	\vdots& &\vdots&\vdots\\
	v_{1}\alpha_{1}^{h-1}&\cdots&v_{n}\alpha_{n}^{h-1}&0\\
	v_{1}\left(\alpha_{1}^{h}+\sum\limits_{j=0}^{\ell}b_{j}\alpha_{1}^{k+j}\right)&\cdots&v_{n}\left(\alpha_{n}^{h}+\sum\limits_{j=0}^{\ell}b_{j}\alpha_{n}^{k+j}\right)&\eta\\
	v_{1}\alpha_{1}^{h+1}&\cdots&v_{n}\alpha_{n}^{h+1}&0\\
	\vdots& &\vdots&\vdots\\
	v_{1}\alpha_{1}^{k-1}&\cdots&v_{n}\alpha_{n}^{k-1}&0\\
\end{pmatrix}_{k\times (n+1)}
$$
To prove that $\H_{h}(\boldsymbol{\alpha},\eta,\boldsymbol{b})$ is the parity-check matrix of $\C_4$, we only need to check that the following two statements are true.

$(1)$ $\rank\left(\H_{h}(\boldsymbol{\alpha},\eta,\boldsymbol{b})\right)=n+1-k$

$(2)$ $\G_{h}(\boldsymbol{\alpha},\eta,\boldsymbol{b})\left(\H_{h}(\boldsymbol{\alpha},\eta,\boldsymbol{b})\right)^{T}=\zero$.

{\bf For the statement (1).} We focus on the $(n+1-k)\times n$ submatrix 
$$\H_{h}^{(1)}(\boldsymbol{\alpha},\eta,\boldsymbol{b})=
\begin{pmatrix}
	\frac{u_{1}}{v_{1}}&\cdots&\frac{u_{n}}{v_{n}}\\
	\frac{u_{1}}{v_{1}}\alpha_{1}&\cdots&\frac{u_{n}}{v_{n}}\alpha_{n}\\
	\vdots& &\\
	\frac{u_{1}}{v_{1}}\alpha_{1}^{n-k-1}&\cdots&\frac{u_{n}}{v_{n}}\alpha_{n}^{n-k-1}\\
	\frac{u_{1}}{v_{1}}\Lambda_{1,h}&\cdots&\frac{u_{n}}{v_{n}}\Lambda_{n,h}
\end{pmatrix}_{(n+1-k)\times n}.$$ 
of $\H_{h}(\boldsymbol{\alpha},\eta,\boldsymbol{b})$. For convenience, we denote $\boldsymbol{h}_{i}^{(1)}(\boldsymbol{\alpha},\eta,\boldsymbol{b})$ be the $(i+1)$-th $(0\leq i\leq n-k)$ row of $\H_{h}^{(1)}(\boldsymbol{\alpha},\eta,\boldsymbol{b})$. Next, we prove that $$\boldsymbol{h}_{0}^{(1)}(\boldsymbol{\alpha},\eta,\boldsymbol{b}),\boldsymbol{h}_{1}^{(1)}(\boldsymbol{\alpha},\eta,\boldsymbol{b}),\ldots, \boldsymbol{h}_{n-k}^{(1)}(\boldsymbol{\alpha},\eta,\boldsymbol{b})$$
are $\F_{q}$-linearly independent. If there exist $\lambda_0,\lambda_1,\ldots,\lambda_{n-k}\in\F_{q}$ such that 
$$\sum\limits_{i=0}^{n-k}\lambda_{i}\boldsymbol{h}_{i}^{(1)}(\boldsymbol{\alpha},\eta,\boldsymbol{b})=\zero,$$
then for each $1\leq j\leq n$, we have
$$\sum\limits_{i=0}^{n-k-1}\lambda_{i}\frac{u_{j}}{v_{j}}\alpha_{j}^{i}+\lambda_{n-k}\frac{u_{j}}{v_{j}}\sum\limits_{s=0}^{n-1-h}(-1)^{s}\sigma_{s}\alpha_{j}^{n-1-h-s}=0.$$
Since $\frac{u_{j}}{v_{j}}\in\F_q^*$, then for each $1\leq j\leq n$,
$$\sum\limits_{i=0}^{n-k-1}\lambda_{i}\alpha_{j}^{i}+\lambda_{n-k}\sum\limits_{s=0}^{n-1-h}(-1)^{s}\sigma_{s}\alpha_{j}^{n-1-h-s}=0.$$
Let $L_1(x)=\sum\limits_{i=0}^{n-k-1}\lambda_{i}x^{i}+\lambda_{n-k}\sum\limits_{s=0}^{n-1-h}(-1)^{s}\sigma_{s}x^{n-1-h-s}$, then $L_1(x)$ has $n$ distinct roots $\alpha_{1},\ldots,\alpha_{n}$. By $0\leq h\leq k-1$, we can get
$$n-k-1<n-k=n-1-(k-1)\leq n-1-h.$$
Furthermore,
$$\deg(L_1(x))\leq\max\left\{n-k-1,n-1-h\right\}=n-1-h\leq n-1<n,$$
it means $L_1(x)\equiv 0$. And by $n-k-1<n-1-h$, we know that the term $x^{n-1-h}$ of $L_1(x)$ has the coefficient $\lambda_{n-k}\sigma_{0}$, then $\lambda_{n-k}=0$. Hence, for each $1\leq j\leq n$, we have
$$\sum\limits_{i=0}^{n-k-1}\lambda_{i}\alpha_{j}^{i}=0,$$
it means that the polynomial $L_2(x)=\sum\limits_{i=0}^{n-k-1}\lambda_ix^i$ has $n$ distinct roots $\alpha_{1},\ldots,\alpha_{n}$. By $n-k-1\leq n-1<n$, $L_2(x)\equiv 0$, so
$$\lambda_0=\cdots=\lambda_{n-k-1}=0.$$ Thus $\boldsymbol{h}_{0}^{(1)}(\boldsymbol{\alpha},\eta,\boldsymbol{b}),\boldsymbol{h}_{1}^{(1)}(\boldsymbol{\alpha},\eta,\boldsymbol{b}),\ldots, \boldsymbol{h}_{n-k}^{(1)}(\boldsymbol{\alpha},\eta,\boldsymbol{b})$
are $\F_{q}$-linearly independent, which means $\rank\left(\H_{h}^{(1)}(\boldsymbol{\alpha},\eta,\boldsymbol{b})\right)=n+1-k$. And so, $\rank\left(\H_{h}(\boldsymbol{\alpha},\eta,\boldsymbol{b})\right)=n+1-k$.

{\bf For the statement (2).} For convenience, we denote $\boldsymbol{g}_{i}^{(4)}$ and $\boldsymbol{h}_{j}^{(4)}$ be the $(i+1)$-th $(0\leq i\leq k-1)$ row and the $(j+1)$-th $(0\leq j\leq n-k)$ row of $\G_{h}(\boldsymbol{\alpha},\eta,\boldsymbol{b})$ and $\H_{h}(\boldsymbol{\alpha},\eta,\boldsymbol{b})$, respectively. 
	
{\bf Case 1.} For $i\in\left\{0,1\ldots,k-1\right\}\backslash\left\{h\right\}$ and $0\le j\le n-k-1$, we have
	\[
	\boldsymbol{g}_{i}^{(4)}\left(\boldsymbol{h}_{j}^{(4)}\right)^{T}
	=\sum_{s=1}^{n}u_{s}\alpha_{s}^{i+j}.
	\]
By $i\in\left\{0,1\ldots,k-1\right\}\backslash\left\{h\right\}$ and $0\le j\le n-k-1$, we can get $$0\le i+j\le(k-1)+(n-k-1)=n-2.$$ Thus by Lemma \ref{usapowersum}, we get
	$\boldsymbol{g}_{i}^{(4)}\left(\boldsymbol{h}_{j}^{(4)}\right)^{T}=0$.
	
{\bf Case 2.} For $i=h$ and $0\le j\le n-k-1$, we have
$$\boldsymbol{g}_{i}^{(4)}\left(\boldsymbol{h}_{j}^{(4)}\right)^{T}=\sum_{s=1}^{n}u_{s}\alpha_{s}^{h+j}
+\sum_{t=0}^{\ell}b_{t}\sum_{s=1}^{n}u_{s}\alpha_{s}^{k+t+j}
+\eta\cdot\left(-\eta^{-1}\sum\limits_{t=0}^{\ell}b_{t}\,S_{k+t-n+1+j}\right).$$
By $0\leq h\leq k-1$, we can get $0\le h+j\le n-2$, then by Lemma
\ref{usapowersum}, we have 
$$\boldsymbol{g}_{i}^{(4)}\left(\boldsymbol{h}_{j}^{(4)}\right)^{T}=\sum_{t=0}^{\ell}b_{t}S_{k+t+j-n+1}
+\eta\cdot\left(-\eta^{-1}\sum\limits_{t=0}^{\ell}b_{t}\,S_{k+t-n+1+j}\right)=0.$$
	
	{\bf Case 3.} For $i\in\left\{0,1\ldots,k-1\right\}\backslash\left\{h\right\}$ and $j=n-k$, by Lemma
	\ref{usapowersum}, we have
$$
\begin{aligned}
\boldsymbol{g}_{i}^{(4)}\left(\boldsymbol{h}_{j}^{(4)}\right)^{T}=&\sum\limits_{s=1}^{n}u_{s}\alpha_{s}^{i}\Lambda_{s,h}=\sum\limits_{s=1}^{n}u_{s}\alpha_{s}^{i}\left(\sum\limits_{t=0}^{n-1-h}(-1)^{t}\sigma_{t}\alpha_{s}^{n-1-h-t} \right)\\
=&\sum\limits_{t=0}^{n-1-h}(-1)^{t}\sigma_{t}\left(\sum\limits_{s=1}^{n}u_{s}\alpha_{s}^{i+n-1-h-t}\right)\\
=&\sum\limits_{t=0}^{n-1-h}(-1)^{t}\sigma_{t}S_{i-h-t}.
\end{aligned}
$$

If $i<h$, then by Definition \ref{completesymmetricpolynomial}, $S_{i-h-t}=0$ for any $t\geq 0$. Hence, $\boldsymbol{g}_{i}^{(4)}\left(\boldsymbol{h}_{j}^{(4)}\right)^{T}=0$.

If $i>h$, then  
$$\sum\limits_{t=0}^{n-1-h}(-1)^{t}\sigma_{t}S_{i-h-t}=\sum\limits_{t=0}^{i-h}(-1)^{t}\sigma_{t}S_{i-h-t}+\sum\limits_{t=i-h+1}^{n-1-h}(-1)^{t}\sigma_{t}S_{i-h-t}.$$
When $t\geq i-h+1$, we can get $i-h-t<0$, then by Definition \ref{completesymmetricpolynomial}, $S_{i-h-t}=0$. Hence, by Remark \ref{sigmaS}, 
$$\boldsymbol{g}_{i}^{(4)}\left(\boldsymbol{h}_{j}^{(4)}\right)^{T}=\sum\limits_{t=0}^{n-1-h}(-1)^{t}\sigma_{t}S_{i-h-t}=\sum\limits_{t=0}^{i-h}(-1)^{t}\sigma_{t}S_{i-h-t}=0.$$
	
And so, for $i\in\left\{0,1\ldots,k-1\right\}\backslash\left\{h\right\}$ and $j=n-k$, we can get $\boldsymbol{g}_{i}^{(4)}\left(\boldsymbol{h}_{j}^{(4)}\right)^{T}=0.$

{\bf Case 4.} For $i=h$ and $j=n-k$, by Lemma
\ref{usapowersum}, we have
$$
\begin{aligned}
\boldsymbol{g}_{i}^{(4)}\left(\boldsymbol{h}_{j}^{(4)}\right)^{T}
=&\sum\limits_{s=1}^{n}u_{s}\left(\alpha_{s}^{h}+\sum\limits_{j=0}^{\ell}b_{j}\alpha_{s}^{k+j}\right)\Lambda_{s,h}+\eta\cdot(-\eta^{-1})\\
=&\sum\limits_{s=1}^{n}u_{s}\alpha_{s}^{h}\left(\sum\limits_{t=0}^{n-1-h}(-1)^{t}\sigma_{t}\alpha_{s}^{n-1-h-t} \right)+\sum\limits_{s=1}^{n}u_{s}\left(\sum\limits_{j=0}^{\ell}b_{j}\alpha_{s}^{k+j}\right)\left(\sum\limits_{t=0}^{n-1-h}(-1)^{t}\sigma_{t}\alpha_{s}^{n-1-h-t} \right)-1\\
=&\sum\limits_{t=0}^{n-1-h}(-1)^{t}\sigma_{t}\left(\sum\limits_{s=1}^{n}u_{s}\alpha_{s}^{n-1-t} \right)+\sum\limits_{j=0}^{\ell}b_{j}\sum\limits_{t=0}^{n-1-h}(-1)^{t}\sigma_{t}\left(\sum\limits_{s=1}^{n}u_{s}\alpha_{s}^{k+j+n-1-t-h} \right)-1\\
=&\sum\limits_{t=0}^{n-1-h}(-1)^{t}\sigma_{t}S_{-t}+\sum\limits_{j=0}^{\ell}b_{j}\sum\limits_{t=0}^{n-1-h}(-1)^{t}\sigma_{t}S_{k+j-t-h}-1\\
=&\sum\limits_{j=0}^{\ell}b_{j}\sum\limits_{t=0}^{n-1-h}(-1)^{t}\sigma_{t}S_{k+j-t-h}\\
=&\sum\limits_{j=0}^{\ell}b_{j}\left(\sum\limits_{t=0}^{k+j-h}(-1)^{t}\sigma_{t}S_{k+j-t-h}+\sum\limits_{t=k+j-h+1}^{n-1-h}(-1)^{t}\sigma_{t}S_{k+j-t-h}\right).
\end{aligned}
$$
When $t\geq k+j-h+1$, we can get $k+j-t-h<0$, then by Definition \ref{completesymmetricpolynomial}, $S_{k+j-t-h}=0$. Hence, by Remark \ref{sigmaS}, 
$$\boldsymbol{g}_{i}^{(4)}\left(\boldsymbol{h}_{j}^{(4)}\right)^{T}=\sum\limits_{j=0}^{\ell}b_{j}\sum\limits_{t=0}^{k+j-h}(-1)^{t}\sigma_{t}S_{k+j-t-h}=0.$$

This proves Theorem \ref{extendedhTGRSH}.
\end{proof} 
\begin{remark}
By taking $h=k-1$, $\ell=0$ and $b_0=\eta\in\F_q^*$
	in Theorem \ref{extendedhTGRSH}, for the corresponding $\H_{k-1}(\boldsymbol{\alpha}_n,\eta,(\eta))$, we have
$$\H_{k-1}(\boldsymbol{\alpha}_n,\eta,(\eta))=-\begin{pmatrix}
			1&0&\cdots&0&0\\
			0&1&\cdots&0&0\\
			\vdots&\vdots&\ddots&\vdots&\vdots\\
			0&0&\cdots&1&0\\
			(-1)^{n-k}\sigma_{n-k}
			&(-1)^{n-k-1}\sigma_{n-k-1}
			&\cdots&-\sigma_1&1
		\end{pmatrix}
		_{(n+1-k)\times(n+1-k)}\cdot\H_{n+1-k}
		\operatorname{diag}(1,\ldots,1,\eta^{-1}),
$$
	where $\H_{n+1-k}$ is given in
	\cite[Theorem~3.1]{Zhu2024extendedTGRS}.
	Hence, $\H_{k-1}(\boldsymbol{\alpha}_n,\eta,(\eta))$
	and $\H_{n+1-k}\operatorname{diag}(1,\ldots,1,\eta^{-1})$
	are row-equivalent to each other.
\end{remark}

Finally, we end this section with an open question.
\begin{openproblem}
Give a parity-check matrix of $\C_{k,\ell}(\L,\P,\A_{m\times m},\B)$ with the most general $\A_{m\times m}$ and the most general $\B$, simultaneously. 
\end{openproblem}

\section{The indirect constructions of parity-check matrix for three classes of non-GRS codes}\label{sec4}

In coding theory, there are several methods to construct new linear codes with interesting parameters and properties from the known ones. Among them are the shortening, puncturing, augmenting and extending techniques. In particular, two distinct extension operations, referred to as the first-kind extension and the second-kind extension, have been widely adopted. To date, most of known families of non-GRS linear codes are constructed via the first-kind extension: given a linear code $\mathcal{C}$ with the generator matrix $\boldsymbol{G}$, one appends an extra column vector $\boldsymbol{g}$ to form the extended code $\underline{\mathcal{C}}\left(\boldsymbol{G},\boldsymbol{g}\right)$.

In this section, motivated by the fact that this first-kind extended code is linked to the second-kind extended code $\overline{\mathcal{C}}(\boldsymbol{w})$ through the linear equation $\boldsymbol{G}\boldsymbol{w}^\top=\boldsymbol{g}^\top$, we use Cramer's rule to derive an explicit expression for a vector $\boldsymbol{w}$ satisfying $\boldsymbol{G}\boldsymbol{w}^\top=\boldsymbol{g}^\top$. This expression is then used to construct parity-check matrices for $(1)$-column TGRS codes, $(2)$-column TGRS codes, a family of $(\mathcal{L},\mathcal{P},s)$ row-column TGRS codes, and modified GRS codes $\mathrm{MGRS}_{n,k}(\boldsymbol{\alpha},\boldsymbol{v},\eta,t)$ with $1\leq t\leq k-1$. Additionally, direct constructions of the vector $\boldsymbol{w}$ satisfying $\boldsymbol{G}\boldsymbol{w}^\top=\boldsymbol{g}^\top$ yield parity-check matrices for $\mathcal{C}_h(\eta,b,\delta)$ and a special class of $\mathcal{C}_{k,\ell}(\mathcal{L},\mathcal{P}, \mathcal{A}_{2\times 2},\boldsymbol{B}_1)$.

The definition of the first kind of
extended codes is given.
\begin{definition}
	Let $\C_1$ be an $[n,k]$ linear code with the generator matrix $\G_1$  over $\mathbb{F}_q$ and $\boldsymbol{g}$ be a $1\times k$ vector over $\mathbb{F}_q$. The linear code $\C_2$ generated by the matrix
	\[
	\G_2=
	\begin{pmatrix}
		\G_1 & \boldsymbol{g}^T
	\end{pmatrix}
	\] is called the first kind of the extended code for $\C_1$, and denote $\C_2$ by $\underline{\C}\left(\G_1,\boldsymbol{g}\right)$.
\end{definition}
The definition of the second kind of
extended codes is given as the following,
\begin{definition}[\cite{SunZH2024MDSextend}]
	Let $\w = (w_1,w_2,\cdots,w_n) \in \F_q^n$ be any vector. For a  given $[n,k,d]$ linear code $\C$, the second kind of
	the extended code $\overline{\C}(\w)$ with the parameters $[n+1,k,\widetilde{d}]$ over $\F_q$ is defined as follows,
	\begin{equation}\label{eq:extend-w}
		\overline{C}(\w) = \left\{ (c_1,\cdots,c_n,c_{n+1})|(c_1,\cdots,c_n) \in \C,\, c_{n+1} = \sum_{i=1}^n w_i c_i \right\}, 
	\end{equation}
	where $\widetilde{d}=d$ or $\widetilde{d}=d+1$.
\end{definition}

In 2024, Sun et al., pointed out that the generator matrix and parity-check matrix of the linear codes $\C$ and $\overline{\C}(\w)$ are closely related given by the following Lemma \ref{CwGH}.  
\begin{lemma}[\cite{SunZH2024MDSextend}]\label{CwGH}
	Let $\C$ be an $[n,k,d]$ linear code over $\F_q$ and $\w\in\F_q^n$. If $\C$ has a generator matrix $\G$ and a parity-check matrix $\H$, then the generator matrix  and the parity-check matrix for the extended code $\overline{\C}(\w)$ defined in \eqref{eq:extend-w} is $
	\begin{pmatrix}
		\G&\G\w^T
	\end{pmatrix}$ and $\begin{pmatrix}
	\H&\zero^T\\
	\w&-1
	\end{pmatrix}$ , respectively.
\end{lemma}

Additionally, they also showed that there exists a connection between two classes of extended codes. For $\underline{\C}(\G,\boldsymbol{g})$, there exist $q^{n-k}$ vectors $\w$ such that $\overline{\C}(\w)=\underline{\C}(\G,\boldsymbol{g})$. The precise statement is given below.

\begin{lemma}\label{firstsecondextened}
	For a given extended code $\underline{\C}(\G,\boldsymbol{g})$ with $\G$ being a $k\times n$ generator matrix of $\C$ and $\boldsymbol{g}$ being a $1\times k$ vector over $\F_q$, there are $q^{n-k}$ vectors $\w\in\mathbb{F}_q^n$ such that $\overline{\C}(\w)=\underline{\C}(\G,\boldsymbol{g})$, where $\rank(\G)=\rank\big((\G\ \boldsymbol{g}^T)\big)=k\le n$. The vector $\w$ is exactly the solution for the system of equations $\G\w^T=\boldsymbol{g}^T$.
\end{lemma}

Up to now, for certain families of linear codes, the first kind of extended codes have been constructed by choosing appropriate vector $\boldsymbol{g}$, and their corresponding properties have been investigated \cite{WangMGRS2026, Zhang2026extendedTGRS, Ma2026extendedTGRS,Liu2026ColumnTGRS,Bhagat2026RowColumnTGRS}. However, explicit parity‑check matrices for these extended codes have not been provided in the existing literature.

 According to the above Lemmas \ref{CwGH}-\ref{firstsecondextened}, to obtain a parity‑check matrix of $\underline{\C}(\G,\boldsymbol{g})$, it suffices to find one solution of the equation $\G\w^T=\boldsymbol{g}^T$. Although Sun et al. theoretically showed that there exist $q^{n-k}$ such vector $\w$, but they did not provide the explicit expression for $\w$. As a consequence, the parity‑check matrix of $\underline{\C}(\G,\boldsymbol{g})$ cannot be directly computed. 
 
 The following Theorem \ref{CextendedCw} presents a special vector $\w$ for an arbitrary $\underline{\C}(\G,\boldsymbol{g})$ based on the generator matrix $\G$ of $\C$.

\begin{theorem}\label{CextendedCw}
Let $\C$ be an $[n,k]$ linear code with the generator matrix $\G$ over $\F_q$ and $\underline{\C}(\G,\boldsymbol{g})$ be an $[n+1,k]$ linear code with the generator matrix $\begin{pmatrix}
\G&\boldsymbol{g}^T
\end{pmatrix}$, where $\boldsymbol{g}\in\F_q^k$ is a $1\times k$ nonzero vector. Choose the $\F_q$-linearly independent column of $\G$ indexed by
$j_1,\ldots,j_k$, and let $\G^{(k)}$ be the corresponding submatrix.
For $i=j_r$ with $1\le r\le k$, let $\G^{(k,i)}$
be the matrix obtained by replacing the $r$-th column of
$\G^{(k)}$ with $\boldsymbol g^T$. Then there exists a nonzero vector $\w$ with $$w_i=\begin{cases}
\frac{ \det\left(\G^{(k,i)}\right)}{\det\left(\G^{(k)}\right)}, & i\in\left\{j_1,j_2,\ldots,j_k\right\};\\
0,&i\in\left\{1,2,\ldots,n\right\}\backslash\left\{j_1,j_2,\ldots,j_k\right\},
\end{cases}$$ such that $\G\w^T=\boldsymbol{g}^T$, i.e., $\underline{\C}(\G,\boldsymbol{g})=\overline{\C}(\w)$.
\end{theorem}
\begin{proof}
By $\rank\left(\G\right)=k$, there exist $k$ columns of $\G$ that are $\mathbb{F}_q$-linearly independent. We assume that the $j_1$-th, $j_2$-th, $\ldots$, $j_k$-th columns of $\G$ are $\mathbb{F}_q$-linearly independent, and denote by $\G^{(k)}$ the submatrix formed by these $k$ columns. Then the $k\times k$ submatrix $\G^{(k)}$ is invertible over $\F_q$, i.e., $\det\left(\G^{(k)}\right)\in\F_q^*$. Furthermore, by $w_{i}=0$ for $i\in\left\{1,2,\ldots,n\right\}\backslash\left\{j_1,j_2,\ldots,j_k\right\}$, we have $$\G\left(\w\right)^{T}=\boldsymbol{g}^T\Longleftrightarrow \G^{(k)}\left(w_{j_1},\ldots,w_{j_k}\right)^{T}=\boldsymbol{g}^T.$$
According to Cramer's rule, $w_{i}=\frac{ \det\left(\G^{(k,i)}\right)}{\det\left(\G^{(k)}\right)}( i\in\left\{j_1,j_2,\ldots,j_k\right\})$ is the solution of the equation $$\G^{(k)}\left(w_{j_1},\ldots,w_{j_k}\right)^{T}=\boldsymbol{g}^T,$$
where $\G^{(k,i)}$ is obtained by replacing the $r$-th
column of $\G^{(k)}$ with $\boldsymbol g^T$ for $i=j_r$ with $1\le r\le k$.

This proves Theorem \ref{CextendedCw}.
\end{proof}

It's worth noting that the above vector $\w$ relies heavily on the complexity of the generation matrix $\G$. For some linear codes with a relatively clear generator matrix structure, Theorem \ref{CextendedCw} is more broadly applicable, for example, GRS codes, GRL codes and column TRS codes. 
\subsection{The parity-check matrices for two classes of column TRS codes}
In this subsection, for the column TGRS code $\C_{s}\left(\boldsymbol{\lambda},\boldsymbol{b},c\right)$, we give the corresponding parity-check matrix. For convenience, we denote $\C_{1}\left(\boldsymbol{\lambda},\boldsymbol{b},c\right)$ be $\C_{5}$. And denote
$$\frac{\det\left(\G_{t}^{(k,i)}(\boldsymbol{\lambda},b_t,c)\right)}{\det\left(\G_{t}^{(k)}(\boldsymbol{\lambda},\boldsymbol{b},c)\right)}=\frac{\prod\limits_{\substack{j=1\\j\neq i}}^{k}(b_t-\alpha_j)-\lambda_t\prod\limits_{\substack{j=1\\j\neq i}}^{k}(c-\alpha_j)}{v_i\prod\limits_{1\leq j\leq k,j\neq i}\left(\alpha_{i}-\alpha_{j}\right)}, 1\leq t\leq s, 1\leq i\leq k,$$ 
where $\G_{t}^{(k)}(\boldsymbol{\lambda},\boldsymbol{b},c)$ represents the matrix formed by the first $k$ columns for  the generator matrix $\G_{t}(\boldsymbol{\lambda},\boldsymbol{b},c)$ of $\C_{t}\left(\boldsymbol{\lambda},\boldsymbol{b},c\right)$.

\begin{theorem}\label{onecolumnTRSH}
The $(1)$-column TGRS code has the parity-check matrix
$$\begin{pmatrix}
	\frac{u_1}{v_1}&\cdots&\frac{u_k}{v_k}&\frac{u_{k+1}}{v_{k+1}}&\cdots&\frac{u_{n}}{v_{n}}&0\\
	\frac{u_1}{v_1}\alpha_{1}&\cdots&\frac{u_k}{v_k}\alpha_{k}&\frac{u_{k+1}}{v_{k+1}}\alpha_{k+1}&\cdots&\frac{u_{n}}{v_{n}}\alpha_{n}&0\\
	\frac{u_1}{v_1}\alpha_{1}^2&\cdots&\frac{u_k}{v_k}\alpha_{k}^2&\frac{u_{k+1}}{v_{k+1}}\alpha_{k+1}^2&\cdots&\frac{u_{n}}{v_{n}}\alpha_{n}^2&0\\
	\vdots& &\vdots&\vdots& &\vdots&\vdots\\
	\frac{u_1}{v_1}\alpha_{1}^{n-k-1}&\cdots&\frac{u_k}{v_k}\alpha_{k}^{n-k-1}&\frac{u_{k+1}}{v_{k+1}}\alpha_{k+1}^{n-k-1}&\cdots&\frac{u_{n}}{v_{n}}\alpha_{n}^{n-k-1}&0\\
	\frac{\det\left(\G_{1}^{(k,1)}(\boldsymbol{\lambda},b_1,c)\right)}{\det\left(\G_{1}^{(k)}(\boldsymbol{\lambda},\boldsymbol{b},c)\right)}&\cdots&\frac{\det\left(\G_{1}^{(k,k)}(\boldsymbol{\lambda},b_1,c)\right)}{\det\left(\G_{1}^{(k)}(\boldsymbol{\lambda},\boldsymbol{b},c)\right)}&0&\cdots&0&-1
\end{pmatrix}.$$
\end{theorem}
\begin{proof}
By Definition \ref{columnTGRSdefinition}, it is easy to know that $\C_5$ has the generator matrix
$$\G_{1}(\boldsymbol{\lambda},\boldsymbol{b},c)=
\begin{pmatrix}
	v_{1} & v_{2} & \cdots & v_{n} & 1-\lambda_1 \\
	v_{1}\alpha_{1} & v_{2}\alpha_{2} & \cdots & v_{n}\alpha_{n} & b_1-\lambda_1 c \\
	v_{1}\alpha_{1}^{2} & v_{2}\alpha_{2}^{2} & \cdots & v_{n}\alpha_{n}^{2} & b_{1}^{2}-\lambda_{1} c^{2} \\
	\vdots & \vdots & \vdots & \vdots & \vdots \\
	v_{1}\alpha_{1}^{k-1} & v_{2}\alpha_{2}^{k-1} & \cdots & v_{n}\alpha_{n}^{k-1} & b_{1}^{k-1}-\lambda_{1} c^{k-1}
\end{pmatrix}$$
It is easy to know that the first $n$ columns of $\G_{1}(\boldsymbol{\lambda},\boldsymbol{b},c)$ form the generator matrix of an $[n,k]$ RS code, then any $k$ columns among the first $n$ columns of $\G_{1}(\boldsymbol{\lambda},\boldsymbol{b},c)$  are  $\F_q$-linearly independent. Without loss of generality, we select the first $k$ columns. Then 
$$\det\left(\G_{1}^{(k)}(\boldsymbol{\lambda},\boldsymbol{b},c)\right)=\det\begin{pmatrix}
	v_{1} & v_{2} & \cdots & v_{k}\\
	v_{1}\alpha_{1} & v_{2}\alpha_{2} & \cdots & v_{k}\alpha_{k}\\
	v_{1}\alpha_{1}^{2} & v_{2}\alpha_{2}^{2} & \cdots & v_{k}\alpha_{k}^{2}\\
	\vdots & \vdots & \vdots & \vdots\\
	v_{1}\alpha_{1}^{k-1} & v_{2}\alpha_{2}^{k-1} & \cdots & v_{k}\alpha_{k}^{k-1}
\end{pmatrix}=\prod\limits_{i=1}^{k}v_i\cdot\prod\limits_{1\leq i<j\leq k}\left(\alpha_{j}-\alpha_{i}\right),$$
and
$$\begin{aligned}
&\det\left(\G_{1}^{(k,i)}(\boldsymbol{\lambda},b_1,c)\right)\\
=&\det\begin{pmatrix}
	v_{1} & \cdots & v_{i-1}&1-\lambda_{1}&v_{i+1}& \cdots & v_{k}\\
	v_{1}\alpha_{1} & \cdots & v_{i-1}\alpha_{i-1}&b_{1}-\lambda_{1} c&v_{i+1}\alpha_{i+1} & \cdots &v_{k}\alpha_{k}\\
	v_{1}\alpha_{1}^2 & \cdots & v_{i-1}\alpha_{i-1}^2&b_{1}^2-\lambda_{1} c^2&v_{i+1}\alpha_{i+1}^2 & \cdots & v_{k}\alpha_{k}^2\\
	\vdots & \cdots & \vdots & \vdots& \vdots & \cdots & \vdots\\
	v_{1}\alpha_{1}^{k-1} & \cdots &v_{i-1}\alpha_{i-1}^{k-1}&b_{1}^{k-1}-\lambda_{1} c^{k-1}&v_{i+1}\alpha_{i+1}^{k-1} & \cdots & v_{k}\alpha_{k}^{k-1}\\
\end{pmatrix}\\
=&\det\begin{pmatrix}
	v_{1} & \cdots & v_{i-1}&1&v_{i+1}& \cdots & v_{k}\\
	v_{1}\alpha_{1} & \cdots & v_{i-1}\alpha_{i-1}&b_{1}&v_{i+1}\alpha_{i+1} & \cdots &v_{k}\alpha_{k}\\
	v_{1}\alpha_{1}^2 & \cdots & v_{i-1}\alpha_{i-1}^2&b_{1}^2&v_{i+1}\alpha_{i+1}^2 & \cdots & v_{k}\alpha_{k}^2\\
	\vdots & \cdots & \vdots & \vdots& \vdots & \cdots & \vdots\\
	v_{1}\alpha_{1}^{k-1} & \cdots &v_{i-1}\alpha_{i-1}^{k-1}&b_{1}^{k-1}&v_{i+1}\alpha_{i+1}^{k-1} & \cdots & v_{k}\alpha_{k}^{k-1}
\end{pmatrix}\\
&-\lambda_{1}\cdot\det\begin{pmatrix}
v_{1} & \cdots & v_{i-1}&1&v_{i+1}& \cdots & v_{k}\\
v_{1}\alpha_{1} & \cdots & v_{i-1}\alpha_{i-1}&c&v_{i+1}\alpha_{i+1} & \cdots &v_{k}\alpha_{k}\\
v_{1}\alpha_{1}^2 & \cdots & v_{i-1}\alpha_{i-1}^2&c^2&v_{i+1}\alpha_{i+1}^2 & \cdots & v_{k}\alpha_{k}^2\\
\vdots & \cdots & \vdots & \vdots& \vdots & \cdots & \vdots\\
v_{1}\alpha_{1}^{k-1} & \cdots &v_{i-1}\alpha_{i-1}^{k-1}&c^{k-1}&v_{i+1}\alpha_{i+1}^{k-1} & \cdots & v_{k}\alpha_{k}^{k-1}
\end{pmatrix}\\
=&(-1)^{k-i}\prod\limits_{s=1,s\neq i}^{k}v_s\cdot\prod_{\substack{1\le s< t\le k\\ s\neq i,\, t\neq i}}(\alpha_t-\alpha_s)\left(\prod_{\substack{j=1\\j\neq i}}^{k}(b_{1}-\alpha_j)-\lambda_{1}\prod_{\substack{j=1\\j\neq i}}^{k}(c-\alpha_j)\right).
\end{aligned}$$
And so, by
$$
\prod\limits_{1\le s<t\le k}(\alpha_t-\alpha_s)
=
(-1)^{k-i}
\prod\limits_{\substack{1\le s<t\le k\\s,t\ne i}}(\alpha_t-\alpha_s)
\prod\limits_{j\ne i}(\alpha_i-\alpha_j),
$$
we have
$$\begin{aligned}
\frac{\det\left(\G_{1}^{(k,i)}(\boldsymbol{\lambda},b_1,c)\right)}{\det\left(\G_{1}^{(k)}(\boldsymbol{\lambda},\boldsymbol{b},c)\right)}=&(-1)^{k-i}v_{i}^{-1}\prod_{\substack{1\le s< t\le k\\ s\neq i,\, t\neq i}}(\alpha_t-\alpha_s)\left(\prod\limits_{\substack{j=1\\j\neq i}}^{k}(b_{1}-\alpha_j)-\lambda_{1}\prod\limits_{\substack{j=1\\j\neq i}}^{k}(c-\alpha_j)\right)\cdot\prod\limits_{1\leq t<s\leq k}\left(\alpha_{s}-\alpha_{t}\right)^{-1}\\
=&\frac{\prod_{\substack{j=1\\j\neq i}}^{k}(b_{1}-\alpha_j)-\lambda_{1}\prod\limits_{\substack{j=1\\j\neq i}}^{k}(c-\alpha_j)}{v_i\prod\limits_{1\leq j\leq k,j\neq i}\left(\alpha_{i}-\alpha_{j}\right)}.
\end{aligned}
$$
Note that $\G_{1}(\boldsymbol{\lambda},\boldsymbol{b},c)=\begin{pmatrix}
\G_{n,k}\left(\boldsymbol{\alpha}\right)&\boldsymbol{g}^{T}
\end{pmatrix}$ with $\boldsymbol{g}=\left(1-\lambda_{1},b_{1}-\lambda_{1} c,\ldots,b_{1}^{k-1}-\lambda_{1} c^{k-1}\right)$, where $\G_{n,k}\left(\boldsymbol{\alpha}\right)$ is the generator matrix of an $[n,k]$ GRS code. It is well known that an $[n,k]$ GRS code has the parity-check matrix
$$\H_{n,k}\left(\boldsymbol{\alpha}\right)=\begin{pmatrix}
\frac{u_1}{v_1}&\cdots&\frac{u_{n}}{v_{n}}\\
\frac{u_1}{v_1}\alpha_{1}&\cdots&\frac{u_{n}}{v_{n}}\alpha_{n}\\
\frac{u_1}{v_1}\alpha_{1}^2&\cdots&\frac{u_{n}}{v_{n}}\alpha_{n}^2\\
\vdots& &\vdots\\
\frac{u_1}{v_1}\alpha_{1}^{n-k-1}&\cdots&\frac{u_{n}}{v_{n}}\alpha_{n}^{n-k-1}
\end{pmatrix}_{(n-k)\times n}.$$
Then by Theorem \ref{CextendedCw},  we prove Theorem \ref{onecolumnTRSH}.
\end{proof}

In the similar proof as that for Theorem \ref{onecolumnTRSH}, for $\C_{s}\left(\boldsymbol{\lambda},\boldsymbol{b},c\right)$, we have 
\begin{theorem}\label{twocolumnTRSH}
The code $\C_{s}\left(\boldsymbol{\lambda},\boldsymbol{b},c\right)$ has the parity-check matrix
	$$\begin{pmatrix}
		\frac{u_1}{v_1}&\cdots&\frac{u_k}{v_k}&\frac{u_{k+1}}{v_{k+1}}&\cdots&\frac{u_{n}}{v_{n}}&0&\cdots&0\\
	\frac{u_1}{v_1}\alpha_{1}&\cdots&\frac{u_k}{v_k}\alpha_{k}&\frac{u_{k+1}}{v_{k+1}}\alpha_{k+1}&\cdots&\frac{u_{n}}{v_{n}}\alpha_{n}&0&\cdots&0\\
	\frac{u_1}{v_1}\alpha_{1}^2&\cdots&\frac{u_k}{v_k}\alpha_{k}^2&\frac{u_{k+1}}{v_{k+1}}\alpha_{k+1}^2&\cdots&\frac{u_{n}}{v_{n}}\alpha_{n}^2&0&\cdots&0\\
	\vdots& &\vdots&\vdots& &\vdots&\vdots& &\vdots\\
	\frac{u_1}{v_1}\alpha_{1}^{n-k-1}&\cdots&\frac{u_k}{v_k}\alpha_{k}^{n-k-1}&\frac{u_{k+1}}{v_{k+1}}\alpha_{k+1}^{n-k-1}&\cdots&\frac{u_{n}}{v_{n}}\alpha_{n}^{n-k-1}&0&\cdots&0\\
		\frac{\det\left(\G_{1}^{(k,1)}(\boldsymbol{\lambda},b_1,c)\right)}{\det\left(\G_{1}^{(k)}(\boldsymbol{\lambda},\boldsymbol{b},c)\right)}&\cdots&\frac{\det\left(\G_{1}^{(k,k)}(\boldsymbol{\lambda},b_1,c)\right)}{\det\left(\G_{1}^{(k)}(\boldsymbol{\lambda},\boldsymbol{b},c)\right)}&0&\cdots&0&-1&\cdots&0\\ 
		\vdots& &\vdots&\vdots& &\vdots&\vdots& &\vdots\\
		\frac{\det\left(\G_{s}^{(k,1)}(\boldsymbol{\lambda},b_s,c)\right)}{\det\left(\G_{s}^{(k)}(\boldsymbol{\lambda},\boldsymbol{b},c)\right)}&\cdots&\frac{\det\left(\G_{s}^{(k,k)}(\boldsymbol{\lambda},b_s,c)\right)}{\det\left(\G_{s}^{(k)}(\boldsymbol{\lambda},\boldsymbol{b},c)\right)}&0&\cdots&0&0&\cdots&-1
	\end{pmatrix}_{(n+s-k)\times (n+s)}.$$
\end{theorem} 

\begin{remark}
	$(1)$ Our result provides a unified formulation of the known parity-check matrices. Specifically, for
	$3\le k\le n-1$, by taking $s=2$ and $v_1=\cdots=v_n=1$
	in Theorem \ref{twocolumnTRSH}, we know that the corresponding matrix is row-equivalent
	to that in \cite[Theorem~5.3]{GuchangCTGRS2026}.
	Additionally, if $b_1=b_2=b$ and
	$\alpha_1,\ldots,\alpha_n,b,c$ are pairwise distinct,
	it is row-equivalent to the parity-check matrix of
	$\C_2(b,c)$ in \cite[Theorem~3(1)]{Liu2026ColumnTGRS}.
	
	$(2)$ By our method, one only computes certain determinants,
	and the parity-check matrix has a clear block structure.
\end{remark}

\subsection{The parity-check matrix for a class of $(\L,\P, 1)$-row-column TRS codes} 
In this subsection, we consider
$\C_1(\boldsymbol\lambda,\boldsymbol b,c,\L,\P,\B)$
with $\ell=0$, $\L=\{0\}$, $\P=\{k-1\}$,
$\eta\in\F_q^*$ and  $$\B=\begin{pmatrix}
	\zero_{(k-1)\times 1}& \zero_{(k-1)\times n-k-1}\\
	\eta &\zero_{1\times (n-k-1)}
\end{pmatrix},$$ we give the parity-check matrix. For convenience, we denote this code be $\C_6$.
\begin{theorem}\label{onerowcolumnTGRSH}
Let
$$S_{k}=\left\{\sum\limits_{i\in I}\alpha_{i}|\forall I\subseteq \left\{1,2,\ldots,n\right\}, \# I=k\right\}$$
If $-\eta^{-1}\notin S_{k}$ and one of the following two conditions holds,

$(1)$ $\sum\limits_{i=1}^{n}\alpha_{i}\neq 0$ and $\eta\neq -\left(\sum\limits_{i=1}^{n}\alpha_{i}\right)^{-1}$;

$(2)$ $\sum\limits_{i=1}^{n}\alpha_{i}=0$ and $\eta\neq 0$,\\ then $\C_6$ has the parity-check matrix
$$\small\begin{pmatrix}
\frac{u_1}{v_1}&\cdots&\frac{u_k}{v_k}&\frac{u_{k+1}}{v_{k+1}}&\cdots&\frac{u_{n}}{v_{n}}&0\\
\frac{u_1}{v_1}\alpha_{1}&\cdots&\frac{u_k}{v_k}\alpha_{k}&\frac{u_{k+1}}{v_{k+1}}\alpha_{k+1}&\cdots&\frac{u_{n}}{v_{n}}\alpha_{n}&0\\
\frac{u_1}{v_1}\alpha_{1}^2&\cdots&\frac{u_k}{v_k}\alpha_{k}^2&\frac{u_{k+1}}{v_{k+1}}\alpha_{k+1}^2&\cdots&\frac{u_{n}}{v_{n}}\alpha_{n}^2&0\\
\vdots& &\vdots\\
\frac{u_1}{v_1}\alpha_{1}^{n-k-2}&\cdots&\frac{u_k}{v_k}\alpha_{k}^{n-k-2}&\frac{u_{k+1}}{v_{k+1}}\alpha_{k+1}^{n-k-2}&\cdots&\frac{u_{n}}{v_{n}}\alpha_{n}^{n-k-2}&0\\
\frac{u_1}{v_1}\left(\alpha_{1}^{n-k-1}-\Delta\alpha_{1}^{n-k}\right)&\cdots&\frac{u_k}{v_k}\left(\alpha_{k}^{n-k-1}-\Delta\alpha_{k}^{n-k}\right)&\frac{u_{k+1}}{v_{k+1}}\left(\alpha_{k+1}^{n-k-1}-\Delta\alpha_{k+1}^{n-k}\right)&\cdots&\frac{u_n}{v_n}\left(\alpha_{n}^{n-k-1}-\Delta\alpha_{n}^{n-k}\right)&0\\
	\frac{\det\left(\G_{k-1,1}^{(k,1)}(\boldsymbol{\alpha},\boldsymbol{b},c,\eta)\right)}{\det\left(\G_{k-1,1}^{(k)}(\boldsymbol{\alpha},\boldsymbol{b},c,\eta)\right)}&\cdots&\frac{\det\left(\G_{k-1,1}^{(k,k)}(\boldsymbol{\alpha},\boldsymbol{b},c,\eta)\right)}{\det\left(\G_{k-1,1}^{(k)}(\boldsymbol{\alpha},\boldsymbol{b},c,\eta)\right)}&0&\cdots&0&-1
\end{pmatrix},$$
where $\Delta=\frac{\eta}{1+\eta\sum\limits_{i=1}^{n}\alpha_{i}}$ and $$\frac{\det\left(\G_{k-1,1}^{(k,i)}(\boldsymbol{\alpha},\boldsymbol{b},c,\eta)\right)}{\det\left(\G_{k-1,1}^{(k)}(\boldsymbol{\alpha},\boldsymbol{b},c,\eta)\right)}=\frac{\left(1+\eta\left(b_{1}+\sum\limits_{j=1,j\neq i}^{k}\alpha_{j}\right)\right)\prod\limits_{\substack{j=1\\j\neq i}}^{k}(b_{1}-\alpha_j)-\lambda_{1}\left(1+\eta\left(c+\sum\limits_{j=1,j\neq i}^{k}\alpha_{j}\right)\right)\prod\limits_{\substack{j=1\\j\neq i}}^{k}(c-\alpha_j)}{v_i\left(1+\eta\sum\limits_{i=1}^{k}\alpha_{i}\right)\prod\limits_{1\leq j\leq k,j\neq i}\left(\alpha_{i}-\alpha_{j}\right)}.$$
\end{theorem}
\begin{proof}
	By Definition of \ref{rowcolumnTGRSdefinition}, $\C_6$ has the generator matrix
	$$\begin{aligned}
&\boldsymbol{G}_{1}(\boldsymbol{\lambda},\boldsymbol{b},c,\L,\P,\B)\\
\triangleq&\G_{k-1,1}(\boldsymbol{\alpha},\boldsymbol{b},c,\eta)\\
=&
	\begin{pmatrix}
		v_{1} & v_{2} & \cdots & v_{n} & 1-\lambda_{1} \\
		v_{1}\alpha_{1} & v_{2}\alpha_{2} & \cdots & v_{n}\alpha_{n} & b_{1}-\lambda_{1} c \\
		v_{1}\alpha_{1}^{2} & v_{2}\alpha_{2}^{2} & \cdots & v_{n}\alpha_{n}^{2} & b_{1}^{2}-\lambda_{1} c^{2} \\
		\vdots & \vdots & \vdots & \vdots & \vdots \\
		v_{1}\alpha_{1}^{k-2} & v_{2}\alpha_{2}^{k-2} & \cdots & v_{n}\alpha_{n}^{k-2} & b_{1}^{k-2}-\lambda_{1} c^{k-2}\\
		v_{1}\left(\alpha_{1}^{k-1}+\eta\alpha_{1}^{k}\right) & v_{2}\left(\alpha_{2}^{k-1}+\eta\alpha_{2}^{k}\right)  & \cdots & v_{n}\left(\alpha_{n}^{k-1}+\eta\alpha_{n}^{k} \right) & b_{1}^{k-1}-\lambda_{1} c^{k-1}+\eta\left(b_{1}^{k}-\lambda_{1} c^{k}\right)
	\end{pmatrix}
\end{aligned}$$
	It is easy to know that the first $n$ columns of $\G_{k-1,1}(\boldsymbol{\alpha},\boldsymbol{b},c,\eta)$ form the generator matrix of an $[n,k]$ $(+)$-TRS code, then any $k$ columns among the first $n$ columns of $\G_{k-1,1}(\boldsymbol{\alpha},\boldsymbol{b},c,\eta)$  are  $\F_q$-linearly independent. Without loss of generality, we select the first $k$ columns. Then 
	$$\begin{aligned} 
	\det\left(\G_{k-1,1}^{(k)}(\boldsymbol{\alpha},\boldsymbol{b},c,\eta)\right)=&\det\begin{pmatrix}
		v_{1} & v_{2} & \cdots & v_{k}\\
		v_{1}\alpha_{1} & v_{2}\alpha_{2} & \cdots & v_{k}\alpha_{k}\\
		v_{1}\alpha_{1}^{2} & v_{2}\alpha_{2}^{2} & \cdots & v_{k}\alpha_{k}^{2}\\
		\vdots & \vdots & \vdots & \vdots\\
		v_{1}\alpha_{1}^{k-2} & v_{2}\alpha_{2}^{k-2} & \cdots & v_{k}\alpha_{k}^{k-2}\\
		v_{1}\left(\alpha_{1}^{k-1}+\eta\alpha_{1}^k\right) & v_{2}\left(\alpha_{2}^{k-1}+\eta\alpha_{2}^k\right) & \cdots & v_{k}\left(\alpha_{k}^{k-1}+\eta\alpha_{k}^k\right)
	\end{pmatrix}\\
	=&\left(1+\eta\sum\limits_{i=1}^{k}\alpha_{i}\right)\prod\limits_{i=1}^{k}v_i\prod\limits_{1\leq i<j\leq k}\left(\alpha_{j}-\alpha_{i}\right),
\end{aligned}$$
	and
	$$\small\begin{aligned}
		&\det\left(\G_{k-1,1}^{(k,i)}(\boldsymbol{\alpha},\boldsymbol{b},c,\eta)\right)\\
		=&\det\begin{pmatrix}
			v_{1} & \cdots & v_{i-1}&1-\lambda_{1}&v_{i+1} & \cdots &v_{k}\\
			v_{1}\alpha_{1} & \cdots & v_{i-1}\alpha_{i-1}&b_{1}-\lambda_{1} c&v_{i+1}\alpha_{i+1} & \cdots &v_{k}\alpha_{k}\\
			v_{1}\alpha_{1}^2 & \cdots &v_{i-1}\alpha_{i-1}^2&b_{1}^2-\lambda_{1} c^2&v_{i+1}\alpha_{i+1}^2 & \cdots &v_{k}\alpha_{k}^2\\
			\vdots & \cdots & \vdots & \vdots& \vdots & \cdots & \vdots\\
			v_{1}\alpha_{1}^{k-2} & \cdots &v_{i-1}\alpha_{i-1}^{k-2}&b_{1}^{k-2}-\lambda_{1} c^{k-2}&v_{i+1}\alpha_{i+1}^{k-2} & \cdots &v_{k}\alpha_{k}^{k-2}\\
			v_{1}\left(\alpha_{1}^{k-1} +\eta\alpha_{1}^k\right)& \cdots & v_{i-1}\left(\alpha_{i-1}^{k-1}+\eta\alpha_{i-1}^k\right)&b_{1}^{k-1}-\lambda_{1} c^{k-1}+\eta\left(b_{1}^{k}-\lambda_{1} c^{k}\right)&v_{i+1}\left(\alpha_{i+1}^{k-1}+\eta\alpha_{i+1}^k \right)& \cdots & v_{k}\left(\alpha_{k}^{k-1}+\eta\alpha_{k}^k\right)
		\end{pmatrix}\\
		=&\det\begin{pmatrix}
			v_{1} & \cdots & v_{i-1}&1-\lambda_{1}&v_{i+1} & \cdots &v_{k}\\
		v_{1}\alpha_{1} & \cdots & v_{i-1}\alpha_{i-1}&b_{1}-\lambda_{1} c&v_{i+1}\alpha_{i+1} & \cdots &v_{k}\alpha_{k}\\
		v_{1}\alpha_{1}^2 & \cdots &v_{i-1}\alpha_{i-1}^2&b_{1}^2-\lambda_{1} c^2&v_{i+1}\alpha_{i+1}^2 & \cdots &v_{k}\alpha_{k}^2\\
		\vdots & \cdots & \vdots & \vdots& \vdots & \cdots & \vdots\\
		v_{1}\alpha_{1}^{k-2} & \cdots &v_{i-1}\alpha_{i-1}^{k-2}&b_{1}^{k-2}-\lambda_{1} c^{k-2}&v_{i+1}\alpha_{i+1}^{k-2} & \cdots &v_{k}\alpha_{k}^{k-2}\\
		v_{1}\alpha_{1}^{k-1}& \cdots & v_{i-1}\alpha_{i-1}^{k-1}&b_{1}^{k-1}-\lambda_{1} c^{k-1}&v_{i+1}\alpha_{i+1}^{k-1}& \cdots & v_{k}\alpha_{k}^{k-1}
		\end{pmatrix}\\
		&+\eta\cdot\det\begin{pmatrix}
			v_{1} & \cdots & v_{i-1}&1-\lambda_{1}&v_{i+1} & \cdots &v_{k}\\
		v_{1}\alpha_{1} & \cdots & v_{i-1}\alpha_{i-1}&b_{1}-\lambda_{1} c&v_{i+1}\alpha_{i+1} & \cdots &v_{k}\alpha_{k}\\
		v_{1}\alpha_{1}^2 & \cdots &v_{i-1}\alpha_{i-1}^2&b_{1}^2-\lambda_{1} c^2&v_{i+1}\alpha_{i+1}^2 & \cdots &v_{k}\alpha_{k}^2\\
		\vdots & \cdots & \vdots & \vdots& \vdots & \cdots & \vdots\\
		v_{1}\alpha_{1}^{k-2} & \cdots &v_{i-1}\alpha_{i-1}^{k-2}&b_{1}^{k-2}-\lambda_{1} c^{k-2}&v_{i+1}\alpha_{i+1}^{k-2} & \cdots &v_{k}\alpha_{k}^{k-2}\\
		v_{1}\alpha_{1}^k& \cdots & v_{i-1}\alpha_{i-1}^k&b_{1}^{k}-\lambda_{1} c^{k}&v_{i+1}\alpha_{i+1}^k& \cdots & v_{k}\alpha_{k}^k
		\end{pmatrix}.
	\end{aligned}$$
By directly computing, we have
$$
\begin{aligned}
&\det\left(\G_{k-1,1}^{(k,i)}(\boldsymbol{\alpha},\boldsymbol{b},c,\eta)\right)\\
=&(-1)^{k-i}\prod\limits_{s=1,s\neq i}^{k}v_s\cdot\prod_{\substack{1\le s< t\le k\\ s\neq i,\, t\neq i}}(\alpha_t-\alpha_s)\left(\prod_{\substack{j=1\\j\neq i}}^{k}(b_{1}-\alpha_j)-\lambda_{1}\prod_{\substack{j=1\\j\neq i}}^{k}(c-\alpha_j)\right)\\
&+(-1)^{k-i}\prod\limits_{s=1,s\neq i}^{k}v_s\cdot\eta\left(\left(b_{1}+\sum\limits_{j=1,j\neq i}^{k}\alpha_{j}\right)\prod_{\substack{1\le s< t\le k\\ s\neq i,\, t\neq i}}(\alpha_t-\alpha_s)\prod_{\substack{j=1\\j\neq i}}^{k}(b_{1}-\alpha_j)\right.\\
&\left.-\lambda_{1}\left(c+\sum\limits_{j=1,j\neq i}^{k}\alpha_{j}\right)\prod_{\substack{1\le s< t\le k\\ s\neq i,\, t\neq i}}(\alpha_t-\alpha_s)\prod_{\substack{j=1\\j\neq i}}^{k}(c-\alpha_j)\right)\\
=&(-1)^{k-i}\prod\limits_{s=1,s\neq i}^{k}v_s\cdot\prod_{\substack{1\le s< t\le k\\ s\neq i,\, t\neq i}}(\alpha_t-\alpha_s)\cdot\mathcal{K}_{i},
\end{aligned}
$$
where
$$\mathcal{K}_{i}=\left(1+\eta\left(b_{1}+\sum\limits_{j=1,j\neq i}^{k}\alpha_{j}\right)\right)\prod_{\substack{j=1\\j\neq i}}^{k}(b_{1}-\alpha_j)-\lambda_{1}\left(1+\eta\left(c+\sum\limits_{j=1,j\neq i}^{k}\alpha_{j}\right)\right)\prod_{\substack{j=1\\j\neq i}}^{k}(c-\alpha_j).$$
	And so, by $$\prod\limits_{1\leq i<j\leq k}\left(\alpha_{j}-\alpha_{i}\right)=(-1)^{k-i}\cdot\prod_{\substack{1\le s< t\le k\\ s\neq i,\, t\neq i}}(\alpha_t-\alpha_s)\cdot\prod\limits_{1\leq j\leq k,j\neq i}\left(\alpha_{i}-\alpha_{j}\right),$$
we have
	$$
\frac{\det\left(\G_{k-1,1}^{(k,i)}(\boldsymbol{\alpha},\boldsymbol{b},c,\eta)\right)}{\det\left(\G_{k-1,1}^{(k)}(\boldsymbol{\alpha},\boldsymbol{b},c,\eta)\right)}=\frac{\mathcal{K}_{i}}{v_i\left(1+\eta\sum\limits_{i=1}^{k}\alpha_{i}\right)\prod\limits_{1\leq j\leq k,j\neq i}\left(\alpha_{i}-\alpha_{j}\right)}.$$
	Note that $\G_{k-1,1}(\boldsymbol{\alpha},\boldsymbol{b},c,\eta)=\begin{pmatrix}
		\G_{n,k}\left(\boldsymbol{\alpha},\eta\right)&\boldsymbol{g}^{T}
	\end{pmatrix}$ with $$\boldsymbol{g}=\left(1-\lambda_{1},b_{1}-\lambda_{1} c,\ldots,b_{1}^{k-2}-\lambda_{1} c^{k-2},b_{1}^{k-1}-\lambda_{1} c^{k-1}+\eta\left(b_{1}^{k}-\lambda_{1} c^{k}\right)\right),$$ where $\G_{n,k}\left(\boldsymbol{\alpha},\eta\right)$ is the generator matrix of an $[n,k]$ $(+)$-TGRS code. By \cite[Theorem 2.4]{Huang2021TGRS}, we know that if $\sum\limits_{i=1}^{n}\alpha_{i}\neq 0$ and $\eta\neq -\left(\sum\limits_{i=1}^{n}\alpha_{i}\right)^{-1}$, or $\sum\limits_{i=1}^{n}\alpha_{i}=0$ and $\eta\neq 0$, then the $[n,k]$ $(+)$-TGRS code has the parity-check matrix
	$$\H_{n,k}\left(\boldsymbol{\alpha}\right)=\begin{pmatrix}
\frac{u_1}{v_1}&\cdots&\frac{u_{n}}{v_{n}}\\
\frac{u_1}{v_1}\alpha_{1}&\cdots&\frac{u_{n}}{v_{n}}\alpha_{n}\\
\frac{u_1}{v_1}\alpha_{1}^2&\cdots&\frac{u_{n}}{v_{n}}\alpha_{n}^2\\
\vdots& &\vdots\\
\frac{u_1}{v_1}\alpha_{1}^{n-k-2}&\cdots&\frac{u_{n}}{v_{n}}\alpha_{n}^{n-k-2}\\
		\frac{u_{1}}{v_{1}}\left(\alpha_{1}^{n-k-1}-\frac{\eta}{1+\eta\sum\limits_{i=1}^{n}\alpha_{i}}\alpha_{1}^{n-k}\right)&\cdots&\frac{u_{n}}{v_{n}}\left(\alpha_{n}^{n-k-1}-\frac{\eta}{1+\eta\sum\limits_{i=1}^{n}\alpha_{i}}\alpha_{n}^{n-k}\right)
	\end{pmatrix}_{(n-k)\times n}.$$
	Now, by Theorem \ref{CextendedCw}, we prove Theorem \ref{onerowcolumnTGRSH}.
\end{proof}
\subsection{Parity-check matrices for a special case of $\C_1$ via two successive extensions}
In this subsection, we consider the special case of $\C_1$
with $m=2$ and
$\A_2=\begin{pmatrix}0&1\\1&\delta\end{pmatrix}$, denote by $\C_{1}^{(1)}$.
We construct the parity-check matrix basing on the parity-check matrix of
$\C(\L,\P,\B_1)$ by two successive extensions.

Firstly, by constructing a  $1\times n$  vector, we prove that $\C_{k,\ell}(\boldsymbol{\alpha}, \boldsymbol{v},\A_1,\B_1)$ is the second kind extended code of $\C(\L,\P,\B_1)$, where $\A_1=\left(1\right)$ and $\B_1$ is given in \eqref{Bi}.  
\begin{theorem}\label{secondextendedw1}
	If $n\geq k+\ell+1$ and $b_{k-1,\ell}\in\F_q^*$, then $\C_{k,\ell}(\boldsymbol{\alpha}, \boldsymbol{v},\A_1,\B_1)=\overline{\C(\L,\P,\B_1)}\left(\w^{(1)}\right)$ with $$w_i^{(1)}=b_{k-1,\ell}^{-1}\frac{u_i}{v_i}\alpha_{i}^{n-k-1-\ell}(1\leq i\leq n).$$
\end{theorem}
\begin{proof}
	By Definition \ref{LPTGRSdefinition}, it is easy to know that $\C(\L,\P,\B_1)$ has the generator matrix
	$$
	\G(\L,\P,\B_1)=\begin{pmatrix}
		v_1&\cdots&v_n\\
		v_1\alpha_{1}&\cdots&v_n\alpha_{n}\\
		\vdots& &\vdots\\
		v_1\alpha_{1}^{k-3}&\cdots&v_n\alpha_{n}^{k-3}\\
		v_1\alpha_{1}^{k-2}&\cdots&v_n\alpha_{n}^{k-2}\\ 
		v_1\left(\alpha_{1}^{k-1}+\sum\limits_{j=0}^{\ell} b_{k-1,j}\alpha_{1}^{k+j} \right)&\cdots&v_n\left(\alpha_{n}^{k-1}+\sum\limits_{j=0}^{\ell} b_{k-1,j}\alpha_{n}^{k+j} \right)
	\end{pmatrix}.
	$$
	Then we only check
	$$\G(\L,\P,\B_1)\left(\w^{(1)}\right)^{T}=\left(0,\ldots,0,1\right)^{T}.$$
	By direct calculation, it's easy to complete the verification as follows.

For convenience, denote $\boldsymbol{g}_{i}^{(4)}(0\leq i\leq k-1)$ be the $(i+1)$-th row vector of $\G(\L,\P,\B_1)$.

For $0\leq i\leq k-2$, note that
$$n-k-1-\ell+i\leq n-\ell-3<n-2,$$
then by Lemma \ref{usapowersum}, we have
$$\boldsymbol{g}_{i}^{(4)}\left(\w^{(1)}\right)^{T}=b_{k-1,\ell}^{-1}\sum\limits_{s=1}^{n}u_s\alpha_{s}^{n-k-1-\ell+i}=0,$$

For $i=k-1$, by Lemma \ref{usapowersum}, we have
$$\begin{aligned}
\boldsymbol{g}_{i}^{(4)}\left(\w^{(1)}\right)^{T}=&b_{k-1,\ell}^{-1}\sum\limits_{s=1}^{n}u_s\left(\alpha_{s}^{i}+\sum\limits_{j=0}^{\ell} b_{k-1,j}\alpha_{s}^{k+j} \right)\alpha_{s}^{n-k-1-\ell}\\
=&b_{k-1,\ell}^{-1}\sum\limits_{s=1}^{n}u_s\alpha_{s}^{n-k-1-\ell+i}+b_{k-1,\ell}^{-1}\sum\limits_{j=0}^{\ell} b_{k-1,j}\sum\limits_{s=1}^{n}u_s\alpha_{s}^{n-k-1-\ell+k+j},
\end{aligned}$$

Note that for $i=k-1$,
$$n-k-1-\ell+i=n-2-\ell\leq n-2,$$
and for $0\leq j\leq \ell-1$, $$n-k-1-\ell+k+j=n-1-\ell+j\leq n-2,$$ then
	$$\begin{aligned}
\boldsymbol{g}_{i}^{(4)}\left(\w^{(1)}\right)^{T}=&b_{k-1,\ell}^{-1}\left(\sum\limits_{j=0}^{\ell-1} b_{k-1,j}\sum\limits_{s=1}^{n}u_s\alpha_{s}^{n-k-1-\ell+k+j}+b_{k-1,\ell}\sum\limits_{s=1}^{n}u_s\alpha_{s}^{n-k-1-\ell+k+\ell}\right)\\
=&b_{k-1,\ell}^{-1}\cdot b_{k-1,\ell}\sum\limits_{i=1}^{n}u_i\alpha_{i}^{n-k-1-\ell+k+\ell}\\
=&1.
	\end{aligned}$$
	
This proves Theorem \ref{secondextendedw1}.	  
\end{proof}

Secondly, by constructing an $1\times (n+1)$ vector, we prove that $\C_{k,\ell}(\boldsymbol{\alpha}, \boldsymbol{v},\A_2,\B_1)$ is the second kind of extended codes of $\C_{k,\ell}(\boldsymbol{\alpha}, \boldsymbol{v},\A_1,\B_1)$, where $\A_1=(1)$, $\A_2=\begin{pmatrix}
0&1\\
1&\delta
\end{pmatrix}$ and $\B_i(i=1,2)$ is given in \eqref{Bi}.
\begin{theorem}\label{secondextendedw2}
	$\C_{k,\ell}(\boldsymbol{\alpha}, \boldsymbol{v},\A_2,\B_1)=\overline{\C_{k,\ell}(\boldsymbol{\alpha}, \boldsymbol{v},\A_1,\B_1)}\left(\w^{(2)}\right)$ with $w_i^{(2)}=\begin{cases}
		\frac{u_i}{v_i}\alpha_{i}^{n-k+1}, & 1\leq i\leq n;\\
		\delta-S_1-\sum\limits_{j=0}^{\ell} b_{k-1,j}S_{j+2}; &i=n+1.
	\end{cases}$
\end{theorem}
\begin{proof}
	By Remark \ref{GRLTGRSG}, it is easy to know that $\C_{k,\ell}(\boldsymbol{\alpha}, \boldsymbol{v},\A_1,\B_1)$ has the generator matrix
	$$
	\G(\B_1)=\begin{pmatrix}
		v_1&\cdots&v_n&0\\
		v_1\alpha_{1}&\cdots&v_n\alpha_{n}&0\\
		\vdots& &\vdots&\vdots \\
		v_1\alpha_{1}^{k-3}&\cdots&v_n\alpha_{n}^{k-3}&0\\
		v_1\alpha_{1}^{k-2}&\cdots&v_n\alpha_{n}^{k-2}&0\\ 
		v_1\left(\alpha_{1}^{k-1}+\sum\limits_{j=0}^{\ell} b_{k-1,j}\alpha_{1}^{k+j} \right)&\cdots&v_n\left(\alpha_{n}^{k-1}+\sum\limits_{j=0}^{\ell} b_{k-1,j}\alpha_{n}^{k+j} \right)&1
	\end{pmatrix}.
	$$
	Then we only check
	$$\G(\B_1)\left(\w^{(2)}\right)^{T}=\left(0,\ldots,0,1,\delta\right)^{T}.$$
	By direct calculation, it's easy to complete the verification as the follows.

For convenience, denote $\boldsymbol{g}_{i}^{(6)}(0\leq i\leq k-1)$ be the $(i+1)$-th row vector of $\G(\B_1)$. 
 
For $0\leq i\leq k-2$, by Lemma \ref{usapowersum}, we have
	$$\boldsymbol{g}_{i}^{(6)}\left(\w^{(2)}\right)^{T}=\begin{cases}
		\sum\limits_{s=1}^{n}u_s\alpha_{s}^{n-k+1+i}=0, & 0\leq i\leq k-3;\\
		\sum\limits_{s=1}^{n}u_s\alpha_{s}^{n-k+1+i}=1, & i=k-2.
	\end{cases}$$

For $i=k-1$, by Lemma \ref{usapowersum}, we have 
	$$
	\begin{aligned}
	\boldsymbol{g}_{k-1}^{(6)}\left(\w^{(2)}\right)^{T}=&\sum\limits_{s=1}^{n}u_s\left(\alpha_{s}^{k-1}+\sum\limits_{j=0}^{\ell} b_{k-1,j}\alpha_{s}^{k+j} \right)\alpha_{s}^{n-k+1}+\left(\delta-S_1-\sum\limits_{j=0}^{\ell} b_{k-1,j}S_{j+2}\right)\\
	=&S_{1}+\sum\limits_{j=0}^{\ell} b_{k-1,j}S_{j+2}+\left(\delta-S_1-\sum\limits_{j=0}^{\ell} b_{k-1,j}S_{j+2}\right)\\
	=&\delta.
\end{aligned}$$

This proves Theorem \ref{secondextendedw2}.	
\end{proof}

Finally, by Lemma \ref{CwGH} and Theorem 3.1 of \cite{Liang+LPTGRS2026}, it is easy to obtain the following
\begin{theorem}\label{C1otherH}
Let $n\geq k+\ell+1$, $\mathcal{A}=\delta-S_1-\sum\limits_{j=0}^{\ell} b_{k-1,j}S_{j+2}$ and $\beta=b_{k-1,\ell}\in\F_q^*$. Then the following two statements are true. 

$(1)$ If $1+\sum\limits_{t=0}^{\ell}b_{k-1,t}S_{t+1}\neq 0$, then 
$$\H_{3}(\delta,\B_1)=\begin{pmatrix}
	\frac{u_{1}}{v_{1}}&\cdots&\frac{u_{n}}{v_{n}}& 0&0\\
	\frac{u_{1}}{v_{1}}\alpha_{1}&\cdots&\frac{u_{n}}{v_{n}}\alpha_{n}&0&0\\
	\vdots& &\vdots&\vdots&\vdots\\
	\frac{u_{1}}{v_{1}}\alpha_{1}^{n-k-(\ell+2)}&\cdots&\frac{u_{n}}{v_{n}}\alpha_{n}^{n-k-(\ell+2)}&0&0\\ 
	\frac{u_1}{v_1}\alpha_{1}^{n-k-1-\ell}&\cdots&\frac{u_n}{v_n}\alpha_{n}^{n-k-1-\ell}&-b_{k-1,\ell}&0\\
	\frac{u_{1}}{v_{1}}\alpha_{1}^{n-k-\ell}&\cdots&\frac{u_{n}}{v_{n}}\alpha_{n}^{n-k-\ell}&-b_{k-1,\ell}\varTheta_{n-k-(\ell+1)}^{-1}\varTheta_{n-k-\ell}&0\\ 
	\vdots& &\vdots&\vdots&\vdots\\
	\frac{u_{1}}{v_{1}}\alpha_{1}^{n-k-1}&\cdots&\frac{u_{n}}{v_{n}}\alpha_{n}^{n-k-1}&-b_{k-1,\ell}\varTheta_{n-k-(\ell+1)}^{-1}\varTheta_{n-k-1}&0\\
	\frac{u_{1}}{v_{1}}\alpha_{1}^{n-k}&\cdots&\frac{u_{n}}{v_{n}}\alpha_{n}^{n-k}&-b_{k-1,\ell}\varTheta_{n-k-(\ell+1)}^{-1}&0\\
	\frac{u_1}{v_1}\alpha_{1}^{n-k+1}&\cdots&\frac{u_n}{v_n}\alpha_{n}^{n-k+1}&\mathcal{A}&-1 
\end{pmatrix}_{(n+2-k)\times (n+2)}$$
is a parity-check matrix of the code $\C_{1}^{(1)}$, where $\varTheta_{i}=\frac{\sum\limits_{t=0}^{\ell}b_{k-1,t}S_{k+t+i-n+1}}{1+\sum\limits_{t=0}^{\ell}b_{k-1,t}S_{t+1}}(n-k-\ell-1\leq i\leq n-k-1)$.

$(2)$ If $\ell\geq 1$ and  $1+\sum\limits_{t=0}^{\ell}b_{k-1,t}S_{t+1}=0$, then the matrix
$$\H_{4}(\delta,\B_1)=\begin{pmatrix}
	\frac{u_{1}}{v_{1}}&\cdots&\frac{u_{n}}{v_{n}}& 0&0\\
	\frac{u_{1}}{v_{1}}\alpha_{1}&\cdots&\frac{u_{n}}{v_{n}}\alpha_{n}&0&0\\
	\vdots& &\vdots&\vdots&\vdots\\
	\frac{u_{1}}{v_{1}}\alpha_{1}^{n-k-(\ell+2)}&\cdots&\frac{u_{n}}{v_{n}}\alpha_{n}^{n-k-(\ell+2)}&0&0\\
	\frac{u_{1}}{v_{1}}\left(\alpha_{1}^{n-k-\ell}-\Omega_{n-k-\ell}\alpha_{1}^{n-k-(\ell+1)}\right)&\cdots&\frac{u_{n}}{v_{n}}\left(\alpha_{n}^{n-k-\ell}-\Omega_{n-k-\ell}\alpha_{n}^{n-k-(\ell+1)}\right)&0&0\\ 
	\vdots& &\vdots&\vdots&\vdots\\
	\frac{u_{1}}{v_{1}}\left(\alpha_{1}^{n-k-1}-\Omega_{n-k-1}\alpha_{1}^{n-k-(\ell+1)}\right)&\cdots&\frac{u_{n}}{v_{n}}\left(\alpha_{n}^{n-k-1}-\Omega_{n-k-1}\alpha_{n}^{n-k-(\ell+1)}\right)&0&0\\ 
	\frac{u_1}{v_1}\alpha_{1}^{n-k}&\cdots&\frac{u_n}{v_n}\alpha_{n}^{n-k}&0&0\\
	b_{k-1,\ell}^{-1}\frac{u_1}{v_1}\alpha_{1}^{n-k-1-\ell}&\cdots&b_{k-1,\ell}^{-1}\frac{u_n}{v_n}\alpha_{n}^{n-k-1-\ell}&-1&0\\
	\frac{u_1}{v_1}\alpha_{1}^{n-k+1}&\cdots&\frac{u_n}{v_n}\alpha_{n}^{n-k+1}&\mathcal{A}&-1 
\end{pmatrix}_{(n+2-k)\times (n+2)}$$
is a parity-check matrix of the code $\C_{1}^{(1)}$, where $\Omega_{i}=b_{k-1,\ell}^{-1}\sum\limits_{t=0}^{\ell}b_{k-1,t}S_{k+t+i-n+1}(n-k-\ell\leq i\leq n-k-1)$. 
\end{theorem}
\begin{proof}
 By Theorem \ref{CextendedCw} and Theorem 3.1 of \cite{Liang+LPTGRS2026}(To save space, this is not presented in the form of a lemma.), it's pretty easy to know that the following two statements are true.

$(i)$ If $1+\sum\limits_{t=0}^{\ell}b_{k-1,t}S_{t+1}\neq 0$, then 
$$\H_{2}(\delta,\B_1)=\begin{pmatrix}
	\frac{u_{1}}{v_{1}}&\cdots&\frac{u_{n}}{v_{n}}& 0&0\\
	\frac{u_{1}}{v_{1}}\alpha_{1}&\cdots&\frac{u_{n}}{v_{n}}\alpha_{n}&0&0\\
	\vdots& &\vdots&\vdots&\vdots\\
	\frac{u_{1}}{v_{1}}\alpha_{1}^{n-k-(\ell+2)}&\cdots&\frac{u_{n}}{v_{n}}\alpha_{n}^{n-k-(\ell+2)}&0&0\\
	\frac{u_{1}}{v_{1}}\left(\alpha_{1}^{n-k-(\ell+1)}-\varTheta_{n-k-(\ell+1)}\alpha_{1}^{n-k}\right)&\cdots&\frac{u_{n}}{v_{n}}\left(\alpha_{n}^{n-k-(\ell+1)}-\varTheta_{n-k-(\ell+1)}\alpha_{n}^{n-k}\right)&0&0\\ 
	\vdots& &\vdots&\vdots&\vdots\\
	\frac{u_{1}}{v_{1}}\left(\alpha_{1}^{n-k-1}-\varTheta_{n-k-1}\alpha_{1}^{n-k}\right)&\cdots&\frac{u_{n}}{v_{n}}\left(\alpha_{n}^{n-k-1}-\varTheta_{n-k-1}\alpha_{n}^{n-k}\right)&0&0\\
	b_{k-1,\ell}^{-1}\frac{u_1}{v_1}\alpha_{1}^{n-k-1-\ell}&\cdots&b_{k-1,\ell}^{-1}\frac{u_n}{v_n}\alpha_{n}^{n-k-1-\ell}&-1&0\\
	\frac{u_1}{v_1}\alpha_{1}^{n-k+1}&\cdots&\frac{u_n}{v_n}\alpha_{n}^{n-k+1}&\mathcal{A}&-1 
\end{pmatrix}_{(n+2-k)\times (n+2)}$$
is a parity-check matrix of the code $\C_{1}^{(1)}$;

$(ii)$ If $\ell\geq 1$ and $1+\sum\limits_{t=0}^{\ell}b_{k-1,t}S_{t+1}=0$, then 
$\H_{4}(\delta,\B_1)$ is a parity-check matrix of the code $\C_{1}^{(1)}$.

For $(i)$, 
for convenience, we denote $\boldsymbol{g}_{i}(\delta,\B_1)(0\leq i\leq n+1-k)$ be the $(i+1)$-th row vector of $\H_{2}(\delta,\B_1)$.  Note that $$\varTheta_{n-k-(\ell+1)}=\frac{\sum\limits_{t=0}^{\ell}b_{k-1,t}S_{t-\ell}}
{1+\sum\limits_{t=0}^{\ell}b_{k-1,t}S_{t+1}}
=
\frac{b_{k-1,\ell}}
{1+\sum\limits_{t=0}^{\ell}b_{k-1,t}S_{t+1}}\in\F_q^*,$$ then by performing the following elementary row operations on $\H_{2}(\delta,\B_1)$, we can obtain $\H_{3}(\delta,\B_1)$.  

{\bf Step 1.} Replace
$\boldsymbol{g}_{n-k-\ell-1}(\delta,\B_1)$ with
$\boldsymbol{g}_{n-k-\ell-1}^{(1)}(\delta,\B_1)
\triangleq
\boldsymbol{g}_{n-k-\ell-1}(\delta,\B_1)
-b_{k-1,\ell}\boldsymbol{g}_{n-k}(\delta,\B_1).$

{\bf Step 2.} Replace
$\boldsymbol{g}_{n-k-\ell-1}^{(1)}(\delta,\B_1)$ with
$\boldsymbol{g}_{n-k-\ell-1}^{(2)}(\delta,\B_1)
\triangleq
-\varTheta_{n-k-\ell-1}^{-1}
\boldsymbol{g}_{n-k-\ell-1}^{(1)}(\delta,\B_1).$

{\bf Step 3.} Replace $\boldsymbol{g}_{i}(\delta,\B_1)$ with
$\boldsymbol{g}_{i}(\delta,\B_1)
+\varTheta_i
\boldsymbol{g}_{n-k-\ell-1}^{(2)}(\delta,\B_1)$
for $n-k-\ell\leq i\leq n-k-1$.

{\bf Step 4.} Replace $\boldsymbol{g}_{n-k}(\delta,\B_1)$
with $b_{k-1,\ell}\boldsymbol{g}_{n-k}(\delta,\B_1)$. 

{\bf Step 5.} Swap the $(n-k-\ell)$-th and
$(n-k+1)$-th rows of the resulting matrix. 

This proves Theorem \ref{C1otherH}.	
\end{proof}
\begin{remark}
	$(1)$ For $n\ge k+\ell+2$ and the respective assumptions
	in Theorem \ref{C1otherH}, the matrices $\H_3(\delta,\B_1)$
	and $\H_4(\delta,\B_1)$ are both row-equivalent to the case $m=2$ and
	$\A_2=\begin{pmatrix}0&1\\1&\delta\end{pmatrix}$ of Theorem \ref{RL+k-1TGRSH}.
	By choosing those extension vectors, which heavily depend on the prescribed
	additional columns, this construction can be continued to
	obtain a parity-check matrix of $\C_1$.
	Whether it is row-equivalent to the matrix in Theorem \ref{RL+k-1TGRSH} relies on the chosen vectors, and this relies on the required row operations.
	
	$(2)$ The extension construction clearly demonstrates the close
	connection between the original code $\C(\L,\P,\B_1)$
	and the resulting extended codes.
\end{remark}
\subsection{The parity matrix for $\C_{h}\left(\eta,b,\delta\right)$}
In this subsection, from the perspective of extended codes, we present the parity-check matrix of $\C_{h}\left(\eta,b,\delta\right)$ basing on the parity-check matrix of $\C_{h}\left(\eta,b\right)$ given in Theorem \ref{extendedhTGRSH}.

\begin{theorem}\label{C4extendeC4}
$\C_{h}(\boldsymbol{\alpha},\eta,\boldsymbol{b},\delta)=\overline{\C_{h}(\boldsymbol{\alpha},\eta,\boldsymbol{b})}\left(\w^{(4)}\right)$ with $$w_i^{(4)}=\begin{cases}
	 \frac{u_i}{v_i}\delta\alpha_{i}^{n-k}, & 1\leq i\leq n;\\
	 \eta^{-1}\left(\eta-\delta\sum\limits_{j=0}^{\ell}b_{j}S_{j+1}\right), &i=n+1.
\end{cases}$$	
\end{theorem}
\begin{proof}
	It is easy to know that $\C_4$ has the generator matrix
$$
\G_{h}(\boldsymbol{\alpha},\eta,\boldsymbol{b})=\begin{pmatrix}
	v_{1}&\cdots&v_{n}&0\\
	v_{1}\alpha_{1}&\cdots&v_{n}\alpha_{n}&0\\
	\vdots& &\vdots&\vdots\\
	v_{1}\alpha_{1}^{h-1}&\cdots&v_{n}\alpha_{n}^{h-1}&0\\
	v_{1}\left(\alpha_{1}^{h}+\sum\limits_{j=0}^{\ell}b_{j}\alpha_{1}^{k+j}\right)&\cdots&v_{n}\left(\alpha_{n}^{h}+\sum\limits_{j=0}^{\ell}b_{j}\alpha_{n}^{k+j}\right)&\eta\\
	v_{1}\alpha_{1}^{h+1}&\cdots&v_{n}\alpha_{n}^{h+1}&0\\
	\vdots& &\vdots&\vdots\\
	v_{1}\alpha_{1}^{k-1}&\cdots&v_{n}\alpha_{n}^{k-1}&0\\
\end{pmatrix}_{k\times (n+1)}
$$
Then we only check
$$\G_{h}(\boldsymbol{\alpha},\eta,\boldsymbol{b})\left(\w^{(4)}\right)^{T}=\left(0,\ldots,0,\eta,0,\ldots,0,\delta\right)^{T}.$$

For $i\in\left\{0,1,\ldots,k-1\right\}\backslash\left\{h\right\}$, 
$$\delta\sum\limits_{s=1}^{n}u_{s}\alpha_{s}^{n-k+i}=\begin{cases}
0, &i\neq k-1;\\
\delta, &i=k-1.\\
\end{cases}$$

For $i=h$,
$$\delta\sum\limits_{s=1}^{n}u_{s}\left(\alpha_{s}^{h}+\sum\limits_{j=0}^{\ell}b_{j}\alpha_{s}^{k+j}\right)\alpha_{s}^{n-k}+\eta\cdot\eta^{-1}\left(\eta-\delta\sum\limits_{j=0}^{\ell}b_{j}S_{j+1}\right)=\eta.$$

This proves Theorem \ref{C4extendeC4}.	\end{proof}
\begin{remark}
By taking $\ell=1$, $b_0=0$ , $b_1\in\F_q^*$ and $\eta=1$ in Theorem \ref{C4extendeC4}, the corresponding result is just Theorem 3.1 of \cite{Zhang2026extendedTGRS} .
\end{remark}
And so, by Theorem \ref{CextendedCw} and Theorem \ref{extendedhTGRSH}, we have the following
\begin{theorem}
	The matrix
$$\H_{h}(\boldsymbol{\alpha},\eta,\boldsymbol{b},\delta)=
\begin{pmatrix}
	\frac{u_{1}}{v_{1}}&\cdots&\frac{u_{n}}{v_{n}}&-\eta^{-1}\sum\limits_{t=0}^{\ell}b_{t}\,S_{k+t-n+1}&0\\
	\frac{u_{1}}{v_{1}}\alpha_{1}&\cdots&\frac{u_{n}}{v_{n}}\alpha_{n}&-\eta^{-1}\sum\limits_{t=0}^{\ell}b_{t}\,S_{k+t-n+2}&0\\
	\vdots& &\vdots&\vdots&\vdots\\
	\frac{u_{1}}{v_{1}}\alpha_{1}^{n-k-1}&\cdots&\frac{u_{n}}{v_{n}}\alpha_{n}^{n-k-1}&-\eta^{-1}\sum\limits_{t=0}^{\ell}b_{t}\,S_{t}&0\\
	\frac{u_{1}}{v_{1}}\Lambda_{1,h}&\cdots&\frac{u_{n}}{v_{n}}\Lambda_{n,h}&-\eta^{-1}&0\\
	\frac{u_1}{v_1}\delta\alpha_{1}^{n-k}&\cdots&\frac{u_n}{v_n}\delta\alpha_{n}^{n-k}& \eta^{-1}\left(\eta-\delta\sum\limits_{j=0}^{\ell}b_{j}S_{j+1}\right)&-1
\end{pmatrix}_{(n+2-k)\times(n+2)}$$
is a parity-check matrix of $\C_{h}\left(\eta,\boldsymbol{b},\delta\right)$, where $
\Lambda_{i,h}=\sum\limits_{j=0}^{n-1-h}(-1)^{j}\sigma_{j}\,\alpha_{i}^{n-1-h-j}.$
\end{theorem} 

\subsection{The parity-check matrix of $\mathrm{MGRS}_{n+1,k}\left(\alpha,v,\eta,t\right)$}
\begin{theorem}\label{tMGRSH}
If $k\leq n$, then for any $1\leq t\leq k-1$, $\mathrm{MGRS}_{n+1,k}\left(\boldsymbol{\alpha},\boldsymbol{v},\eta,t\right)$ has the following parity-check matrix 
$$\begin{pmatrix}
	\frac{u_1}{v_1}&\cdots&\frac{u_k}{v_k}&\frac{u_{k+1}}{v_{k+1}}&\cdots&\frac{u_{n}}{v_{n}}&0\\
	\frac{u_1}{v_1}\alpha_{1}&\cdots&\frac{u_{k}}{v_k}\alpha_{k}&\frac{u_{k+1}}{v_{k+1}}\alpha_{k+1}&\cdots&\frac{u_{n}}{v_{n}}\alpha_{n}&0\\
	\frac{u_1}{v_1}\alpha_{1}^2&\cdots&\frac{u_{k}}{v_k}\alpha_{k}^2&\frac{u_{k+1}}{v_{k+1}}\alpha_{k+1}^2&\cdots&\frac{u_{n}}{v_{n}}\alpha_{n}^2&0\\
	\vdots& &\vdots&\vdots& &\vdots&\vdots\\
	\frac{u_1}{v_1}\alpha_{1}^{n-k-1}&\cdots&\frac{u_{k}}{v_k}\alpha_{k}^{n-k-1}&\frac{u_{k+1}}{v_{k+1}}\alpha_{k+1}^{n-k-1}&\cdots&\frac{u_{n}}{v_{n}}\alpha_{n}^{n-k-1}&0\\
	\frac{\det\left(\G_{\v,n+1,t}^{(k,1)}\right)}{\det\left(\G_{\v,n+1,t}^{(k)}\right)}&\cdots&\frac{\det\left(\G_{\v,n+1,t}^{(k,k)}\right)}{\det\left(\G_{\v,n+1,t}^{(k)}\right)}&0&\cdots&0&-1
\end{pmatrix},
$$
where $\frac{\det\left(\G_{\v,n+1,t}^{(k,i)}\right)}{\det\left(\G_{\v,n+1,t}^{(k)}\right)}$ is given in \eqref{detGkiGk}.
\end{theorem}
\begin{proof} 
By Definition \ref{MGRSdefinition}, we know that $\mathrm{MGRS}_{n+1,k}\left(\boldsymbol{\alpha},\boldsymbol{v},\eta,t\right)$ has the following generator matrix 
$$\G_{\boldsymbol{v},n+1,t} =\begin{pmatrix}
	v_{1}&v_{2}&\cdots&v_{n}&v_{n+1}\\
	v_{1}\alpha_{1}&v_{2}\alpha_{2}&\cdots&v_{n}\alpha_{n}&0\\
	
	\vdots&\vdots& &\vdots&\vdots\\
	v_{1}\alpha_{1}^{t-1}&v_{2}\alpha_{2}^{t-1}&\cdots&v_{n}\alpha_{n}^{t-1}&0\\
	
	v_{1}\alpha_{1}^{t}&v_{2}\alpha_{2}^{t}&\cdots&v_{n}\alpha_{n}^{t}&v_{n+1}\eta\\
	
	v_{1}\alpha_{1}^{t+1}&v_{2}\alpha_{2}^{t+1}&\cdots&v_{n}\alpha_{n}^{t+1}&0\\
	\vdots&\vdots& &\vdots&\vdots\\ 
	v_{1}\alpha_{1}^{k-1}&v_{2}\alpha_{2}^{k-1}&\cdots&v_{n}\alpha_{n}^{k-1}&0
\end{pmatrix}_{k\times (n+1)}.$$
It is easy to know that the first $n$ columns of $\G_{\v,n+1,t}$ form the generator matrix of an $[n,k]$ GRS code, then any $k$ columns among the first $n$ columns of $\G_{\v,n+1,t}$ are  $\F_q$-linearly independent. Without loss of generality, we select the first $k$ columns. Then 
$$\det\left(\G_{\v,n+1,t}^{(k)}\right)=\det\begin{pmatrix}
	v_{1} & v_{2} & \cdots & v_{k}\\
	v_{1}\alpha_{1} & v_{2}\alpha_{2} & \cdots & v_{k}\alpha_{k}\\
	v_{1}\alpha_{1}^{2} & v_{2}\alpha_{2}^{2} & \cdots & v_{k}\alpha_{k}^{2}\\
	\vdots & \vdots & \vdots & \vdots\\
	v_{1}\alpha_{1}^{k-1} & v_{2}\alpha_{2}^{k-1} & \cdots & v_{k}\alpha_{k}^{k-1}
\end{pmatrix}=\prod\limits_{i=1}^{k}v_i\cdot\prod\limits_{1\leq i<j\leq k}\left(\alpha_{j}-\alpha_{i}\right),$$
and
$$\begin{aligned}
	&\det\left(\G_{\v,n+1,t}^{(k,i)}\right)\\
	=&\det\begin{pmatrix}
		v_{1}&\cdots&v_{i-1}&v_{n+1}&v_{i+1}&\cdots&v_{k}\\
		v_{1}\alpha_{1}&\cdots&v_{i-1}\alpha_{i-1}&0&v_{i+1}\alpha_{i+1}&\cdots&v_{k}\alpha_{k}\\
		
		\vdots&\vdots& &\vdots&\vdots\\
		v_{1}\alpha_{1}^{t-1}&\cdots&v_{i-1}\alpha_{i-1}^{t-1}&0&v_{i+1}\alpha_{i+1}^{t-1}&\cdots&v_{k}\alpha_{k}^{t-1}\\
		
		v_{1}\alpha_{1}^{t}&\cdots&v_{i-1}\alpha_{i-1}^{t}&v_{n+1}\eta&v_{i+1}\alpha_{i+1}^{t}&\cdots&v_{k}\alpha_{k}^{t}\\
		
		v_{1}\alpha_{1}^{t+1}&\cdots&v_{i-1}\alpha_{i-1}^{t+1}&0&v_{i+1}\alpha_{i+1}^{t+1}&\cdots&v_{k}\alpha_{k}^{t+1}\\
		\vdots& &\vdots&\vdots&\vdots& &\vdots\\ 
		v_{1}\alpha_{1}^{k-1}&\cdots&v_{i-1}\alpha_{i-1}^{k-1}&0&v_{i+1}\alpha_{i+1}^{k-1}&\cdots&v_{k}\alpha_{k}^{k-1}
	\end{pmatrix}\\
	=&(-1)^{k-i}\cdot v_{n+1}\cdot \prod\limits_{j=1,j\neq i}^{k}v_{j}\left((-1)^{k+1}\prod\limits_{j=1,j\neq i}^{k}\alpha_{j}\cdot\det\left(\boldsymbol{M}_{1,i}\right)+(-1)^{k+t+1}\eta\cdot\det\left(\boldsymbol{M}_{2,i}\right)\right),
	\end{aligned}
	$$
where
$$\boldsymbol{M}_{1,i}=\begin{pmatrix}
	1&\cdots&1&1&\cdots&1\\
	\alpha_{1}&\cdots&\alpha_{i-1}&\alpha_{i+1}&\cdots&\alpha_{k}\\ 
	\vdots&\vdots& &\vdots&\vdots\\ 
	\alpha_{1}^{k-2}&\cdots&\alpha_{i-1}^{k-2}&\alpha_{i+1}^{k-2}&\cdots&\alpha_{k}^{k-2}
\end{pmatrix},\boldsymbol{M}_{2,i}=\begin{pmatrix}
	1&\cdots&1&1&\cdots&1\\
	\alpha_{1}&\cdots&\alpha_{i-1}&\alpha_{i+1}&\cdots&\alpha_{k}\\ 
	\vdots& &\vdots&\vdots& &\vdots\\
	\alpha_{1}^{t-1}&\cdots&\alpha_{i-1}^{t-1}&\alpha_{i+1}^{t-1}&\cdots&\alpha_{k}^{t-1}\\ 
	\alpha_{1}^{t+1}&\cdots&\alpha_{i-1}^{t+1}&\alpha_{i+1}^{t+1}&\cdots&\alpha_{k}^{t+1}\\
	\vdots&\vdots& &\vdots&\vdots\\ 
	\alpha_{1}^{k-1}&\cdots&\alpha_{i-1}^{k-1}&\alpha_{i+1}^{k-1}&\cdots&\alpha_{k}^{k-1}
\end{pmatrix}.$$
It is easy to get 
$$\det\left(\boldsymbol{M}_{1,i}\right)=\prod\limits_{\substack{1\le r<s\le k\\ r,s\ne i}}
(\alpha_s-\alpha_r),$$
and by Lemma \ref{DtVandermonde} and the ordinary
Vandermonde formula, we have
$$\det\left(\boldsymbol{M}_{2,i}\right)=\prod\limits_{\substack{1\le r<s\le k\\ r,s\ne i}}
(\alpha_s-\alpha_r)\cdot \sigma_{k-1-t}(\alpha_1,\ldots,\alpha_{i-1},\alpha_{i+1},\ldots,\alpha_{k}).$$
Furthermore, 
$$\det\left(\G_{\v,n+1,t}^{(k,i)}\right)=(-1)^{k-i}v_{n+1}\cdot \prod\limits_{j=1,j\neq i}^{k}v_{j}\cdot\prod\limits_{\substack{1\le r<s\le k\\ r,s\ne i}}
(\alpha_s-\alpha_r)\cdot\left((-1)^{k+1}\prod\limits_{j=1,j\neq i}^{k}\alpha_{j}+(-1)^{k+t+1}\eta\sigma_{k-1-t}(\boldsymbol{\alpha_{k,i}})\right),$$
where $\boldsymbol{\alpha_{k,i}}=\left\{\alpha_1,\ldots,\alpha_{i-1},\alpha_{i+1},\ldots,\alpha_{k} \right\}$. Then
\begin{equation}\label{detGkiGk}
\frac{\det\left(\G_{\v,n+1,t}^{(k,i)}\right)}{\det\left(\G_{\v,n+1,t}^{(k)}\right)}=\frac{v_{n+1}\left((-1)^{k+1}\prod\limits_{j=1,j\neq i}^{k}\alpha_{j}+(-1)^{k+t+1}\eta\sigma_{k-1-t}(\boldsymbol{\alpha_{k,i}})\right)}{v_i\prod\limits_{j=1,j\neq i}^{k}
	(\alpha_i-\alpha_j)}
\end{equation}

And so, by Theorem \ref{CextendedCw}, 
we prove Theorem \ref{tMGRSH}.	 
\end{proof}

\section{The MDS or NMDS property for $(1)$-column TGRS codes and a class of $(\L,\P)$-TGRL codes }\label{sec5}
\subsection{The MDS property for a class of $(\L,\P)$-TGRL codes}\label{sec5.1}
In this subsection, we establish a necessary and sufficient condition for a class of $(\L,\P)$-TGRL codes  to be
MDS.

\begin{theorem}\label{MDSTGRL}
The $(\L,\P)$-TGRL code $\C_{k,\ell}(\L,\P,\A_{2\times 2},\B)$ with $k\geq 3$, $\L=[\ell+1]$ and $\P=[k]$ is MDS if and only if the following three statements are true.

$(1)$ $\det\left(\boldsymbol{I}_k+\widetilde{\B_1}\boldsymbol{F}_{\mathcal{I}_{k}}\right)\in\F_{q}^*$ for any $k$-subset $\mathcal{I}_{k}\subseteq\left\{1,2,\ldots,n\right\}$,  where $\widetilde{\B_1}$ is given in \eqref{B1F1} and $\boldsymbol{F}_{\mathcal{I}_{k}}=\begin{pmatrix}
	f_{0,0}^{(\mathcal{I}_{k})}&\cdots&f_{0,k-1}^{(\mathcal{I}_{k})}\\
	\vdots& &\vdots\\
	f_{\ell,0}^{(\mathcal{I}_{k})}&\cdots&f_{\ell,k-1}^{(\mathcal{I}_{k})}
\end{pmatrix}.$

$(2)$ $\left(a_{2i}\cdot\det\left(\boldsymbol{I}_{k-1}+\widetilde{\B_2}\boldsymbol{F}_{\mathcal{I}_{k-1}}\right)-a_{1i}\det\left(\boldsymbol{T}_{\mathcal{I}_{k-1}}+\widetilde{\B_3}\boldsymbol{F}_{\mathcal{I}_{k-1}}\right)\right)\in\F_{q}^*$ for any $(k-1)$-subset $\mathcal{I}_{k-1}\subseteq\left\{1,2,\ldots,n\right\}$ and $i\in\left\{1,2\right\}$,  where $\widetilde{\B_2},\widetilde{\B_3}$ is given in \eqref{TB2F2}, and $$\boldsymbol{T}_{\mathcal{I}_{k-1}}=\begin{pmatrix}
	1&0&\cdots&0&0\\
	0&1&\cdots&0&0\\
	\vdots&\vdots& &\vdots&\vdots\\
	0&0&\cdots&1&0\\
	f_{0,0}^{(\mathcal{I}_{k-1})}&f_{0,1}^{(\mathcal{I}_{k-1})}&\cdots&f_{0,k-3}^{(\mathcal{I}_{k-1})}&f_{0,k-2}^{(\mathcal{I}_{k-1})}\\
\end{pmatrix},\ \boldsymbol{F}_{\mathcal{I}_{k-1}}=\begin{pmatrix}
	f_{1,0}^{(\mathcal{I}_{k-1})}&\cdots&f_{1,k-2}^{(\mathcal{I}_{k-1})}\\
	\vdots& &\vdots\\
	f_{\ell+1,0}^{(\mathcal{I}_{k-1})}&\cdots&f_{\ell+1,k-2}^{(\mathcal{I}_{k-1})}
\end{pmatrix}.$$

$(3)$ $\det\left(\boldsymbol{I}_{k-2}+\widetilde{\B_4}\boldsymbol{F}_{\mathcal{I}_{k-2}}\right)\in\F_{q}^*$ for any $(k-2)$-subset $\mathcal{I}_{k-2}\subseteq\left\{1,2,\ldots,n\right\}$, where $\widetilde{\B_4}$ is given in \eqref{B4F4} and $$\boldsymbol{F}_{\mathcal{I}_{k-2}}=\begin{pmatrix}
	f_{2,0}^{(\mathcal{I}_{k-2})}&\cdots&f_{2,k-3}^{(\mathcal{I}_{k-2})}\\
	\vdots& &\vdots\\
	f_{\ell+2,0}^{(\mathcal{I}_{k-2})}&\cdots&f_{\ell+2,k-3}^{(\mathcal{I}_{k-2})}
\end{pmatrix}.$$
\end{theorem}
\begin{proof}
By Definition \ref{LPTGRLdefinition}, it is easy to know that the $(\L,\P)$-TGRL code $\C_{k,\ell}(\L,\P,\A_{2\times 2},\B)$ with $\L=[\ell+1]$ and $\P=[k]$ has the following generator matrix
$$
\G\left(\L,\P,\A_{2\times 2},\B\right)=\begin{pmatrix}
	v_{1}\left(1+\sum\limits_{j=0}^{\ell}b_{0,j}\alpha_{1}^{k+j}\right)&\cdots&v_{n}\left(1+\sum\limits_{j=0}^{\ell}b_{0,j}\alpha_{n}^{k+j}\right)&0&0\\
	v_{1}\left(\alpha_{1}+\sum\limits_{j=0}^{\ell}b_{1,j}\alpha_{1}^{k+j}\right)&\cdots&v_{n}\left(\alpha_{n}+\sum\limits_{j=0}^{\ell}b_{1,j}\alpha_{n}^{k+j}\right)&0&0\\ 
	\vdots& &\vdots&\vdots&\vdots\\
	v_{1}\left(\alpha_{1}^{k-3}+\sum\limits_{j=0}^{\ell}b_{k-3,j}\alpha_{1}^{k+j}\right)&\cdots&v_{n}\left(\alpha_{n}^{k-3}+\sum\limits_{j=0}^{\ell}b_{k-3,j}\alpha_{n}^{k+j}\right)&0&0\\
	v_{1}\left(\alpha_{1}^{k-2}+\sum\limits_{j=0}^{\ell}b_{k-2,j}\alpha_{1}^{k+j}\right)&\cdots&v_{n}\left(\alpha_{n}^{k-2}+\sum\limits_{j=0}^{\ell}b_{k-2,j}\alpha_{n}^{k+j}\right)&a_{11}&a_{12}\\ 
	v_{1}\left(\alpha_{1}^{k-1}+\sum\limits_{j=0}^{\ell}b_{k-1,j}\alpha_{1}^{k+j}\right)&\cdots&v_{n}\left(\alpha_{n}^{k-1}+\sum\limits_{j=0}^{\ell}b_{k-1,j}\alpha_{n}^{k+j}\right)&a_{21}&a_{22}
\end{pmatrix}_{k\times(n+2)}.
$$
By Lemma \ref{MDSorNMDSequivalent} (1), $\C_{k,\ell}(\L,\P,\A_{2\times 2},\B)$ is MDS if and only if all $k\times k$ minors of $\G\left(\L,\P,\A_{2\times 2},\B\right)$ are nonzero, i.e., the determinant of the $k\times k$ submatrix formed
by any $k$ columns of $\G\left(\L,\P,\A_{2\times 2},\B\right)$ is nonzero. For convenience, $\widetilde{\boldsymbol{g}}_i$ denotes the $i$-th column of $\G\left(\L,\P,\A_{2\times 2},\B\right)$ with $1\leq i\leq n$ and
$$\boldsymbol{a}_i=\left(0,\ldots,0,a_{1i},a_{2i}\right)^T,i=1,2.$$
Without loss of generality, the $k\times k$ submatrix of $\G\left(\L,\P,\A_{2\times 2},\B\right)$ only has three forms, i.e.,
$$\G_1=\left[\widetilde{\boldsymbol{g}}_1,\ldots,\widetilde{\boldsymbol{g}}_k\right], \G_{2i}=\left[\widetilde{\boldsymbol{g}}_1,\ldots,\widetilde{\boldsymbol{g}}_{k-1},\boldsymbol{a}_i\right], \G_3=\left[\widetilde{\boldsymbol{g}}_1,\ldots,\widetilde{\boldsymbol{g}}_{k-2},\boldsymbol{a}_1,\boldsymbol{a}_2\right].$$
Next, we only compute $\det(\G_i)(i=1,2,3)$.

For $\det(\G_1)$, by taking $\mathcal{T}=\left\{1,2,\ldots,k\right\}$ in Lemma \ref{alphak+j}, we know that $\alpha_{i}^{k+j}$ can be expressed as $\sum\limits_{s=0}^{k-1}f_{j,s}^{(\mathcal{T})}\alpha_{i}^{s}\triangleq\sum\limits_{s=0}^{k-1}f_{j,s}^{(k)}\alpha_{i}^{s}$ for $1\leq i\leq k$ and $0\leq j\leq \ell$, furthermore,
$$\det(\G_1)=\prod\limits_{i=1}^{k}v_i\cdot\det\left(\boldsymbol{I}_k+\widetilde{\B_1}\boldsymbol{F}_{1}\right)\prod\limits_{1\leq j<i\leq k}\left(\alpha_{i}-\alpha_{j}\right), $$
where
\begin{equation}\label{B1F1} 
\widetilde{\B_1}=\begin{pmatrix}
	b_{0,0}&\cdots&b_{0,\ell}\\
	\vdots& &\vdots\\
	b_{k-1,0}&\cdots&b_{k-1,\ell}
\end{pmatrix}, \boldsymbol{F}_{1}=\begin{pmatrix}
	f_{0,0}^{(k)}&\cdots&f_{0,k-1}^{(k)}\\
	\vdots& &\vdots\\
	f_{\ell,0}^{(k)}&\cdots&f_{\ell,k-1}^{(k)}
\end{pmatrix}.
\end{equation}  
And so, it is easy to know that (1) of Theorem \ref{MDSTGRL} holds.

For $\det(\G_{2i})$, by taking $\mathcal{T}=\left\{1,2,\ldots,k-1\right\}$ in Lemma \ref{alphak+j}, we know that $\alpha_{i}^{k+j}=\alpha_{i}^{k-1+j+1}$ can be expressed as $$\sum\limits_{s=0}^{k-2}f_{j+1,s}^{(\mathcal{T})}\alpha_{i}^{s}\triangleq\sum\limits_{s=0}^{k-2}f_{j+1,s}^{(k-1)}\alpha_{i}^{s}$$ for $1\leq i\leq k-1$ and $0\leq j\leq \ell$, furthermore,
$$\det(\G_{2i})=\left(a_{2i}\cdot\det\left(\boldsymbol{I}_{k-1}+\widetilde{\B_2}\boldsymbol{F}_{2}\right)-a_{1i}\cdot\det\left(\boldsymbol{T}+\widetilde{\B_3}\boldsymbol{F}_{2}\right)\right)\cdot\prod\limits_{i=1}^{k-1}v_i\cdot\prod\limits_{1\leq j<i\leq k-1}\left(\alpha_{i}-\alpha_{j}\right)$$
	with 
$$\boldsymbol{T}=\begin{pmatrix}
	1&0&\cdots&0&0\\
	0&1&\cdots&0&0\\
	\vdots&\vdots& &\vdots&\vdots\\
	0&0&\cdots&1&0\\
	f_{0,0}^{(k-1)}&f_{0,1}^{(k-1)}&\cdots&f_{0,k-3}^{(k-1)}&f_{0,k-2}^{(k-1)}\\
\end{pmatrix},  \boldsymbol{F}_{2}=\begin{pmatrix}
	f_{1,0}^{(k-1)}&\cdots&f_{1,k-2}^{(k-1)}\\
	\vdots& &\vdots\\
	f_{\ell+1,0}^{(k-1)}&\cdots&f_{\ell+1,k-2}^{(k-1)}
\end{pmatrix},$$
\begin{equation}\label{TB2F2}  
\widetilde{\B_2}=\begin{pmatrix}
	b_{0,0}&\cdots&b_{0,\ell}\\
	\vdots& &\vdots\\
	b_{k-2,0}&\cdots&b_{k-2,\ell}
\end{pmatrix},\widetilde{\B_3}=\begin{pmatrix}
	b_{0,0}&\cdots&b_{0,\ell}\\
	\vdots& &\vdots\\
	b_{k-3,0}&\cdots&b_{k-3,\ell}\\
	b_{k-1,0}&\cdots&b_{k-1,\ell}
\end{pmatrix}.
\end{equation} 
And so, it is easy to know that (2) of Theorem \ref{MDSTGRL} holds.

For $\det(\G_3)$, by taking $\mathcal{T}=\left\{1,2,\ldots,k-2\right\}$ in Lemma \ref{alphak+j}, we know that $\alpha_{i}^{k+j}$ can be expressed as $$\sum\limits_{s=0}^{k-3}f_{j+2,s}^{(\mathcal{T})}\alpha_{i}^{s}\triangleq\sum\limits_{s=0}^{k-3}f_{j+2,s}^{(k-2)}\alpha_{i}^{s}$$ for $1\leq i\leq k-2$ and $0\leq j\leq \ell$, furthermore,
$$\det(\G_3)=\prod\limits_{i=1}^{k-2}v_i\cdot\det\left(\boldsymbol{I}_{k-2}+\widetilde{\B_4}\boldsymbol{F}_{3}\right)\cdot\prod\limits_{1\leq j<i\leq k-2}\left(\alpha_{i}-\alpha_{j}\right)\cdot\det\left(\A_{2\times 2}\right), $$
where
\begin{equation}\label{B4F4} 
	\widetilde{\B_4}=\begin{pmatrix}
		b_{0,0}&\cdots&b_{0,\ell}\\
		\vdots& &\vdots\\
		b_{k-3,0}&\cdots&b_{k-3,\ell}
	\end{pmatrix}, \boldsymbol{F}_{3}=\begin{pmatrix}
		f_{2,0}^{(k-2)}&\cdots&f_{2,k-3}^{(k-2)}\\
		\vdots& &\vdots\\
		f_{\ell+2,0}^{(k-2)}&\cdots&f_{\ell+2,k-3}^{(k-2)}
	\end{pmatrix}.
\end{equation}
And so, it is easy to know that (3) of Theorem \ref{MDSTGRL} holds.

This completes the proof of Theorem \ref{MDSTGRL}.
\end{proof}
\subsection{The NMDS property of the $(1)$-column TGRS code and weight distribution}
\begin{theorem}\label{oneCTGRSMDSNMDS}
If $k\geq 3$ and $\alpha_{1},\ldots,\alpha_{n},b,c$ are pairwise distinct, then the $(1)$-column TGRS code is either MDS or NMDS.
\end{theorem}
\begin{proof}
By Definition \ref{columnTGRSdefinition} and the definition of the monomially equivalent, we only focus on $\C_1(\boldsymbol{\lambda},b,c)$ which has the generator matrix
$$\G_{1}(\boldsymbol{\lambda},b,c)=\begin{pmatrix}
	1&1&\cdots&1&1-\lambda_{1}\\
	\alpha_{1}&\alpha_{2}&\cdots&\alpha_{n}&b-\lambda_{1}c\\
	\alpha_{1}^{2}&\alpha_{2}^{2}&\cdots&\alpha_{n}^{2}&b^{2}-\lambda_{1}c^{2}\\
	\vdots&\vdots& &\vdots&\vdots\\
	\alpha_{1}^{k-2}&\alpha_{2}^{k-2}&\cdots&\alpha_{n}^{k-2}&b^{k-2}-\lambda_{1}c^{k-2}\\
	\alpha_{1}^{k-1}&\alpha_{2}^{k-1}&\cdots&\alpha_{n}^{k-1}&b^{k-1}-\lambda_{1}c^{k-1}
\end{pmatrix}.
$$

If the code $\C_1(\boldsymbol{\lambda},b,c)$ is not MDS, then by Lemma \ref{MDSorNMDSequivalent}, there exist $k$ columns in $\G_{1}(\boldsymbol{\lambda},b,c)$ that are  $\mathbb{F}_{q}$-linearly dependent, i.e., Lemma \ref{MDSorNMDSequivalent} $(ii)$ holds.

Next, we prove Lemma \ref{MDSorNMDSequivalent} $(i)$ and $(iii)$.

For Lemma \ref{MDSorNMDSequivalent} $(i)$, we only prove that there exists some $(k-1)\times (k-1)$ non-zero minor for the following two classes of submatrix 
$$\G_{1}^{(1)}(\boldsymbol{\lambda},b,c)=\begin{pmatrix}
	1&1&\cdots&1\\
	\alpha_{1}&\alpha_{2}&\cdots&\alpha_{k-1}\\
	\alpha_{1}^{2}&\alpha_{2}^{2}&\cdots&\alpha_{k-1}^{2}\\
	\vdots&\vdots& &\vdots\\
	\alpha_{1}^{k-2}&\alpha_{2}^{k-2}&\cdots&\alpha_{k-1}^{k-2}\\
	\alpha_{1}^{k-1}&\alpha_{2}^{k-1}&\cdots&\alpha_{k-1}^{k-1}
\end{pmatrix}, \G_{1}^{(2)}(\boldsymbol{\lambda},b,c)=\begin{pmatrix}
1&1&\cdots&1&1-\lambda_{1}\\
\alpha_{1}&\alpha_{2}&\cdots&\alpha_{k-2}&b-\lambda_{1}c\\
\alpha_{1}^{2}&\alpha_{2}^{2}&\cdots&\alpha_{k-2}^{2}&b^{2}-\lambda_{1}c^{2}\\
\vdots&\vdots& &\vdots&\vdots\\
\alpha_{1}^{k-2}&\alpha_{2}^{k-2}&\cdots&\alpha_{k-2}^{k-2}&b^{k-2}-\lambda_{1}c^{k-2}\\
\alpha_{1}^{k-1}&\alpha_{2}^{k-1}&\cdots&\alpha_{k-2}^{k-1}&b^{k-1}-\lambda_{1}c^{k-1}
\end{pmatrix}.
$$
For the submatrix $\G_{1}^{(1)}(\boldsymbol{\lambda},b,c)$, by deleting the last row of $\G_{1}^{(1)}(\boldsymbol{\lambda},b,c)$, it is easy to obtain $(k-1)\times (k-1)$ non-zero minor.

For the submatrix $\G_{1}^{(2)}(\boldsymbol{\lambda},b,c)$, we consider the submatrix $\boldsymbol{R}_1$ given by deleting the last row of $\G_{1}^{(2)}(\boldsymbol{\lambda},b,c)$ and the submatrix $\boldsymbol{R}_2$ given by deleting the penultimate row of $\G_{1}^{(2)}(\boldsymbol{\lambda},b,c)$. It is easy to know that for $\det\left(\boldsymbol{R}_1\right)\in\F_{q}^{*}$, $\boldsymbol{R}_1$ is a  $(k-1)\times (k-1)$ non-zero minor of $\G_{1}^{(2)}(\boldsymbol{\lambda},b,c)$.

Next, we prove that for $\det\left(\boldsymbol{R}_1\right)=0$, $\det\left(\boldsymbol{R}_2\right)\in\F_{q}^{*}$, i.e., $\boldsymbol{R}_2$ is a  $(k-1)\times (k-1)$ non-zero minor of $\G_{1}^{(2)}(\boldsymbol{\lambda},b,c)$.

By directly computing, we have
$$\det\left(\boldsymbol{R}_1\right)=\prod\limits_{1\leq j<i\leq k-2} \left(\alpha_{i}-\alpha_{j}\right)\left(\prod\limits_{i=1}^{k-2} \left(b-\alpha_{i}\right)-\lambda_{1}\prod\limits_{i=1}^{k-2}\left(c-\alpha_{i}\right)\right)$$
and
$$\det\left(\boldsymbol{R}_2\right)=\prod\limits_{1\leq j<i\leq k-2} \left(\alpha_{i}-\alpha_{j}\right)\left[\left(\sum\limits_{i=1}^{k-2}\alpha_{i}+b\right)\prod\limits_{i=1}^{k-2} \left(b-\alpha_{i}\right)-\lambda_{1}\left(\sum\limits_{i=1}^{k-2}\alpha_{i}+c\right)\prod\limits_{i=1}^{k-2}\left(c-\alpha_{i}\right)\right].$$

If $\det\left(\boldsymbol{R}_1\right)=0$, i.e.,  $\prod\limits_{i=1}^{k-2} \left(b-\alpha_{i}\right)=\lambda_{1}\prod\limits_{i=1}^{k-2}\left(c-\alpha_{i}\right)$, then  $$\det\left(\boldsymbol{R}_2\right)=\lambda_{1}\prod\limits_{1\leq j<i\leq k-2} \left(\alpha_{i}-\alpha_{j}\right)\prod\limits_{i=1}^{k-2}\left(c-\alpha_{i}\right)\left(b-c\right).$$
Note that $\alpha_{1},\ldots,\alpha_{n},b,c$ are pairwise distinct, then we have $\det\left(\boldsymbol{R}_2\right)\in\F_{q}^{*}$.

And so, there exists some $(k-1)\times (k-1)$ non-zero minor for $\G_{1}^{(2)}(\boldsymbol{\lambda},b,c)$.

For Lemma \ref{MDSorNMDSequivalent} $(iii)$, by deleting the last column of the submatrix formed any $k+1$ columns of $\G_{1}(\boldsymbol{\lambda},b,c)$, it is easy to obtain a $k\times k$ non-zero minor, furthermore, since the submatrix has exactly $k$ rows, it follows that any submatrix formed by $k+1$ columns of $\G_{1}(\boldsymbol{\lambda},b,c)$ is of full rank.  

This completes the proof of Theorem \ref{oneCTGRSMDSNMDS}.
\end{proof}
 
Next, we recall a necessary and sufficient condition for The $(1)$-column TGRS code to be MDS.
\begin{lemma}[\cite{Liu2026ColumnTGRS}, Lemma 1]\label{MDSCTGRS}
The $(1)$-column TGRS code is MDS if and only if for any $(k-1)$-subset $\mathcal{J}\subseteq \left\{1,2,\ldots,n\right\}$,
$$
\prod\limits_{i\in \mathcal{J}} (b - \alpha_{i}) - \lambda \prod\limits_{i\in \mathcal{J}} (c - \alpha_{i}) \neq 0.
$$
\end{lemma}
Furthermore, by Theorem \ref{oneCTGRSMDSNMDS} and Lemma \ref{MDSCTGRS}, we have
\begin{theorem}\label{oneCTGRSNMDS}
If  $k\geq 3$ and $\alpha_{1},\ldots,\alpha_{n},b,c$ are pairwise distinct, then the $(1)$-column TGRS code is NMDS if and only if there exists a $(k-1)$-subset $\mathcal{J}\subseteq \left\{1,2,\ldots,n\right\}$ such that
	$$
	\prod\limits_{i\in \mathcal{J}} (b - \alpha_{i}) - \lambda \prod\limits_{i\in \mathcal{J}} (c - \alpha_{i})=0.
	$$
\end{theorem}

Finally, for  a class of NMDS $(1)$-column TGRS codes, we determine the weight distribution.

\begin{theorem}\label{NMDS1CRS}
Let $\F_q^*=\langle\omega\rangle$ and $\lambda=\omega^s, s\in\left\{0,1,\ldots, q-2\right\}$. If $n=q-2$ and $\left\{\alpha_{1},\alpha_{2},\ldots,\alpha_{n}\right\}=\mathbb{F}_{q}\backslash\left\{b,c\right\}$, then for any NMDS $(1)$-column TGRS code, the total number $A_{q-1-k}$ of codewords with Hamming weight $q-1-k$ is
$$A_{q-1-k}= \sum_{i\mid q-1}
(-1)^{k-1+\left\lfloor \frac{k-1}{i}\right\rfloor}
\binom{\frac{q-1}{i}-1}{\left\lfloor \frac{k-1}{i}\right\rfloor}
\sum_{d\mid \gcd(i,\gcd(s,q-1))}
d\mu\left(\frac{i}{d}\right).$$
\end{theorem}
\begin{proof}
By Definition \ref{columnTGRSdefinition}, it is easy to know that $$\C_1(\boldsymbol{\lambda},b,c)=\left\{\c_f=\left(f(\alpha_{1}),f(\alpha_{2}),\ldots,f(\alpha_{q-2}),f(b)-\lambda f(c)\right)|f(x)\in\F_{q}^{k}[x]\right\},$$
where $\lambda\in\F_q^*.$ For the convenience, we set $\mathcal{I}=\left\{\alpha_{1},\alpha_{2},\ldots,\alpha_{q-2}\right\}$. Then $\wt\left(\c_f\right)=q-1-k$ if and only if
$$\#\left\{\alpha_{i}\in \mathcal{I}|f(\alpha_{i})=0\right\}=k\ \text{and}\ f(b)-\lambda f(c)\neq 0$$
or
$$\#\left\{\alpha_{i}\in \mathcal{I}|f(\alpha_{i})=0\right\}=k-1\ \text{and}\ f(b)-\lambda f(c)= 0.$$

If $f(b)-\lambda f(c)\neq 0$, then 
$$\#\left\{\alpha_{i}\in \mathcal{I}|f(\alpha_{i})=0\right\}=k,$$
it means that $f(x)\in\F_{q}^{k}[x]$ has $k$ distinct roots, which is contradict, thus, $A_{q-1-k}=0$ in this case.

If $f(b)-\lambda f(c)= 0$, then 
$$\#\left\{\alpha_{i}\in \mathcal{I}|f(\alpha_{i})=0\right\}=k-1,$$
thus,
$$
\begin{aligned}
A_{q-1-k}=&\#\left\{\F_{q}^{k}[x]\cap\left\{f(x)=\rho\prod\limits_{\alpha_{i}\in \mathcal{J}}\left(x-\alpha_{i}\right)|\# \mathcal{J}=k-1,  \mathcal{J}\subseteq \mathcal{I},f(b)=\lambda f(c),\rho\in\F_q^*\right\} \right\}\\
=&\#\left\{\F_{q}^{k}[x]\cap\left\{f(x)=\rho\prod\limits_{\alpha_{i}\in \mathcal{J}}\left(x-\alpha_{i}\right)|\# \mathcal{J}=k-1,  \mathcal{J}\subseteq \mathcal{I},\prod\limits_{\alpha_{i}\in \mathcal{J}}\frac{b-\alpha_{i}}{c-\alpha_{i}}=\lambda,\rho\in\F_q^*\right\} \right\}\\
=&\#\left\{(\rho,\mathcal{J})|\# \mathcal{J}=k-1,  \mathcal{J}\subseteq \mathcal{I},\prod\limits_{\alpha_{i}\in \mathcal{J}}\frac{b-\alpha_{i}}{c-\alpha_{i}}=\lambda,\rho\in\F_q^*\right\}\\
=&(q-1)\#\left\{\mathcal{J}|\# \mathcal{J}=k-1,  \mathcal{J}\subseteq \mathcal{I},\prod\limits_{\alpha_{i}\in \mathcal{J}}\frac{b-\alpha_{i}}{c-\alpha_{i}}=\lambda\right\}.
\end{aligned}
$$ 
For $\alpha_{i}\in \mathcal{J}$, by $\mathcal{J}\subseteq \mathcal{I}=\left\{\alpha_{1},\alpha_{2},\ldots,\alpha_{n}\right\}=\mathbb{F}_{q}\backslash\left\{b,c\right\}$, we have $$\left\{\frac{b-\alpha_{i}}{c-\alpha_{i}}|\alpha_{i}\in \mathcal{J}\right\}\subseteq\F_q^*\backslash\left\{1\right\}.$$
Additionally, it is easy to prove that the map
$$\begin{aligned}
	\varphi:\mathcal{I}&\longrightarrow\F_q^*\backslash\left\{1\right\}\\
\alpha&\longmapsto \frac{b-\alpha}{c-\alpha}	 
\end{aligned}$$
is bijection. And so, by $\F_q^*=\langle\omega\rangle$, we can set $\frac{b-\alpha_{i}}{c-\alpha_{i}}=\omega^{t},t\in\left\{1,2,\ldots,q-2\right\},$ and $\lambda=\omega^{s},s\in\left\{0,1,\ldots,q-2\right\}$. Furthermore, 
$$
\begin{aligned}
	A_{q-1-k}=&(q-1)\#\left\{\mathcal{J}|\# \mathcal{J}=k-1,  \mathcal{J}\subseteq \mathcal{I},\prod\limits_{\alpha_{i}\in \mathcal{J}}\frac{b-\alpha_{i}}{c-\alpha_{i}}=\lambda\right\}\\
	=&(q-1)\#\left\{\mathcal{M}|\# \mathcal{M}=k-1,  \mathcal{M}\subseteq \left\{1,2,\ldots,q-2\right\},\prod\limits_{t\in \mathcal{M}}\omega^{t}=\omega^{s}\right\}\\
	=&(q-1)\#\left\{\mathcal{M}|\# \mathcal{M}=k-1,  \mathcal{M}\subseteq \left\{1,2,\ldots,q-2\right\},\omega^{\sum\limits_{t\in \mathcal{M}}t}=\omega^{s}\right\}\\
	=&(q-1)\#\left\{\mathcal{M}|\# \mathcal{M}=k-1,  \mathcal{M}\subseteq \left\{1,2,\ldots,q-2\right\},\sum\limits_{t\in \mathcal{M}}t\equiv s\pmod{q-1}\right\}.
\end{aligned}
$$
And so, by Lemma \ref{subsumGneq0}, we have
$$
\begin{aligned}
A_{q-1-k}=&(q-1) N\left(\mathbb{Z}_{q-1}\backslash\left\{0\right\},k-1,s\right)\\
=& \sum_{i\mid \exp(\mathbb{Z}_{q-1})}
(-1)^{k-1+\left\lfloor \frac{k-1}{i}\right\rfloor}
\binom{\frac{q-1}{i}-1}{\left\lfloor \frac{k-1}{i}\right\rfloor}
\sum_{d\mid \gcd(i,e(s))}
\mu\left(\frac{i}{d}\right)\# G[d].
\end{aligned}$$ 
Note that $\exp(\mathbb{Z}_{q-1})=q-1$,
$$e(s)=\max\left\{d : d\mid q-1,\, s\in d\mathbb{Z}_{q-1}\right\}=\gcd(s,q-1).$$
By $i\mid q-1$ and $d\mid i$, we have $d\mid q-1$, thus, 
$$\# \mathbb{Z}_{q-1}[d]=\# \{h\in \mathbb Z_{q-1}\ |\  dh\equiv 0 \pmod{q-1}\}=\gcd(d,q-1)=d.$$
Hence,
$$A_{q-1-k}= \sum_{i\mid q-1}
(-1)^{k-1+\left\lfloor \frac{k-1}{i}\right\rfloor}
\binom{\frac{q-1}{i}-1}{\left\lfloor \frac{k-1}{i}\right\rfloor}
\sum_{d\mid \gcd(i,\gcd(s,q-1))}
d\mu\left(\frac{i}{d}\right).$$
\end{proof}

\section{The non-GRS property for several classes of linear codes}\label{sec6}

In this section, basing on the non-GRS property of $(\L,\P)$-TGRS codes, we determine the corresponding non-GRS property for the following four classes of linear codes $\C_{k,\ell}(\L,\P,\A_{m\times m},\B)$, $\C_{h}\left(\eta,\boldsymbol{b}\right)$, $\C_{h}\left(\eta,\boldsymbol{b},\delta\right)$ and $\C_{s}\left(\boldsymbol{\lambda},\boldsymbol{b},c,\L,\P,\B\right)$ under many conditions. 
\begin{lemma}\label{shortpuncturGRS}
Both the shortened and punctured codes
of a GRS code at any coordinate position are GRS codes.
\end{lemma}

Lemma \ref{shortpuncturGRS} implies that the (column) extended code of a non-GRS
code cannot be a GRS code. 

Notice that the codes $\C_{k,\ell}(\L,\P,\A_{m\times m},\B)$, $\C_{h}\left(\eta,\boldsymbol{b}\right)$, $\C_{h}\left(\eta,\boldsymbol{b},\delta\right)$ and $\C_{s}\left(\boldsymbol{\lambda},\boldsymbol{b},c,\L,\P,\B\right)$ are all obtained by adding some column vectors to the code  $\C\left(\L,\P,\B\right)$, and so, we only focus on the non-GRS property of $\C\left(\L,\P,\B\right)$.

Next, we summarize some existing works on the non-GRS property of $\C\left(\L,\P,\B\right)$, and then get the non-GRS properties of the codes $\C_{k,\ell}(\L,\P,\A_{m\times m},\B)$, $\C_{h}\left(\eta,\boldsymbol{b}\right)$, $\C_{h}\left(\eta,\boldsymbol{b},\delta\right)$ and $\C_{s}\left(\boldsymbol{\lambda},\boldsymbol{b},c,\L,\P,\B\right)$.

\begin{lemma}\label{LPTGRSnonRS}
Assume that $b_{i,j}=0$ whenever
$(i,j)\notin\P\times\L$. If one of the
following conditions holds, then $\C\left(\L,\P,\B\right)$ is non-GRS.
\begin{enumerate}
\item [$(1)$] \rm(Theorem 5, \cite{Hu2025LPTGRS}) $n\geq 2k$, $\varepsilon\geq2k$ and $\B= \begin{pmatrix}
	\zero_{(k-\ell)\times \ell} & \zero_{(k-\ell)\times (n-k-\ell)} \\
	\B_{\ell\times \ell}^{(3)} & \zero_{\ell\times (n-k-\ell)}
\end{pmatrix},$
where $
\B_{\ell\times \ell}^{(3)}=\begin{pmatrix}
	b_{k-\ell,0} & 0 & \cdots & 0 \\
	b_{k-\ell+1,0} & b_{k-\ell+1,1} & \cdots & 0 \\
	\vdots & \vdots & \ddots & \vdots \\
	b_{k-1,0} & b_{k-1,1} & \cdots & b_{k-1,\ell-1}
\end{pmatrix},$ $b_{k-j,\ell-j} \in \mathbb{F}_q^*$ for $1 \le j \le \ell$, $0 < \ell < k$ and
$$\varepsilon=\begin{cases}
	2k + 2\ell - 2, & \text{if } k \ge 2\ell + 2 \text{ and } n > 2k + 2\ell - 2; \\
	n - 1, & \text{if } k \ge 2\ell + 2 \text{ and } n \le 2k + 2\ell - 2; \\
	3k - 3, & \text{if } k < 2\ell + 2 \text{ and } n > 2k + 2\ell - 2; \\
	n + k - 2\ell - 2, & \text{if } k < 2\ell + 2 \text{ and } n \le 2k + 2\ell - 2.
\end{cases}$$
\item [$(2)$] \rm(Theorem 5.3, \cite{Liang+LPTGRS2026}) $\B=\begin{pmatrix}
\zero_{(k-1)\times (\ell+1)}&\zero_{(k-1)\times (n-k-\ell-1)}\\
\boldsymbol{\eta}_{\ell+1}&\zero_{1\times(n-k-\ell-1)}
\end{pmatrix}$ with $\boldsymbol{\eta}_{\ell+1}=\left(\eta_{0},\cdots,\eta_{\ell}\right)\in\F_q^{\ell+1}\backslash\left\{\zero\right\}$, and one of the
following conditions holds,

$(2.1)$ $2k>n>k+\ell+2$ and $1+\sum\limits_{t=0}^{\ell}\eta_{t}S_{t+1}\neq 0$;

$(2.2)$ $2k>n\geq k+\ell+2\geq k+3$ and $1+\sum\limits_{t=0}^{\ell}\eta_{t}S_{t+1}=0$.  
\item [$(3)$] \rm(Theorem 5.1, \cite{SuiJZ2022DCCTGRS}) $2<k\leq k+t<\frac{n}{2}$ and $\B=\begin{pmatrix}
\zero_{h\times t}&0&\zero_{h\times (n-k-t-1)}\\
\zero_{1\times t}&b_{h,t}&\zero_{1\times (n-k-t-1)}\\
\zero_{(k-h-1)\times t}&0&\zero_{(k-h-1)\times (n-k-t-1)}
\end{pmatrix}$ with $b_{h,t}\in\F_q^*$;
\item [$(4)$] \rm(Theorems 6.2-6.3, \cite{SuiJZ2022TITTGRS}) $3\leq k\leq n-3$ with $k<\lfloor \frac{n}{2} \rfloor-1$ or $k>\lceil \frac{n}{2} \rceil+1$, $b_{k-2,1},b_{k-1,0}\in\F_q^*$ and $$\B=\begin{pmatrix}
	\zero_{(k-2)\times 1}&\zero_{(k-2)\times 1}&\zero_{(k-2)\times n-k-2}\\
	0&b_{k-2,1}&\zero_{1\times( n-k-2)}\\  
	b_{k-1,0} &0 &\zero_{1\times (n-k-2)}
\end{pmatrix};$$
\item [$(5)$] \rm(Theorem 3.3, \cite{ZhuCZFFATGRS}) $4\leq k<n-4$ and $\B=\begin{pmatrix}
	\zero_{(k-2)\times 1}&\zero_{(k-2)\times 1}&\zero_{(k-2)\times n-k-2}\\
	b_{k-2,0}&b_{k-2,1}&\zero_{1\times( n-k-2)}\\  
	0&0 &\zero_{1\times (n-k-2)}
\end{pmatrix}$ with $b_{k-2,0},b_{k-2,1}\in\F_q^*$;
\item [$(6)$] \rm(Theorems 4-5, \cite{Cheng2026TGRS}) $\B=\begin{pmatrix}
	\zero_{h\times 1}&\zero_{h\times 1}&\zero_{h\times n-k-2}\\
	b_{h,0}&b_{h,1}&\zero_{1\times( n-k-2)}\\
	\zero_{(k-h-1)\times 1}&\zero_{(k-h-1)\times 1}&\zero_{(k-h-1)\times n-k-2}\\
\end{pmatrix}$ with $b_{h,0},b_{h,1}\in\F_q^*$ and one of the following conditions holds,

$(6.1)$  $8\leq 2k<n$;

$(6.2)$  $2k\geq n\geq k+4$;

\item [$(7)$] \rm(Theorem 1, \cite{ZhuCZarxivnongrsTRS}) $0 \leq h \leq k - 1$, $k + t \leq n \leq q$, $\frac{n}{2} \leq k \leq n-3$, $$\B=\begin{pmatrix}
	\zero_{h\times (t-1)}&\zero_{h\times 1}&\zero_{h\times n-k-t}\\
	\zero_{1\times (t-1)}&b_{h,t-1}&\zero_{1\times( n-k-t)}\\
	\zero_{(k-h-1)\times (t-1)}&\zero_{(k-h-1)\times 1}&\zero_{(k-h-1)\times (n-k-t)}\\
\end{pmatrix},b_{h,t-1}\in\F_q^*$$ and one of the following conditions holds,

$(7.1)$ $n = k + t, 2k \geq n + 2$ and $3 \leq h \leq k - 3$;

$(7.2)$ $n = k + t + 1, 2k \geq n + 2$ and $2 \leq h \leq k - 3$;

$(7.3)$ $n \geq k + t + 2, 2k \geq n, t = 1$ and $2 \leq h \leq k - 2$;

$(7.4)$ $n \geq k + t + 2, 2k \geq n + 1, t = 2$ and $1 \leq h \leq k - 3$;

$(7.5)$ $n \geq k + t + 2, 2k \geq n, t \geq 3$ and $h \in \{0,\dots,k-1\}\setminus\{k-t\}$;

$(7.6)$ $n \geq k + t + 2, 2k \geq n + 1, t \geq 3$ and $h = k - t$. 
\end{enumerate}
\end{lemma}

Firstly, since $\C_{k,\ell}(\L,\P,\A_{m\times m},\B)$ and $\C_{s}\left(\boldsymbol{\lambda},\boldsymbol{b},c,\L,\P,\B\right)$ are obtained by adding column vectors from $\C\left(\L,\P,\B\right)$, and then by Lemma \ref{shortpuncturGRS} and Lemma \ref{LPTGRSnonRS} (1)-(7), we have the following 
\begin{theorem}\label{nonGRScodes1}
If one of Lemma \ref{LPTGRSnonRS} $(1)$-$(7)$ holds, then $\C_{k,\ell}(\L,\P,\A_{m\times m},\B)$ and $\C_{s}\left(\boldsymbol{\lambda},\boldsymbol{b},c,\L,\P,\B\right)$ are both non-GRS.
\end{theorem} 
\begin{remark}
	For $s=1$ and $v_1=\cdots=v_n=1$,
	by Lemma \ref{LPTGRSnonRS} $(3)$ with $t=0$
	and Theorem \ref{nonGRScodes1},
	the corresponding row-column TRS codes are non-GRS
	for $3\le k<n/2$.
	This provides a direct proof via puncturing,
	without the additional Schur-square computations used in
	\cite[Theorem~6.2]{Bhagat2026RowColumnTGRS}.
\end{remark}
Secondly, since $\C_{h}\left(\eta,\boldsymbol{b}\right)$ and $\C_{h}\left(\eta,\boldsymbol{b},\delta\right)$ are obtained by adding column vectors from $\C\left(\L,\P,\B\right)$ with $$\B=\begin{pmatrix}
	\zero_{h\times (\ell+1)}&\zero_{h\times (n-k-\ell-1)}\\
	\boldsymbol{b}&\zero_{1\times (n-k-\ell-1)}\\
	\zero_{(k-h-1)\times (\ell+1)}&\zero_{(k-h-1)\times (n-k-\ell-1)}
\end{pmatrix},$$
where $\boldsymbol{b}=\left(b_0,b_1,\ldots,b_{\ell}\right)\in\F_{q}^{\ell+1}\backslash\left\{\zero_{\ell+1}\right\}$. And then by Lemma \ref{shortpuncturGRS} and Lemma \ref{LPTGRSnonRS} (2)-(3) and (5)-(7), we have the following
\begin{theorem}\label{nonGRScodes2}
	If one of Lemma \ref{LPTGRSnonRS} $(2)$-$(3)$ and $(5)$-$(7)$ holds, then $\C_{h}\left(\eta,\boldsymbol{b}\right)$ and $\C_{h}\left(\eta,\boldsymbol{b},\delta\right)$ are both non-GRS.
\end{theorem}
\begin{remark}
For $\ell=1$, $b_0=0$, $b_1\in\F_q^*$ and $\eta=1$,
	the non-GRS property of $\C_h(\eta,\boldsymbol b)$ follows
	from Lemma \ref{LPTGRSnonRS} $(3)$ with $t=1$, or
	$(7.4)$ with $t=2$, under the corresponding conditions.
	For $h\le k-2$, its twice extended code is also non-GRS.
	These conclusions follow directly by puncturing, without
	the Schur-square computations used in
	\cite[Theorems~3.2 and~3.3]{Zhang2026extendedTGRS}.
\end{remark}
\section{Conclusions}
In this paper, we mainly focus on the constructions of the parity-check matrices for the known non-GRS type linear codes and introduce a class of non-GRS type linear codes, i.e., $(\L,\P)$-TGRL codes. The main contributions are summarized as follows.
\begin{itemize}
\item Give the parity-check matrix for three subclasses of $(\L,\P)$-TGRL codes (Theorem \ref{RL+k-1TGRSH}, Theorem \ref{RL+k-2k-1TGRSH}, Theorem \ref{RL+k-3k-2k-1TGRSH}, Theorem \ref{C1otherH}), the generation and the extended code of the codes introduced by \cite{LiY2025extendedTGRS} (Theorem \ref{extendedhTGRSH}, Theorem \ref{C4extendeC4}), two classes of column TGRS codes introduced by \cite{Liu2026ColumnTGRS}  (Theorem \ref{onecolumnTRSH}, Theorem \ref{twocolumnTRSH}, ), a special subclass of $(\L,\P, 1)$-row-column TRS codes (Theorem \ref{onerowcolumnTGRSH}), MGRS codes introduced by \cite{WangMGRS2026} (Theorem \ref{tMGRSH}), .
\item Give a necessary and sufficient condition for a general class of TGRL codes to be MDS (Theorem \ref{MDSTGRL}).
\item Prove that the $(1)$-column TGRS code introduced by \cite{Liu2026ColumnTGRS} is either MDS or NMDS, furthermore, give a necessary and sufficient condition for this code be NMDS (Theorem \ref{oneCTGRSNMDS}). Additionally, by using the subset sums of finite abelian groups, we determine the weight distribution for a special subclass of NMDS $(1)$-column TGRS codes. 
\item Determine the non-GRS property for several classes of linear codes (Theorem \ref{nonGRScodes1}, Theorem \ref{nonGRScodes2}). 
\end{itemize}

As future work, it would be interesting to exploit these parity-check matrices to design decoding algorithms, investigate the hull dimensions, determine deep holes, and characterize the automorphism groups of the considered codes.

\appendices
\section{The calculation for the determinant of $G_i(i=1,2,3)$ in Theorem \ref{MDSTGRL}}
{\bf The calculation of $\det(\G_1)$.}
$$\begin{aligned}
	\det(\G_1)=&\prod\limits_{i=1}^{k}v_i\cdot\det\begin{pmatrix}
		1+\sum\limits_{j=0}^{\ell}b_{0,j}\alpha_{1}^{k+j}&\cdots&1+\sum\limits_{j=0}^{\ell}b_{0,j}\alpha_{k}^{k+j}\\
		\alpha_{1}+\sum\limits_{j=0}^{\ell}b_{1,j}\alpha_{1}^{k+j}&\cdots&\alpha_{k}+\sum\limits_{j=0}^{\ell}b_{1,j}\alpha_{k}^{k+j}\\ 
		\vdots& &\vdots\\   
		\alpha_{1}^{k-1}+\sum\limits_{j=0}^{\ell}b_{k-1,j}\alpha_{1}^{k+j}&\cdots&\alpha_{k}^{k-1}+\sum\limits_{j=0}^{\ell}b_{k-1,j}\alpha_{k}^{k+j}
	\end{pmatrix}\\
	=&\prod\limits_{i=1}^{k}v_i\cdot\det\begin{pmatrix}
		1+\sum\limits_{j=0}^{\ell}b_{0,j}\sum\limits_{s=0}^{k-1}f_{j,s}^{(k)}\alpha_{1}^{s}&\cdots&1+\sum\limits_{j=0}^{\ell}b_{0,j}\sum\limits_{s=0}^{k-1}f_{j,s}^{(k)}\alpha_{k}^{s}\\
		\alpha_{1}+\sum\limits_{j=0}^{\ell}b_{1,j}\sum\limits_{s=0}^{k-1}f_{j,s}^{(k)}\alpha_{1}^{s}&\cdots&\alpha_{k}+\sum\limits_{j=0}^{\ell}b_{1,j}\sum\limits_{s=0}^{k-1}f_{j,s}^{(k)}\alpha_{k}^{s}\\ 
		\vdots& &\vdots\\   
		\alpha_{1}^{k-1}+\sum\limits_{j=0}^{\ell}b_{k-1,j}\sum\limits_{s=0}^{k-1}f_{j,s}^{(k)}\alpha_{1}^{s}&\cdots&\alpha_{k}^{k-1}+\sum\limits_{j=0}^{\ell}b_{k-1,j}\sum\limits_{s=0}^{k-1}f_{j,s}^{(k)}\alpha_{k}^{s}
	\end{pmatrix}\\
		=&\prod\limits_{i=1}^{k}v_i\cdot\det\left(\begin{pmatrix}
		1+\sum\limits_{j=0}^{\ell}b_{0,j}f_{j,0}^{(k)}&\sum\limits_{j=0}^{\ell}b_{0,j}f_{j,1}^{(k)}&\cdots&\sum\limits_{j=0}^{\ell}b_{0,j}f_{j,k-1}^{(k)}\\
		\sum\limits_{j=0}^{\ell}b_{1,j}f_{j,0}^{(k)}&1+\sum\limits_{j=0}^{\ell}b_{1,j}f_{j,1}^{(k)}&\cdots&\sum\limits_{j=0}^{\ell}b_{1,j}f_{j,k-1}^{(k)}\\ 
		\vdots&\vdots& &\vdots\\   
		\sum\limits_{j=0}^{\ell}b_{k-1,j}f_{j,0}^{(k)}&\sum\limits_{j=0}^{\ell}b_{k-1,j}f_{j,1}^{(k)}&\cdots&1+\sum\limits_{j=0}^{\ell}b_{k-1,j}f_{j,k-1}^{(k)}
	\end{pmatrix}\begin{pmatrix}
		1&\cdots&1\\
		\alpha_{1}&&\alpha_{k}\\
		\vdots& &\vdots\\
		\alpha_{1}^{k-1}&&\alpha_{k}^{k-1}\\
	\end{pmatrix}\right)\\
	=&\prod\limits_{i=1}^{k}v_i\cdot
	\det\left(\boldsymbol{I}_k+\widetilde{\B_1}\boldsymbol{F}_{1}\right)\cdot\prod\limits_{1\leq j<i\leq k}\left(\alpha_{i}-\alpha_{j}\right),
\end{aligned}$$
where
$$
\widetilde{\B_1}=\begin{pmatrix}
	b_{0,0}&\cdots&b_{0,\ell}\\
	\vdots& &\vdots\\
	b_{k-1,0}&\cdots&b_{k-1,\ell}
\end{pmatrix}, \boldsymbol{F}_{1}=\begin{pmatrix}
	f_{0,0}^{(k)}&\cdots&f_{0,k-1}^{(k)}\\
	\vdots& &\vdots\\
	f_{\ell,0}^{(k)}&\cdots&f_{\ell,k-1}^{(k)}
\end{pmatrix}.
$$

{\bf The calculation of $\det(\G_{2i})$.}
$$\begin{aligned}
	\det(\G_{2i})=&\prod\limits_{i=1}^{k-1}v_i\cdot\det\begin{pmatrix}
		1+\sum\limits_{j=0}^{\ell}b_{0,j}\alpha_{1}^{k+j}&\cdots&1+\sum\limits_{j=0}^{\ell}b_{0,j}\alpha_{k-1}^{k+j}&0\\
		\alpha_{1}+\sum\limits_{j=0}^{\ell}b_{1,j}\alpha_{1}^{k+j}&\cdots&\alpha_{k-1}+\sum\limits_{j=0}^{\ell}b_{1,j}\alpha_{k-1}^{k+j}&0\\ 
		\vdots& &\vdots\\ 
		\alpha_{1}^{k-3}+\sum\limits_{j=0}^{\ell}b_{k-3,j}\alpha_{1}^{k+j}&\cdots&\alpha_{k-1}^{k-3}+\sum\limits_{j=0}^{\ell}b_{k-3,j}\alpha_{k-1}^{k+j}&0\\  
		\alpha_{1}^{k-2}+\sum\limits_{j=0}^{\ell}b_{k-2,j}\alpha_{1}^{k+j}&\cdots&\alpha_{k-1}^{k-2}+\sum\limits_{j=0}^{\ell}b_{k-2,j}\alpha_{k-1}^{k+j}&a_{1i}\\
		\alpha_{1}^{k-1}+\sum\limits_{j=0}^{\ell}b_{k-1,j}\alpha_{1}^{k+j}&\cdots&\alpha_{k-1}^{k-1}+\sum\limits_{j=0}^{\ell}b_{k-1,j}\alpha_{k-1}^{k+j}&a_{2i}
	\end{pmatrix}\\
	=&\prod\limits_{i=1}^{k-1}v_i\cdot\left(a_{2i}\cdot\det(\G_{2i}^{(1)})-a_{1i}\cdot\det(\G_{2i}^{(2)})\right),\end{aligned}$$
where
$$\begin{aligned}
	\det(\G_{2i}^{(1)})=&\det\begin{pmatrix}
		1+\sum\limits_{j=0}^{\ell}b_{0,j}\alpha_{1}^{k+j}&\cdots&1+\sum\limits_{j=0}^{\ell}b_{0,j}\alpha_{k-1}^{k+j}\\
		\alpha_{1}+\sum\limits_{j=0}^{\ell}b_{1,j}\alpha_{1}^{k+j}&\cdots&\alpha_{k-1}+\sum\limits_{j=0}^{\ell}b_{1,j}\alpha_{k-1}^{k+j}\\ 
		\vdots& &\vdots\\ 
		\alpha_{1}^{k-3}+\sum\limits_{j=0}^{\ell}b_{k-3,j}\alpha_{1}^{k+j}&\cdots&\alpha_{k-1}^{k-3}+\sum\limits_{j=0}^{\ell}b_{k-3,j}\alpha_{k-1}^{k+j}\\  
		\alpha_{1}^{k-2}+\sum\limits_{j=0}^{\ell}b_{k-2,j}\alpha_{1}^{k+j}&\cdots&\alpha_{k-1}^{k-2}+\sum\limits_{j=0}^{\ell}b_{k-2,j}\alpha_{k-1}^{k+j} 
	\end{pmatrix}\\
	=& \det\begin{pmatrix}
		1+\sum\limits_{j=0}^{\ell}b_{0,j}\sum\limits_{s=0}^{k-2}f_{j+1,s}^{(k-1)}\alpha_{1}^{s}&\cdots&1+\sum\limits_{j=0}^{\ell}b_{0,j}\sum\limits_{s=0}^{k-2}f_{j+1,s}^{(k-1)}\alpha_{k-1}^{s}\\
		\alpha_{1}+\sum\limits_{j=0}^{\ell}b_{1,j}\sum\limits_{s=0}^{k-2}f_{j+1,s}^{(k-1)}\alpha_{1}^{s}&\cdots&\alpha_{k-1}+\sum\limits_{j=0}^{\ell}b_{1,j}\sum\limits_{s=0}^{k-2}f_{j+1,s}^{(k-1)}\alpha_{k-1}^{s}\\ 
		\vdots& &\vdots\\   
		\alpha_{1}^{k-2}+\sum\limits_{j=0}^{\ell}b_{k-2,j}\sum\limits_{s=0}^{k-2}f_{j+1,s}^{(k-1)}\alpha_{1}^{s}&\cdots&\alpha_{k-1}^{k-2}+\sum\limits_{j=0}^{\ell}b_{k-2,j}\sum\limits_{s=0}^{k-2}f_{j+1,s}^{(k-1)}\alpha_{k-1}^{s}
	\end{pmatrix}\\
	=&\det\left(\begin{pmatrix}
		1+\sum\limits_{j=0}^{\ell}b_{0,j}f_{j+1,0}^{(k-1)}&\sum\limits_{j=0}^{\ell}b_{0,j}f_{j+1,1}^{(k-1)}&\cdots&\sum\limits_{j=0}^{\ell}b_{0,j}f_{j+1,k-2}^{(k-1)}\\
		\sum\limits_{j=0}^{\ell}b_{1,j}f_{j+1,0}^{(k-1)}&1+\sum\limits_{j=0}^{\ell}b_{1,j}f_{j+1,1}^{(k-1)}&\cdots&\sum\limits_{j=0}^{\ell}b_{1,j}f_{j+1,k-2}^{(k-1)}\\ 
		\vdots&\vdots& &\vdots\\   
		\sum\limits_{j=0}^{\ell}b_{k-2,j}f_{j+1,0}^{(k-1)}&\sum\limits_{j=0}^{\ell}b_{k-2,j}f_{j+1,1}^{(k-1)}&\cdots&1+\sum\limits_{j=0}^{\ell}b_{k-2,j}f_{j+1,k-2}^{(k-1)}
	\end{pmatrix}\begin{pmatrix}
	1&\cdots&1\\
	\alpha_{1}&&\alpha_{k-1}\\
	\vdots& &\vdots\\
	\alpha_{1}^{k-2}&&\alpha_{k-1}^{k-2}\\
	\end{pmatrix}\right)\\ 
	=&
	\det\left(\boldsymbol{I}_{k-1}+\widetilde{\B_2}\boldsymbol{F}_{2}\right)\cdot\prod\limits_{1\leq j<i\leq k-1}\left(\alpha_{i}-\alpha_{j}\right)
\end{aligned}$$
with
$$
\widetilde{\B_2}=\begin{pmatrix}
	b_{0,0}&\cdots&b_{0,\ell}\\
	\vdots& &\vdots\\
	b_{k-2,0}&\cdots&b_{k-2,\ell}
\end{pmatrix}, \boldsymbol{F}_{2}=\begin{pmatrix}
	f_{1,0}^{(k-1)}&\cdots&f_{1,k-2}^{(k-1)}\\
	\vdots& &\vdots\\
	f_{\ell+1,0}^{(k-1)}&\cdots&f_{\ell+1,k-2}^{(k-1)}
\end{pmatrix},
$$
and
$$\begin{aligned} 	&\det(\G_{2i}^{(2)})\\
=&\det\begin{pmatrix}
		1+\sum\limits_{j=0}^{\ell}b_{0,j}\alpha_{1}^{k+j}&\cdots&1+\sum\limits_{j=0}^{\ell}b_{0,j}\alpha_{k-1}^{k+j}\\
		\alpha_{1}+\sum\limits_{j=0}^{\ell}b_{1,j}\alpha_{1}^{k+j}&\cdots&\alpha_{k-1}+\sum\limits_{j=0}^{\ell}b_{1,j}\alpha_{k-1}^{k+j}\\ 
		\vdots& &\vdots\\ 
		\alpha_{1}^{k-3}+\sum\limits_{j=0}^{\ell}b_{k-3,j}\alpha_{1}^{k+j}&\cdots&\alpha_{k-1}^{k-3}+\sum\limits_{j=0}^{\ell}b_{k-3,j}\alpha_{k-1}^{k+j}\\  
		\alpha_{1}^{k-1}+\sum\limits_{j=0}^{\ell}b_{k-1,j}\alpha_{1}^{k+j}&\cdots&\alpha_{k-1}^{k-1}+\sum\limits_{j=0}^{\ell}b_{k-1,j}\alpha_{k-1}^{k+j} 
	\end{pmatrix}\\
	=& \det\begin{pmatrix}
		1+\sum\limits_{j=0}^{\ell}b_{0,j}\sum\limits_{s=0}^{k-2}f_{j+1,s}^{(k-1)}\alpha_{1}^{s}
		&\cdots&
		1+\sum\limits_{j=0}^{\ell}b_{0,j}\sum\limits_{s=0}^{k-2}f_{j+1,s}^{(k-1)}\alpha_{k-1}^{s}\\
		\alpha_{1}+\sum\limits_{j=0}^{\ell}b_{1,j}\sum\limits_{s=0}^{k-2}f_{j+1,s}^{(k-1)}\alpha_{1}^{s}
		&\cdots&
		\alpha_{k-1}+\sum\limits_{j=0}^{\ell}b_{1,j}\sum\limits_{s=0}^{k-2}f_{j+1,s}^{(k-1)}\alpha_{k-1}^{s}\\
		\vdots&&\vdots\\ 
		\alpha_{1}^{k-3}+\sum\limits_{j=0}^{\ell}b_{k-3,j}\sum\limits_{s=0}^{k-2}f_{j+1,s}^{(k-1)}\alpha_{1}^{s}
		&\cdots&
		\alpha_{k-1}^{k-3}+\sum\limits_{j=0}^{\ell}b_{k-3,j}\sum\limits_{s=0}^{k-2}f_{j+1,s}^{(k-1)}\alpha_{k-1}^{s}\\
		\sum\limits_{s=0}^{k-2}f_{0,s}^{(k-1)}\alpha_{1}^{s}
		+\sum\limits_{j=0}^{\ell}b_{k-1,j}\sum\limits_{s=0}^{k-2}f_{j+1,s}^{(k-1)}\alpha_{1}^{s}
		&\cdots&
		\sum\limits_{s=0}^{k-2}f_{0,s}^{(k-1)}\alpha_{k-1}^{s}
		+\sum\limits_{j=0}^{\ell}b_{k-1,j}\sum\limits_{s=0}^{k-2}f_{j+1,s}^{(k-1)}\alpha_{k-1}^{s}
	\end{pmatrix}\\
	=&\det\begin{pmatrix}
		1+\sum\limits_{j=0}^{\ell}b_{0,j}f_{j+1,0}^{(k-1)}
		&\sum\limits_{j=0}^{\ell}b_{0,j}f_{j+1,1}^{(k-1)}
		&\cdots&
		\sum\limits_{j=0}^{\ell}b_{0,j}f_{j+1,k-3}^{(k-1)}&
		\sum\limits_{j=0}^{\ell}b_{0,j}f_{j+1,k-2}^{(k-1)}\\
		\sum\limits_{j=0}^{\ell}b_{1,j}f_{j+1,0}^{(k-1)}
		&1+\sum\limits_{j=0}^{\ell}b_{1,j}f_{j+1,1}^{(k-1)}
		&\cdots&
		\sum\limits_{j=0}^{\ell}b_{1,j}f_{j+1,k-3}^{(k-1)}&
		\sum\limits_{j=0}^{\ell}b_{1,j}f_{j+1,k-2}^{(k-1)}\\
		\vdots&\vdots& & \vdots&\vdots\\ 
		\sum\limits_{j=0}^{\ell}b_{k-3,j}f_{j+1,0}^{(k-1)}
		&\sum\limits_{j=0}^{\ell}b_{k-3,j}f_{j+1,1}^{(k-1)}
		&\cdots
		& 1+\sum\limits_{j=0}^{\ell}b_{k-3,j}f_{j+1,k-3}^{(k-1)}&\sum\limits_{j=0}^{\ell}b_{k-3,j}f_{j+1,k-2}^{(k-1)}\\ 
		f_{0,0}^{(k-1)}+\sum\limits_{j=0}^{\ell}b_{k-1,j}f_{j+1,0}^{(k-1)}
		&f_{0,1}^{(k-1)}+\sum\limits_{j=0}^{\ell}b_{k-1,j}f_{j+1,1}^{(k-1)}
		&\cdots&
		f_{0,k-3}^{(k-1)}+\sum\limits_{j=0}^{\ell}b_{k-1,j}f_{j+1,k-3}^{(k-1)}&
		f_{0,k-2}^{(k-1)}+\sum\limits_{j=0}^{\ell}b_{k-1,j}f_{j+1,k-2}^{(k-1)}
	\end{pmatrix}\\
	&\cdot\det\begin{pmatrix}
		1&\cdots&1\\
		\alpha_{1}&&\alpha_{k-1}\\
		\vdots&&\vdots\\
		\alpha_{1}^{k-3}&&\alpha_{k-1}^{k-3}\\
		\alpha_{1}^{k-2}&&\alpha_{k-1}^{k-2}
	\end{pmatrix}\\
	=&\det\left(\boldsymbol{T}+\widetilde{\B_3}\boldsymbol{F}_{2}\right)\cdot\prod\limits_{1\leq j<i\leq k-1}\left(\alpha_{i}-\alpha_{j}\right)
\end{aligned}$$
with
$$
\boldsymbol{T}=\begin{pmatrix}
1&0&\cdots&0&0\\
0&1&\cdots&0&0\\
\vdots&\vdots& &\vdots&\vdots\\
0&0&\cdots&1&0\\
f_{0,0}^{(k-1)}&f_{0,1}^{(k-1)}&\cdots&f_{0,k-3}^{(k-1)}&f_{0,k-2}^{(k-1)}\\
\end{pmatrix},\widetilde{\B_3}=\begin{pmatrix}
	b_{0,0}&\cdots&b_{0,\ell}\\
	\vdots& &\vdots\\
	b_{k-3,0}&\cdots&b_{k-3,\ell}\\
	b_{k-1,0}&\cdots&b_{k-1,\ell}
\end{pmatrix}, \boldsymbol{F}_{2}=\begin{pmatrix}
	f_{1,0}^{(k-1)}&\cdots&f_{1,k-2}^{(k-1)}\\
	\vdots& &\vdots\\
	f_{\ell+1,0}^{(k-1)}&\cdots&f_{\ell+1,k-2}^{(k-1)}
\end{pmatrix}.
$$

{\bf The calculation of $\det(\G_3)$.}
$$\begin{aligned}
	\det(\G_3)=&\prod\limits_{i=1}^{k-2}v_i\cdot\det\begin{pmatrix}
		1+\sum\limits_{j=0}^{\ell}b_{0,j}\alpha_{1}^{k+j}&\cdots&1+\sum\limits_{j=0}^{\ell}b_{0,j}\alpha_{k-2}^{k+j}&0&0\\
		\alpha_{1}+\sum\limits_{j=0}^{\ell}b_{1,j}\alpha_{1}^{k+j}&\cdots&\alpha_{k-2}+\sum\limits_{j=0}^{\ell}b_{1,j}\alpha_{k-2}^{k+j}&0&0\\ 
		\vdots& &\vdots\\   
		\alpha_{1}^{k-3}+\sum\limits_{j=0}^{\ell}b_{k-3,j}\alpha_{1}^{k+j}&\cdots&\alpha_{k-2}^{k-3}+\sum\limits_{j=0}^{\ell}b_{k-3,j}\alpha_{k-2}^{k+j}&0&0\\
		\alpha_{1}^{k-2}+\sum\limits_{j=0}^{\ell}b_{k-2,j}\alpha_{1}^{k+j}&\cdots&\alpha_{k-2}^{k-2}+\sum\limits_{j=0}^{\ell}b_{k-2,j}\alpha_{k-2}^{k+j}&a_{11}&a_{12}\\
		\alpha_{1}^{k-1}+\sum\limits_{j=0}^{\ell}b_{k-1,j}\alpha_{1}^{k+j}&\cdots&\alpha_{k-2}^{k-1}+\sum\limits_{j=0}^{\ell}b_{k-1,j}\alpha_{k-2}^{k+j}&a_{21}&a_{22}
	\end{pmatrix}\\
	=&\prod\limits_{i=1}^{k-2}v_i\cdot\det(\A_{2\times 2})\cdot\det\begin{pmatrix}
		1+\sum\limits_{j=0}^{\ell}b_{0,j}\sum\limits_{s=0}^{k-3}f_{j+2,s}^{(k-2)}\alpha_{1}^{s}&\cdots&1+\sum\limits_{j=0}^{\ell}b_{0,j}\sum\limits_{s=0}^{k-3}f_{j+2,s}^{(k-2)}\alpha_{k-2}^{s}\\
		\alpha_{1}+\sum\limits_{j=0}^{\ell}b_{1,j}\sum\limits_{s=0}^{k-3}f_{j+2,s}^{(k-2)}\alpha_{1}^{s}&\cdots&\alpha_{k-2}+\sum\limits_{j=0}^{\ell}b_{1,j}\sum\limits_{s=0}^{k-3}f_{j+2,s}^{(k-2)}\alpha_{k-2}^{s}\\ 
		\vdots& &\vdots\\   
		\alpha_{1}^{k-3}+\sum\limits_{j=0}^{\ell}b_{k-3,j}\sum\limits_{s=0}^{k-3}f_{j+2,s}^{(k-2)}\alpha_{1}^{s}&\cdots&\alpha_{k-2}^{k-3}+\sum\limits_{j=0}^{\ell}b_{k-3,j}\sum\limits_{s=0}^{k-3}f_{j+2,s}^{(k-2)}\alpha_{k-2}^{s}
	\end{pmatrix}\\
	=&\prod\limits_{i=1}^{k-2}v_i\cdot\det(\A_{2\times 2})\\
	&\cdot\det\left(\begin{pmatrix}
		1+\sum\limits_{j=0}^{\ell}b_{0,j}f_{j+2,0}^{(k-2)}&\sum\limits_{j=0}^{\ell}b_{0,j}f_{j+2,1}^{(k-2)}&\cdots&\sum\limits_{j=0}^{\ell}b_{0,j}f_{j+2,k-3}^{(k-2)}\\
		\sum\limits_{j=0}^{\ell}b_{1,j}f_{j+2,0}^{(k-2)}&1+\sum\limits_{j=0}^{\ell}b_{1,j}f_{j+2,1}^{(k-2)}&\cdots&\sum\limits_{j=0}^{\ell}b_{1,j}f_{j+2,k-3}^{(k-2)}\\ 
		\vdots&\vdots& &\vdots\\   
		\sum\limits_{j=0}^{\ell}b_{k-3,j}f_{j+2,0}^{(k-2)}&\sum\limits_{j=0}^{\ell}b_{k-3,j}f_{j+2,1}^{(k-2)}&\cdots&1+\sum\limits_{j=0}^{\ell}b_{k-3,j}f_{j+2,k-3}^{(k-2)}
	\end{pmatrix}\begin{pmatrix}
		1&\cdots&1\\
		\alpha_{1}&&\alpha_{k-2}\\
		\vdots& &\vdots\\
		\alpha_{1}^{k-3}&&\alpha_{k-2}^{k-3}\\
	\end{pmatrix}\right)\\
	=&
	\det\left(\boldsymbol{I}_{k-2}+\widetilde{\B_4}\boldsymbol{F}_{3}\right)\cdot\prod\limits_{i=1}^{k-2}v_i\cdot\prod\limits_{1\leq j<i\leq k-2}\left(\alpha_{i}-\alpha_{j}\right)\cdot\det(\A_{2\times 2}),
\end{aligned}$$
where
$$
\widetilde{\B_4}=\begin{pmatrix}
	b_{0,0}&\cdots&b_{0,\ell}\\
	\vdots& &\vdots\\
	b_{k-3,0}&\cdots&b_{k-3,\ell}
\end{pmatrix}, \boldsymbol{F}_{3}=\begin{pmatrix}
	f_{2,0}^{(k-2)}&\cdots&f_{2,k-3}^{(k-2)}\\
	\vdots& &\vdots\\
	f_{\ell+2,0}^{(k-2)}&\cdots&f_{\ell+2,k-3}^{(k-2)}
\end{pmatrix}.
$$
\section{Examples of Section \ref{sec5.1}}
\begin{example}
There exist numerous MDS $(\L,\P)$-TGRL codes. For brevity, we only list several parameter choices in the following Table \ref{MDSexamples} and fix $\v=\boldsymbol{1}$, which can be verified by Magma program. And $\omega$ denotes the default
primitive element of $\F_q$. 
\begin{table}[H]
\centering 
\caption{Examples for MDS $(\L,\P)$-TGRL codes}
\begin{tabular}{|c|c|c|c|c|c|c|c|}
	\hline
$q$&$n$&$k$&$\ell$&$\A_{2\times 2}$&$\B$&$\boldsymbol{\alpha}$&Parameters\\
	\hline
$3^4$&$7$&$4$&$1$&$ \begin{pmatrix}
1&1\\0&1
\end{pmatrix}$&$\begin{pmatrix}
1&1&0\\
0&1&0\\
2&1&0\\
2&1&0
\end{pmatrix}$&$\left(\omega^{77}, 0, \omega^{43}, \omega^{17}, \omega^{20}, \omega^{59}, \omega^{19}\right)$& $[9,4,6]_{3^4}$ \\
	\hline
$5^3$&$8$&$5$&$2$&$ \begin{pmatrix}
	2&2\\4&2
\end{pmatrix}$&$\begin{pmatrix}
	1&1&1\\
	1&1&1\\
	1&1&1\\
	1&1&1\\
	1&1&1
\end{pmatrix}$&$\left(\omega^{38},\omega^{46},\omega^{6},\omega^{123},\omega^{102},\omega^{97},\omega^{22},\omega^{111} \right)$&$[10,5,6]_{5^3}$ \\
	\hline
$7^3$&$7$&$4$&$2$&$ \begin{pmatrix}
	2&5\\4&1
\end{pmatrix}$&$\begin{pmatrix}
	1&1&1\\
	1&2&1\\
	1&1&2\\
	1&1&3
\end{pmatrix}$&$\left(\omega^{185}, \omega^{303}, \omega^{240}, \omega^{193}, \omega^{127}, \omega^{248}, \omega^{246} \right)$&$[9,4,6]_{7^3}$ \\
	\hline
$11^2$&$8$&$4$&$3$&$ \begin{pmatrix}
	1&5\\4&1
\end{pmatrix}$&$\begin{pmatrix}
2&1&1&2\\
1&2&1&2\\
1&1&2&3\\
1&1&3&5 
\end{pmatrix}$&$\left(\omega^{111},\omega^{19},\omega^{62},\omega^{83},\omega^{103}, 2 ,\omega^{17},\omega^{85} \right)$&$[10,4,7]_{11^2}$\\
	\hline
$13^2$&$8$&$4$&$3$&$ \begin{pmatrix}
	1&5\\2&1
\end{pmatrix}$&$\begin{pmatrix}
	2&1&1&2\\
	1&2&1&2\\
	1&1&2&3\\
	1&1&3&5 
\end{pmatrix}$&$\left(\omega^{90},\omega^{148},\omega^{118},\omega^{95},\omega^{63},\omega^{122},\omega^{121},\omega^{137} \right)$&$[10,4,7]_{13^2}$\\
	\hline
$17^3$&$8$&$4$&$3$&$ \begin{pmatrix}
	2&5\\2&7
\end{pmatrix}$&$\begin{pmatrix}
	2&1&1&2\\
	3&2&1&2\\
	1&4&2&3\\
	1&2&3&5 
\end{pmatrix}$&$\left(\omega^{2674}, \omega^{4355}, \omega^{3459}, \omega^{2774}, \omega^{1820}, \omega^{3551}, \omega^{3518}, \omega^{3983}  \right)$&$[10,4,7]_{17^3}$\\
	\hline
$19^2$&$8$&$4$&$3$&$ \begin{pmatrix}
	4&5\\2&4
\end{pmatrix}$&$\begin{pmatrix}
	2&1&1&2\\
	3&2&1&2\\
	1&4&2&3\\
	1&2&3&5
\end{pmatrix}$&$\left(\omega^{216}, \omega^{76}, \omega^{283}, \omega^{190}, \omega^{47}, \omega^{134}, \omega^{352}, \omega^{113} \right)$&$[10,4,7]_{19^2}$\\
	\hline
$23^2$&$8$&$4$&$3$&$ \begin{pmatrix}
	4&5\\2&4
\end{pmatrix}$&$\begin{pmatrix}
	2&1&1&2\\
	3&2&1&2\\
	1&4&2&3\\
	1&2&3&5 
\end{pmatrix}$&$\left(\omega^{22},\omega^{177},\omega^{135},\omega^{357},\omega^{40},\omega^{478},\omega^{502}, 13 \right)$&$[10,4,7]_{23^2}$ \\
	\hline
\end{tabular}
\label{MDSexamples} 
\end{table}
\end{example}
\section{An example for Theorem \ref{NMDS1CRS}}
\begin{example}
	Let $q=3^2$, $\F_q^*=\langle\omega\rangle$ with $\omega^2+2\omega+2=0$, $k=5$, $\lambda=\omega$,  $b=0$, $c=1$ and $\alpha_{i}=\omega^{i}(1\leq i\leq q-2)$. Then the corresponding $(1)$-column TGRS code has the following generator matrix
	$$\G_{1}(\omega,0,1)=\begin{pmatrix}
		1&1&1&1&1&1&1&1-\omega\\
		\omega&\omega^{2}&\omega^{3}&\omega^{4}&\omega^{5}&\omega^{6}&\omega^{7}&-\omega\\
		\omega^2&\left(\omega^{2}\right)^{2}&\left(\omega^{3}\right)^{2}&\left(\omega^{4}\right)^{2}&\left(\omega^{5}\right)^{2}&\left(\omega^{6}\right)^{2}&\left(\omega^{7}\right)^{2}&-\omega\\
		\omega^3&\left(\omega^{2}\right)^{3}&\left(\omega^{3}\right)^{3}&\left(\omega^{4}\right)^{3}&\left(\omega^{5}\right)^{3}&\left(\omega^{6}\right)^{3}&\left(\omega^{7}\right)^{3}&-\omega\\
		\omega^4&\left(\omega^{2}\right)^{4}&\left(\omega^{3}\right)^{4}&\left(\omega^{4}\right)^{4}&\left(\omega^{5}\right)^{4}&\left(\omega^{6}\right)^{4}&\left(\omega^{7}\right)^{4}&-\omega\\
	\end{pmatrix}.
	$$
	Based on Magma program, the linear code generated by $\G_{1}(\omega,0,1)$ is NMDS with the parements $[8, 5, 3]_{9}$ and the weight distribution is
	$$A(x)=1+32x^3+400x^4+2560x^5+9984x^6+23072x^7+23000x^8.$$  which is consistent with the result given in Theorem \ref{NMDS1CRS}, i.e.,
	$$
	\begin{aligned}
		A_{3}=&\sum_{i\mid 8}
		(-1)^{4+\left\lfloor \frac{4}{i}\right\rfloor}
		\binom{\frac{8}{i}-1}{\left\lfloor \frac{4}{i}\right\rfloor}
		\sum_{d\mid \gcd(i,\gcd(1,8))}
		d\mu\left(\frac{i}{d}\right)\\
		=&\sum_{i\mid 8}
		(-1)^{4+\left\lfloor \frac{4}{i}\right\rfloor}
		\binom{\frac{8}{i}-1}{\left\lfloor \frac{4}{i}\right\rfloor}\mu\left(i\right)\\
		=&(-1)^{4+\left\lfloor \frac{4}{1}\right\rfloor}
		\binom{\frac{8}{1}-1}{\left\lfloor \frac{4}{1}\right\rfloor}\mu\left(1\right)+(-1)^{4+\left\lfloor \frac{4}{2}\right\rfloor}
		\binom{\frac{8}{2}-1}{\left\lfloor \frac{4}{2}\right\rfloor}\mu\left(2\right)+(-1)^{4+\left\lfloor \frac{4}{4}\right\rfloor}
		\binom{\frac{8}{4}-1}{\left\lfloor \frac{4}{4}\right\rfloor}\mu\left(4\right)+(-1)^{4+\left\lfloor \frac{4}{8}\right\rfloor}
		\binom{\frac{8}{8}-1}{\left\lfloor \frac{4}{8}\right\rfloor}\mu\left(8\right)\\
		=&(-1)^{8}
		\binom{7}{4}\mu\left(1\right)+(-1)^{6}
		\binom{3}{2}\mu\left(2\right)+(-1)^{5}
		\binom{1}{1}\mu\left(4\right)+(-1)^{4}
		\binom{-1}{0}\mu\left(8\right)\\
		=&32.
	\end{aligned}
	$$ 
\end{example}
\end{document}